\documentclass{article}
\usepackage{graphicx}
\usepackage{amssymb}

\usepackage{makeidx}
\usepackage{graphicx}
\usepackage{multicol}
\newcommand{\be}{\begin{eqnarray}}

\newcommand{\ee}{\end{eqnarray}}
\newcommand{\x}{{\rm x}}
\newcommand{\y}{{\rm y}}

\makeindex

\newcommand{\nn}{\nonumber\\ }
\newcommand{\beq}{\begin{eqnarray}}
\newcommand{\eeq}{\end{eqnarray}}
\def\labe{\label}
\def\simge{\mathrel{%
   \rlap{\raise 0.511ex \hbox{$>$}}{\lower 0.511ex \hbox{$\sim$}}}}
\def\simle{\mathrel{
   \rlap{\raise 0.511ex \hbox{$<$}}{\lower 0.511ex \hbox{$\sim$}}}}
\def\bigs{\mathrel{
   \rlap{\raise 0.531ex \hbox{$>$}}{\lower 0.531ex \hbox{$<$}}}}

\def\grad{\nabla}                               
\def\del{\partial}                              

\begin{document}

\setcounter{footnote}{0}
\centerline{\hfill\protect\raisebox{10pt}{\tt SACLAY-T02/024}}
\renewcommand{\thefootnote}{\fnsymbol{footnote}}
\begin{flushleft}\large\bf
{THE COLOUR GLASS CONDENSATE: \\ AN 
INTRODUCTION\footnote{Lectures given at the NATO Advanced Study
Institute ``QCD perspectives on hot and dense matter'', August 6--18, 2001,
in Carg\`ese, Corsica, France} } 
\end{flushleft}
\vspace{5pt}
\begin{quote}
{\large\sc Edmond Iancu}\\[2pt]
{\small Service de Physique Th\'eorique\\[-1pt] CE Saclay,
        F-91191 Gif-sur-Yvette, France}
\vspace{5pt}

{\large\sc Andrei Leonidov}\\[2pt]
{\small P. N. Lebedev Physical Institute, Moscow, Russia}
\vspace{5pt}

{\large\sc Larry McLerran}\\[2pt]
{\small Nuclear Theory Group, Brookhaven National Laboratory\\[-1pt]
        Upton, NY 11793, USA} 
\end{quote}
\vspace{5pt}
\begin{quote}\small
{\bf Abstract:}
In these lectures, we develop the theory of the Colour Glass
Condensate. This is the matter made of gluons in the high density
environment characteristic of deep inelastic scattering or
hadron-hadron collisions at very high energy.
The lectures are self contained and comprehensive.  They start with a
phenomenological introduction, develop the theory of classical gluon
fields appropriate for the Colour Glass, and end with a derivation and
discussion of the renormalization group equations which determine this
effective theory. 
\end{quote}

\vspace{5pt}

\renewcommand{\thefootnote}{\arabic{footnote}}
\setcounter{footnote}{0}

\section{General Considerations}
\setcounter{equation}{0}
\label{INTRO}

\subsection{Introduction}

The goal of these lectures 
is to convince you that the  average properties of hadronic
interactions at very high energies are
controlled by a new form
of matter, a dense condensate of gluons.
This is called the Colour Glass Condensate since
\begin{itemize}
\item {\sf Colour}:  The gluons are coloured.
\item {\sf Glass}:  The associated fields
evolve very slowly relative to natural time scales, and are disordered.  This 
is like a glass which is disordered and is a liquid on long time scales but
seems to be a solid on short time scales.
\item {\sf Condensate}:
 There is a very high density of massless gluons.  These 
gluons can be packed until their phase space density is so high that 
interactions prevent more gluon occupation. With increasing energy,
this forces the gluons to occupy higher momenta, so that the coupling
becomes weak.  The gluon density saturates at a value of order
$1/\alpha_s \gg 1$, corresponding to a multiparticle state which
is a Bose condensate.
\end{itemize}

In these lectures, we will try to explain why the above is very plausible.

Before doing this, however, it is useful to review some of the
typical features of hadronic interactions, and some unanswered
theoretical questions  which are associate with these phenomena.  This will
motivate much of the later discussion.

\subsection{Total Cross Sections at Asymptotic Energy}

Computing total cross sections as $E \rightarrow \infty$ is one of the
great unsolved problems of QCD.  
Unlike for processes which are computed in perturbation theory,
it is not required that any energy transfer become large as the total 
collision energy $E \rightarrow \infty$.  Computing a total cross section for 
hadronic scattering therefore appears to be intrinsically non-perturbative.
In the 60's and early 70's, Regge
theory was extensively developed in an attempt to understand the total
cross section.  The results of this analysis were to our mind
inconclusive, and certainly can not be claimed to be a first principles
understanding from QCD.

\begin{figure}
\begin{center}
\includegraphics[width=\textwidth] {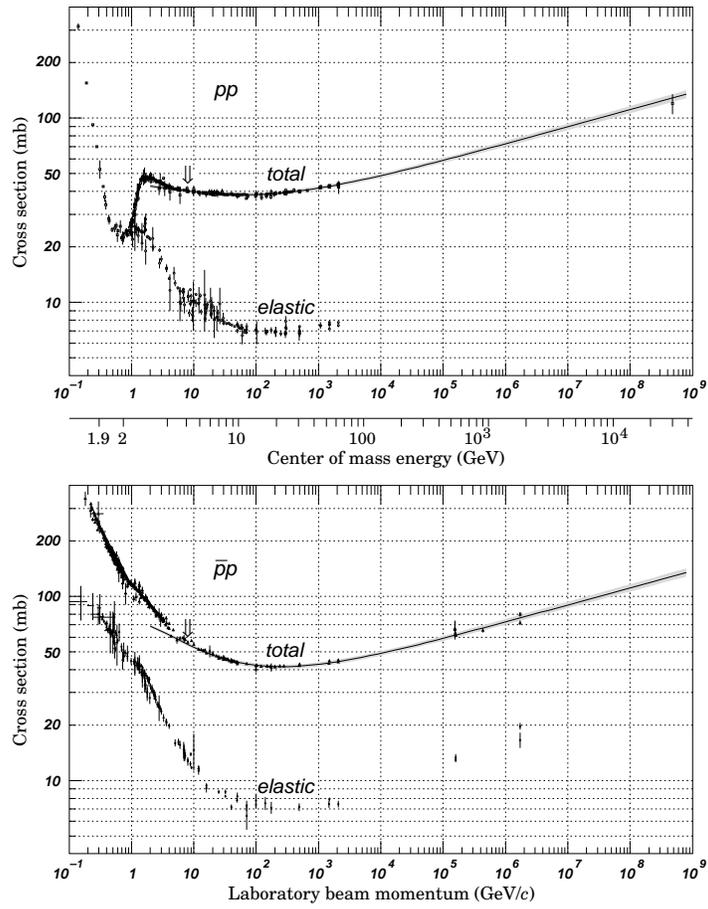}
\vspace{-1in}
\caption{The cross sections for $pp$ and $p\overline p$ scattering.}
\label{crosssection}
\end{center}
\end{figure}
The total cross section for  $pp$ and 
$\overline p p$ collisions is shown in Fig.
\ref{crosssection}.  
Typically, it is assumed that the total cross section grows as $\ln^2
E$ as $E \rightarrow \infty$.  This is the so called Froissart bound,
which corresponds to the maximal growth allowed by the unitarity of the
scattering matrix. Is this correct?  Is the coefficient of $\ln^2 E$ universal
for all hadronic precesses?  Why is the unitarity limit saturated?  Can
we understand the total cross section from first principles in QCD?  Is
it understandable in weakly coupled QCD, or is it an intrinsically
non-perturbative phenomenon?

\subsection{Particle Production in High Energy Collisions}

In order to discuss particle production, 
it is useful to introduce some  kinematical variables
adapted for high energy collisions:
the light cone coordinates. Let $z$ be the longitudinal 
axis of the collision. For an arbitrary 4-vector
$v^\mu=(v^0,v^1,v^2,v^3)$ ($v^3=v_z$, etc.), 
we define its light-cone (LC) coordinates as
\be
v^+\equiv {1 \over \sqrt{2}} (v^0+v^3),\qquad
v^-\equiv {1 \over \sqrt{2}} (v^0-v^3),\qquad v_\perp\equiv (v^1,v^2).
\ee
In particular, we shall refer to $x^+=(t+z)/\sqrt{2}$ 
as the LC ``time'', and to $x^-=(t-z)/\sqrt{2}$ 
as the LC ``longitudinal coordinate''. The
invariant dot product reads:
\be
        p \cdot x \,= \,p^- x^++ p^+x^-  - p_\perp \cdot x_\perp,
\ee
which suggests that $p^-$ --- the momentum
variable conjugate to the ``time'' $x^+$ --- should be
interpreted as the LC energy,
and $p^+$ as the (LC) longitudinal momentum. In particular,
for particles on the mass-shell: $p^\pm = (E \pm p_z)/\sqrt{2}$,
with $E=(m^2+{\bf p}^2)^{1/2}$, and therefore:
\be
p^+p^- \,= \, {1 \over 2} (E^2 - p_z^2) = {1 \over 2} (p_\perp^2 +m^2) \,= \,
 {1 \over 2} \, m_\perp^2 \,.
\ee 
This equation defines the transverse mass $m_\perp$. 
We shall also need the {\it rapidity} :
\be\label{y-DEF}
        \y \,\equiv\, {1 \over 2} 
\ln\frac{p^+}{p^-} = {1 \over 2} \ln{2p^{+2}\over m_\perp^2}.
\ee
These definitions are useful, among other reasons, because of their simple
properties under longitudinal Lorentz boosts:  $p^+ \rightarrow
\kappa p^+$, $p^- \rightarrow (1/\kappa)p^-$,
where $\kappa$ is a constant.  Under boosts,
the rapidity is just shifted by a constant: $\y\rightarrow
\y+ \kappa$.

\begin{figure} 
\begin{center} 
\includegraphics[width=0.5\textwidth]
{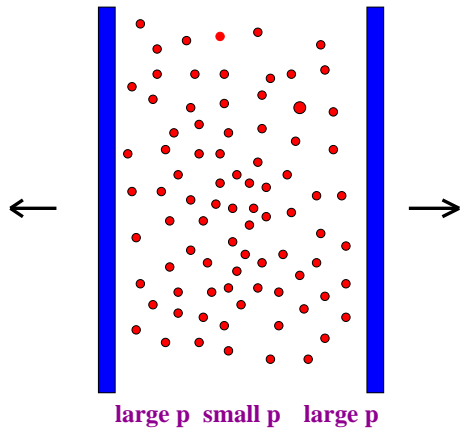} 
\caption{A hadron-hadron collision.  The produced particles are shown
as circles. } 
\label{collision} 
\end{center}
\end{figure}

Consider now the collision of two identical hadrons
in the center of mass frame, as shown in Fig. 
\ref{collision}.  In this figure, we have assumed that the colliding hadrons
have a transverse extent which is large compared to the size of the 
produced particles.  This is true for
nuclei, or if the typical transverse momenta of the produced particles is
large compared to $\Lambda_{QCD}$, since the corresponding size will be
much smaller than a Fermi.  We have also assumed that the colliding particles
have an energy which is large enough so that they pass through one another
and produce mesons in their wake.  This is known to happen experimentally:
the particles which carry the quantum numbers of the colliding particles
typically lose only some finite fraction of their momenta in the collision.
Because of their large energy, the incoming hadrons propagate nearly 
at the speed of light, and therefore are Lorentz contracted in the 
longitudinal direction, as suggested by the figure.

In LC coordinates, the right moving particle (``the projectile'') has 
a 4-momentum $p_1^\mu=(p_1^+,p_1^-,0_\perp)$
with $p^+_1\simeq \sqrt{2} p_z$  and
$p_1^- = M^2/2p^+_1$
(since $p_z\gg M$, with $M=$ the projectile mass). 
Similarly, for the left moving hadron (``the target''),
we have $p^+_2 = p^-_1$ and $p^-_2 = p^+_1$.
The invariant energy squared is $s=(p_1+p_2)^2=2p_1\cdot p_2
\simeq 2 p^+_1 p^-_2 \simeq 4p_z^2$, and coincides, at it should, with
the total energy squared $(E_1+E_2)^2$ in the center of mass frame.

We define the longitudinal momentum fraction,
or Feynman's $\x$, of a produced pion as
\be
       \x \,\equiv\,\frac{p^+_\pi}{p_1^+}
\ee
(with $0 < \x \le 1$). The {rapidity} of the pion is then
\be
        \y \,=\, {1 \over 2} 
\ln{p^+_\pi \over p^-_\pi} = {1 \over 2} \ln
{2p^{+2}_\pi\over m_\perp^2}\,=\,
\y_{proj} - \ln{1\over \x} + \ln{M\over m_\perp},
\ee
where $\y_{proj} = \ln(\sqrt{2}{p^+_1/M})=\ln(\sqrt{s}/M)$.
The pion rapidity is in the range $-\y_{proj} \simle \y \simle \y_{proj}$
(up to an overall shift by $\Delta y =\ln(M/m_\perp)$).

\begin{figure} 
\begin{center} 
\includegraphics[width=0.5\textwidth]
{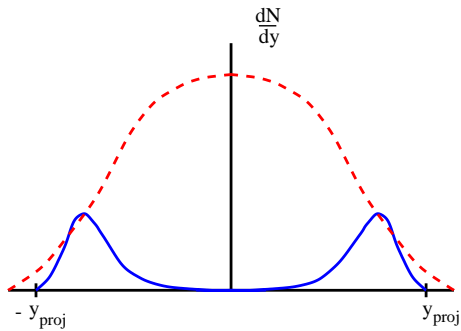} 
\caption{The rapidity distribution of particles produced in a hadronic 
collision. } 
\label{dndy} 
\end{center}
\end{figure}

A typical distribution of produced particles (say,
pions) in a hadronic collision is shown in 
Fig. \ref{dndy}. We denote by $dN/d{\rm y}$ 
the number of produced particles per unit rapidity.
The leading particles are shown by the solid line and are clustered
around the projectile and target rapidities.  For example, in a heavy ion
collision, this is where the nucleons would be.  The dashed
line is the distribution of produced mesons.
Several theoretical issues arise in multiparticle production:

Can we
compute $dN/d{\rm y}$ ?  Or even $dN/d{\rm y}$ at $\y = 0$ (``central
rapidity'') ? How does the average transverse
momentum of produced particles $\langle p_\perp\rangle$ 
behave with energy?  What is the
ratio of produced strange/nonstrange mesons, and corresponding ratios of
charm, top, bottom etc at $\y = 0$ as the center of mass energy
approaches infinity?
Does multiparticle production as $s \rightarrow \infty$ at
$\y = 0$ become simple, understandable and computable?

{ Note that $\y = 0$ corresponds to particles with $p_z=0$ or
$p^+=m_\perp/\sqrt{2}$, for which 
$\x=m_\perp/(\sqrt{2}p^+_1) = m_\perp/\sqrt{s}$ 
is small, $\x\ll 1$, in the high-energy limit of interest.
Thus, presumably, the multiparticle production at central
rapidity reflects properties of the  small-x degrees of freedom
in the colliding hadron wavefunctions.}

\begin{figure} 
\begin{center} 
\includegraphics[width=0.5\textwidth]
{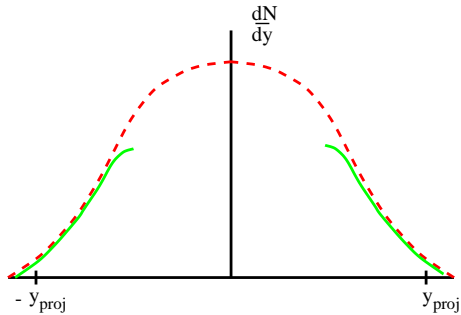} 
\caption{Feynman scaling of rapidity distributions. The two different lines
correspond to rapidity distributions at different energies.} 
\label{feynman} 
\end{center}
\end{figure}

There is a remarkable feature of rapidity distributions of produced 
hadrons, 
which we shall refer to as Feynman scaling.  If we plot rapidity distributions
of produced hadrons at different energies, then as function of 
$\y-\y_{proj}$,
the rapidity distributions are to a good
approximation independent of energy.  
This is illustrated in Fig. \ref{feynman}, where the rapidity distribution
measured at one energy is shown with a solid line and the rapidity
distribution at a different, higher,
 energy is shown with a dotted line.  (In this plot,
the rapidity distribution at the lower energy has been shifted by an amount
so that particles of positive rapidity begin their distribution at 
the same $\y_{proj}$ as the high energy particles, and correspondingly
for the negative rapidity particles.  This of course leads to a gap
in the center for the low energy particles due to this mapping.)

This means that as we go to higher and higher energies, the new physics is 
associated with the additional degrees of freedom at small rapidities
in the center of mass frame (small-x degrees of freedom).  
The large x degrees of freedom do not change much.
This suggests that there may be some sort of renormalization group description
in rapidity where the degrees of freedom at larger x are held fixed as we go to
smaller values of x.  We shall see that in fact these large x degrees of
freedom act as sources for the small x degrees of freedom, and the
renormalization group is generated by integrating out degrees of 
freedom at relatively large x to generate these sources.

\subsection{Deep Inelastic Scattering}
\label{SECT-DIS}

\begin{figure}
\begin{center}
\includegraphics[width=0.5\textwidth] {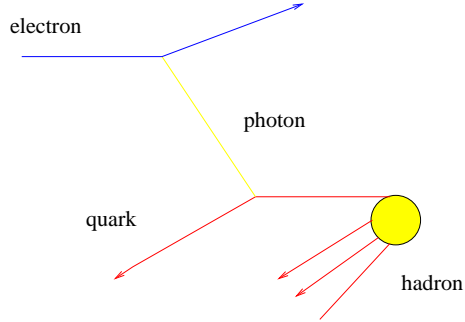}   
\caption{Deep inelastic scattering of an electron on a hadron.}
\label{electron}
\end{center}
\end{figure}

In Fig. \ref{electron},  deep inelastic scattering is shown. 
Here an
electron emits a virtual photon which scatters 
from a quark in a hadron.  The momentum and energy
transfer of the electron is measured, but the results of the 
hadron break up
are not.  In these lectures, we do not have sufficient time to
develop the theory of deep
inelastic scattering (see, e.g., \cite{TB-DIS} for more
details).  For the present purposes, it is enough to say that, 
at large momentum transfer $Q^2\gg \Lambda_{QCD}^2$, 
this experiment can be used
to measure the distributions of quarks in the hadron.

To describe the quark distributions, it is convenient to work in a
reference frame where the hadron has a large light-cone
longitudinal momentum
$P^+\gg M$ (``infinite momentum frame'').  In this frame, one can
describe the hadron as a collection of constituents (``partons''),
which are nearly on-shell excitations carrying some fraction
x of the total longitudinal momentum $P^+$.
Thus, the longitudinal momentum of a parton is $p^+=\x P^+$, 
with $\,0\le \x  <1$.

For the struck quark in Fig. \ref{electron}, this x variable
(``Feynman's x'') is equal to the Bjorken variable $\x_{Bj}$, 
which is defined in a frame independent way as
$\x_{Bj} = Q^2/2P\cdot q$, and is directly measured in the experiment.
In this definition, $Q^2=-q^\mu q_\mu$, with $q^\mu$ the 
(space-like) 4-momentum of the exchanged photon. 
The condition that $\x=\x_{Bj}$ is what maximizes the spatial
overlap between the struck quark and the  virtual photon,
thus making the interaction favourable.

The Bjorken variable scales like $\x_{Bj}\sim Q^2/s$, with
$s=$ the invariant energy squared. Thus,  in deep inelastic
scattering at high energy (large $s$ at fixed $Q^2$)
one measures quark distributions $dN_{quark}/d\x$ at small x
($\x\ll 1$).

\begin{figure} 
\begin{center} 
\includegraphics[width=0.5\textwidth]
{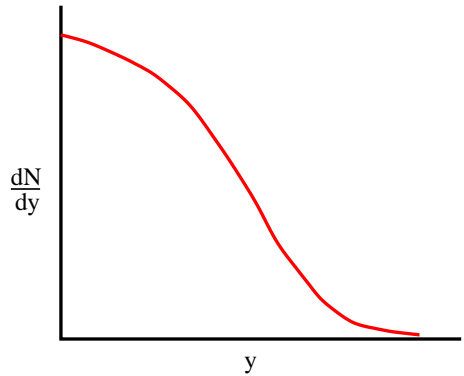} 
\caption{The rapidity distribution of gluons inside of a hadron. } 
\label{dndya} 
\end{center}
\end{figure}

It is useful to think about these distributions as a function of rapidity.
We define the rapidity in deep inelastic scattering as
\be
        \y = \y_{hadron} - \ln(1/\x),
\ee
and the invariant rapidity distribution as
\be
        {dN\over d{\rm y}}\, = {\,\x {dN\over d{\rm x}}}\,.
\ee
In Fig. \ref{dndya},  
a typical $dN/d{\rm y}$ distribution for  constituent gluons
of a hadron
is shown.  This plot is similar to the rapidity distribution of produced
particles in hadron-hadron collisions (see Fig.~\ref{dndy}). 
The main difference is that, now, we
have only half of the plot, corresponding to the right moving hadron in
a collision in the center of mass frame.

One may in fact argue that there is indeed a relationship between the
structure functions as measured in deep inelastic scattering and the
rapidity distributions for particle production.  We expect,
for instance, the gluon distribution function to be proportional
to the pion rapidity distribution. This is what comes out in many 
models of particle production. It is further plausible,
since the degrees of freedom of the gluons should not be lost, but
rather converted into the degrees of freedom of the produced hadrons. 

The small x problem is that in experiments at HERA, the rapidity
distributions for quarks and gluons
grow rapidly as the rapidity difference
\be\label{tau-Intro}
\tau\,\equiv\,\ln(1/\x)\,=\,\y_{hadron} - \y\ee
between the quark and the hadron increases \cite{z}.  
This growth appears to be more rapid
than $\tau$ or $\tau^2$,
and various theoretical models based on the
original considerations by Lipatov and colleagues \cite{BFKL}
suggest it may grow as an exponential in $\tau$ \cite{BFKL,TB-BFKL}. 
The  more established DGLAP evolution equation \cite{DGLAP}
predicts a less rapide growth, like an exponential in $\sqrt{\tau}$,
but this is still exceeding the Froissart
unitarity bound, which requires
rapidity distributions to grow at most as $\tau^2$ (since
$\tau\sim\ln s$).
\begin{figure} 
\begin{center} 
\includegraphics[width=0.5\textwidth]
{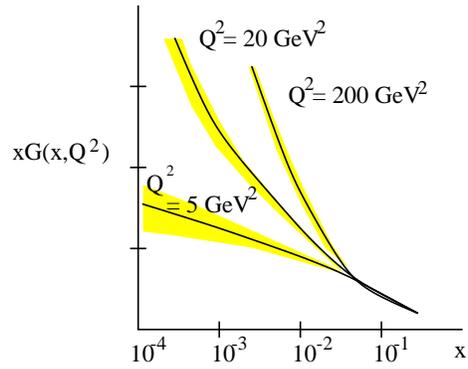} 
\caption{The Zeus data for the gluon structure functions. } 
\label{gluon} 
\end{center}
\end{figure}

In Fig. \ref{gluon}, the ZEUS data for the gluon distribution are
plotted for $Q^2 = 5~ {\rm GeV}^2$, 
$20~ {\rm GeV}^2$ and $200~{\rm GeV}^2$ \cite{z}.
The gluon distribution is the number of gluons per unit rapidity
in the hadron wavefunction, $\x G(\x,Q^2)=dN_{gluons}/d\y\;$.
Experimentally, it is extracted from the data for the quark structure
functions, by exploiting the dependence of the latter upon the resolution of
the probe, that is, upon the transferred momentum $Q^2$. 
Note the rise of $\x G(\x,Q^2)$ at small x: this is
the small x problem.  
If one had plotted the total multiplicity of
produced particles in $pp$ and $\overline p p$ collisions on the same plot,
one would have found rough agreement in the shape of the curves.  
\begin{figure} 
\begin{center} 
\includegraphics[width=0.5\textwidth]
{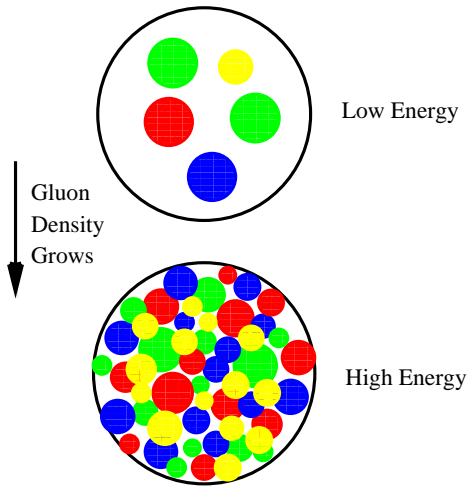} 
\caption{Saturation of gluons in a hadron.  A view of a hadron head on as
x decreases.} 
\label{saturation} 
\end{center}
\end{figure}

Why is the small x rise in the gluon distribution a problem?  
Consider Fig. \ref{saturation}, 
where we view the hadron head on. The constituents are the
valence quarks, gluons and sea quarks
shown as coloured circles.  As we add more and more constituents,
the hadron becomes more and more crowded.  If we were to try to measure
these constituents with say an elementary photon probe, as we do in deep
inelastic scattering, we might expect that the hadron would become so
crowded that we could not ignore the shadowing effects of constituents as
we make the measurement.  (Shadowing means that some of the partons are
obscured by virtue of having another parton in front of them.  This 
would result in a decrease of the scattering
cross section relative to what is expected from incoherent independent
scattering.) 

We shall later argue that the distribution functions at fixed
$Q^2$ {\em saturate}, which means that they
cease growing so rapidly at high energy \cite{GLR,MQ,MV94,AM2,SAT}. 
(See also Refs. \cite{AM1,Larry01,Levin,AMCARGESE} for recent reviews
and more references.) This saturation will be seen to occur at 
transverse momenta below some intrinsic scale, the ``saturation
scale'', which is estimated as:
\be\label{Qs-INTRO}
        Q_s^2\, =\, \alpha_sN_c\,{1 \over {\pi R^2}} {{dN} \over {d{\rm y}}}\,,
\ee 
where $dN/d\y$ is the gluon distribution. Only gluons matter since,
at small x, the gluon density grows faster then the quark density,
and is the driving force towards saturation. This is why in
the forthcoming considerations we shall ignore
the (sea) quarks, but focus on the gluons alone.
Furthermore, $\pi R^2$ --- with $R$ the hadron radius --- is the area of the
hadron in the transverse plane. (This is well defined as long as
the wavelengths of the external probes are small compared to $R$.)
Finally, $\alpha_sN_c$ is the colour charge squared of a single
gluon. Thus, the ``saturation scale'' (\ref{Qs-INTRO}) has the meaning
of the average colour charge squared of the gluons in
the hadron wavefunction per unit transverse area.

Since the gluon distribution increases rapidly with the energy,
as shown by the HERA data, so does the saturation scale. We shall use
the rapidity difference $\tau=\ln(1/\x)\sim \ln s$, eq.~(\ref{tau-Intro}),
to characterize this increase, and write $Q_s^2 \equiv Q_s^2(\tau)$.
For sufficiently large $\tau$ (i.e., high enough energy, or small enough x), 
 \be
        Q_s^2(\tau)  \gg \Lambda^2_{QCD},
\ee
and $\alpha_s( Q_s^2) \ll 1$. Then
we are dealing with {\em weakly coupled} QCD, so we should be 
able to perform a first principle calculation of, e.g.,
\begin{itemize}

\item the gluon distribution function;

\item the quark and heavy quark distribution functions;

\item the intrinsic $p_\perp$ distributions of quarks and gluons.

\end{itemize}

But weak coupling does not necessarily mean that the
physics is perturbative.  There are many examples of nonperturbative
phenomena at weak coupling.  An example is instantons in electroweak
theory, which lead to the violation of baryon number.   Another example
is the atomic physics of highly charged nuclei, where the electron
propagates in the background of a strong nuclear Coulomb field.
Also, at very high temperature, QCD becomes
a weakly coupled quark-gluon plasma, but it exhibits nonperturbative
phenomena on large distances $r\gg 1/T$ (with $T$ the temperature),
due to the collective behaviour of many quanta \cite{BI01}.

Returning to our small-x gluons, we notice that, at low
transverse momenta $Q^2\le Q_s^2(\tau)$, they make a high
density system, in which the interaction probability 
\be  \label{QSGLR}
\frac{\alpha_s(Q^2)}{Q^2}\,{1 \over {\pi R^2}} \,{{dN} \over {d{\rm y}}}
\,\,\sim\,\,
1 \quad {\rm when} \quad Q^2 \sim Q^2_s(\tau)\ee
is of order one \cite{GLR,MQ,BM87}. That is, although the coupling is
small, $\alpha_s( Q^2) \ll 1$, the effects of the interactions
are amplified by the large gluon density (we shall see that
$dN/d\y \sim 1/\alpha_s $
at saturation), and ordinary perturbation theory breaks down.

To cope with this, a resummation of the high density effects is necessary.
Our strategy to do so --- to be described at length in these
lectures --- will be to construct an {\it effective theory} in which
the small-x gluons are described as the classical colour
fields radiated by ``colour sources'' at higher rapidity. Physically,
these sources are the ``fast'' partons, i.e., the hadron constituents 
with larger longitudinal momenta $p^+\gg \x P^+$. The properties
of the colour sources
will be obtained via a 
renormalization group analysis, in which the ``fast'' partons 
are integrated out in steps of rapidity and in the background
of the classical field generated at the previous steps.

The advantage of this strategy
is that the non-linear effects are dealt with in a classical context,
which makes exact calculations possible. Specifically, (a) the
classical field problem will be solved exactly, and  (b) 
at each step in the renormalization group analysis,
the non-linear effects associated with the classical fields will
be treated exactly. On the other hand, the mutual interactions of
the fast partons will be treated in perturbation theory, in a
``leading-logarithmic'' approximation which resums the most important
quantum corrections at high energy (namely, those which are 
enhanced by the large logarithm $\ln(1/\x)$).

As we shall see, the resulting effective theory describes
the saturated gluons as a {\it Colour Glass Condensate}.
The classical field approximation is appropriate for these
saturated gluons, because of the large occupation
number $N_k\sim 1/\alpha_s \gg 1$ of their true quantum state.
In this limit, the Heisenberg commutators between particle 
creation and annihilation operators become negligible:
\be
[a_k,\,a^\dagger_k]\,=\,1\,\ll\,a^\dagger_ka_k=N_k\,,\ee
which corresponds indeed to a classical regime. The
classical field language is also well adapted to describe
the {\it coherence} of these small-x gluons, which  overlap with each other
because of their large longitudinal wavelengths.

The phenomenon of saturation provides also a natural solution
to the unitarity problem alluded to before. We shall see that,
with increasing energy, the new partons are produced preponderently
at momenta $p_\perp\simge Q_s$. Thus, these new partons have a
typical transverse size $\sim 1/p_\perp \simle 1/Q_s$. Smaller
is x (i.e., larger is $\tau$), 
larger is $Q_s(\tau)$, and therefore smaller are the 
newly produced partons. An external probe of transverse
resolution $\Delta x_\perp \sim 1/Q$ 
will not see partons smaller than this resolution size. For
$\tau$ large enough, $Q^2<Q_s^2(\tau)$, so that the partons produced 
when further increasing the energy will not contribute to the  cross
section at fixed $Q^2$. Thus, although the gluon distribution
keeps increasing with $\tau$, there is nevertheless
no contradiction with unitarity.

\subsection{Geometrical Scaling}
\label{Sect-Scaling}

\begin{figure} 
\begin{center} 
\includegraphics[width=0.8\textwidth]
{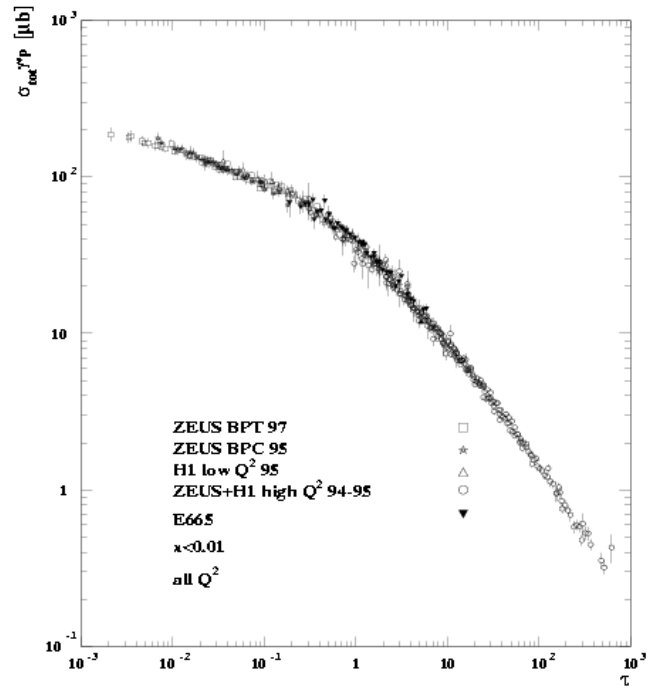} 
\caption{Experimental data on the cross section for virtual photon-proton
deep inelastic scattering from the region $\x < 0.01$ plotted verses the 
scaling variable ${\cal T} = Q^2 R^2(\x)$ \cite{gb}. }
\label{gb} 
\end{center}
\end{figure}

Another striking feature of the experimental data at HERA 
is {\it geometrical scaling} at Bjorken $\x < 0.01$ \cite{gb}.
In general, one expects the structure functions extracted from
deep inelastic scattering to depend upon two dimensionless kinematical
variables, x and $Q^2/\Lambda^2 $, where $\Lambda^2$ is some arbitrary
momentum scale of reference, which is fixed.
The striking feature alluded to before is the observation that the
x dependence measured at HERA at  $\x < 0.01$ and for a
broad region of $Q^2$ (between $0.045$ and $450\,{\rm Gev}^2$)
can be entirely accounted for by a corresponding
dependence of the reference scale $\Lambda^2\rightarrow 1/R^2(\x)$
alone.
That is, rather than being functions of two independent variables
x and $Q^2/\Lambda^2 $, the measured structure functions at 
$\x < 0.01$ depend effectively only upon the scaling variable
\be
	{\cal T} \,\equiv\, Q^2 R^2(\x)
\ee
where $R^2(\x) \sim \x^\lambda$ and $\lambda \sim 0.3-0.4$ in order
to fit the data. This is illustrated in Fig. \ref{gb} \cite{gb}.
Such a scaling behaviour is consistent with the saturation scenario
\cite{GBW99,SAT,LT99}, as we shall discuss towards the end
of these lectures. Note however that the experimentally observed
scaling extends to relatively large values of x and $Q^2$,
above all the estimates for the saturation scale. Thus, this feature
seems to be more general than the phenomenon of saturation.

\subsection{Universality}

There are two separate formulations of universality which are important
in understanding small x physics.

a) The first is a weak universality \cite{MV94,SAT}. 
This is the statement that at
sufficiently high energy, physics should depend upon the specific
properties of the hadron at hand (like its size or atomic number $A$)
only via the saturation scale $Q_s(\tau, A)$. 
Thus, at high energy,
there should be some equivalence between nuclei and
protons:  When their $Q_s^2$ values are the same, their properties
must be the same.  An empirical parameterization of the
gluon structure function in eq.~(\ref{Qs-INTRO}) is
\be
{1 \over {\pi R^2}} {{dN} \over {d{\rm y}}} \sim {{A^{1/3}} \over x^\delta}
\ee
where $\delta \sim 0.2 - 0.3$ \cite{z}.
This suggests the following correspondences:
\begin{itemize}

\item RHIC with nuclei $\sim$ HERA with protons;

\item LHC with nuclei $\sim$ HERA with nuclei.

\end{itemize}

Estimates of the saturation scale for nuclei at RHIC energies give
$\sim 1-2 ~{\rm Gev}$, and at LHC $Q_s \sim 2-3 ~{\rm Gev}$.

b) The second is a strong universality which is meant in a statistical
mechanical sense. This is the statement that the effective action
which describes small x
distribution function is critical and at a fixed point of some
renormalization group.  This means that the behavior of correlation
functions is given by universal critical exponents, which
depend only on general properties of the theory such
as its symmetries and dimensionality.

\subsection{Some applications}

We conclude these introductory considerations with a (non-exhaustive)
enumeration of recent applications of the concept of saturation and
the Colour Glass Condensate (CGC) to phenomenology.

Consider deep inelastic scattering first. It has been shown in
Refs. \cite{GBW99} that the HERA data for (both inclusive and
diffractive) structure functions can be well accounted for by
a phenomenological model which incorporates saturation.
The same model has motivated the search for geometrical
scaling in the data, as explained in Sect. \ref{Sect-Scaling}.

Coming to ultrarelativistic heavy ion collisions,
as experimentally realized at RHIC and, in perspective, at LHC,
we note that the CGC should be the appropriate description
of the initial conditions. Indeed, most of the multiparticle
production at central rapidities is from the small-x
($\x \le 10^{-3}$) partons in the nuclear wavefunctions,
which are in a high-density, semi-classical, regime.
The early stages of a nuclear collision, up to times
$\sim 1/Q_s$, can thus be described as the melting of the 
Colour Glass Condensates in the two nuclei.
In Refs. \cite{KV00}, this melting has been systematically
studied, and the multiparticle production computed,
via numerical simulations of the classical effective theory
\cite{MV94,KMW95}.
After they form, the particles scatter with each other,
and their subsequent evolution can be described by transport theory
\cite{bottom}.

The first experimental data at RHIC \cite{RHIC} have been analyzed from the
perspective of the CGC in Refs. \cite{KN01,KL01,MSB01}. 
Specifically, the multiparticle production
has been studied with respect to its dependence upon centrality 
(``number of participants'') \cite{KN01}, rapidity \cite{KL01}
and transverse momentum distribution \cite{MSB01}.

The charm production from the CGC in peripheral heavy-ion collisions
has been investigated in \cite{GP01}.

Electron-nucleus ($eA$) deeply inelastic scattering 
has been recently summarized in \cite{RV01}.
Some implications of the Colour Glass Condensate for the central region 
of $p+A$ collisions have been explored in Refs. \cite{MD01,YK01}.

Instantons in the saturation environment have been considered in
Ref. \cite{KKL01}.

\section{The classical effective theory}
\label{EFT}
\setcounter{equation}{0}

With this section, we start the study of an effective theory for the
small x component of the hadron wavefunction 
\cite{MV94,SAT,K96,JKMW97,KM98,LM00,JKLW97,PI} (see also the previous
review papers \cite{Larry01,Raju99}). Motivated by the
physical arguments exposed before, in particular, by the separation
of scales between {\it fast} partons and {\it soft} (i.e., small-x) gluons,
in the infinite momentum frame, this effective theory
admits a rigourous derivation from QCD, to be described in 
Sect. \ref{RGE-CGC}. Here, we shall rather rely on simple kinematical
considerations to motivate its general structure.

\subsection{A stochastic Yang-Mills theory}
\label{EFT-YM}

In brief, the effective theory is a classical Yang-Mills theory with
a random colour source which has only a ``plus'' component 
\footnote{Written as it stands, eq.~(\ref{cleq0})
is correct only for field configurations having $A^-=0$; 
when $A^-\ne 0$, the source $\rho$ in its r.h.s. gets rotated
by Wilson lines built from $A^-$ \cite{PI}.} :
\beq
(D_{\nu} F^{\nu \mu})_a(x)\, =\, \delta^{\mu +} \rho_a(x)\,.
\label{cleq0}
\eeq
The classical gauge fields $A^\mu_a$ represent the {\it soft}
gluons in the hadron wavefunction, i.e., the gluons with small 
longitudinal momenta ($k^+={\rm x}P^+$ with $\x\ll 1$).
For these gluons, the classical approximation should be appropriate 
since they are in a multiparticle state with large occupation numbers.

\begin{figure}[htb]
\centering
\resizebox{.9\textwidth}{!}{%
\includegraphics*{{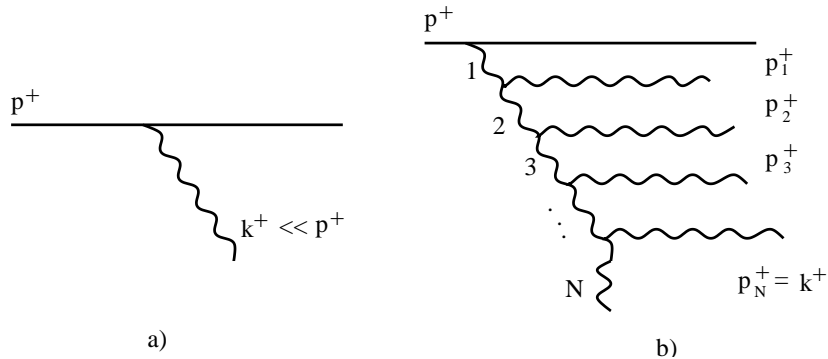}}}
   \caption{a) Soft gluon emission by a fast parton;
b) a gluon cascade.}
\label{cascade}
\end{figure}

The {\it fast}
partons, with momenta $p^+ \gg k^+$, are not dynamical fields
anylonger, but they have been rather replaced by the
colour current $J^\mu_a=\delta^{\mu +}\!\rho_a$ which acts as a
source for the soft gluon fields. This is quite intuitive: 
the soft gluons in the hadron wavefunction are radiated
by typically fast partons, via the parton cascades shown in Fig. \ref{cascade}.
It is in fact well known that, for the tree-level radiative process 
shown in Fig. \ref{cascade}.a, classical and quantum calculations give
identical results in the limit where the emitted gluon is soft \cite{TB-DIS}.
What is less obvious, but will be demonstrated by the analysis
in Sect. \ref{RGE-CGC}, is that quantum corrections like those displayed in
Fig. \ref{cascade}.b do not invalidate this classical description,
but simply renormalize the properties of the classical source,
in particular, its correlations.

The gross properties of this source follow from 
kinematics. The fast partons move along the $z$ axis 
at nearly the speed of light. They can emit, or absorb, soft gluons,
but in a first approximation they preserve straightline 
trajectories along the light-cone ($z=t$). In terms of LC coordinates,
they propagate in the positive $x^+$ direction, while sitting
at $x^-=0$. Their colour current is proportional to their
velocity, which implies $J_a^\mu=\delta^{\mu+}\!\rho_a$,
with a charge density $\rho_a(x)$ which is localized near $x^-=0$.
More precisely, as quantum fields, the fast partons
are truly delocalized over a longitudinal distance 
$\Delta x^- \sim 1/p^+$,
as required by the uncertainty principle.
But since $1/p^+\ll 1/k^+$, they still look as sharply
localized when ``seen'' by the soft gluons, which have long
wavelengths and therefore a poor longitudinal resolution.

The separation of scales in longitudinal momenta 
implies a corresponding separation in time:
Softer partons have larger energies, and
therefore shorter lifetimes. Consider indeed
the radiative process in Fig. \ref{cascade}.a, where $k^+\ll p^+$. 
This is a virtual excitation whose lifetime (in units of LC time $x^+$)
can be estimated from the uncertainty principle as
\beq\label{lifetime}
\Delta x^+\,=\,\frac{1}{\varepsilon_{p-k}+\varepsilon_k-
\varepsilon_p}\,\simeq\,\frac{1}{\varepsilon_k}\,\ll\,
\frac{1}{\varepsilon_p}\,.\eeq
This is small as compared to the typical time scale 
$1/\varepsilon_p$ for the dynamics of the fast partons.
[In eq.~(\ref{lifetime}), $\varepsilon_p\equiv p_\perp^2/2p^+$ is the LC
energy of the on-shell gluon with momentum ${\vec p}=(p^+,
{\bf p}_\perp)$, and we have used the fact that, for $k^+ \ll p^+$
and comparable transverse momenta $k_\perp$ and $p_\perp$, 
$\varepsilon_k\gg\varepsilon_p,\varepsilon_{p-k}$.]
Thus, the ``fast'' degrees of freedom are effectively frozen
over the short lifetime of the soft gluon, and can be described by
a {\it time-independent} (i.e., independent of $x^+$) colour source
$\rho_a(x^-,x_\perp)$.

Still, this colour source is eventually changing over the 
larger time scale $1/\varepsilon_p$. Thus, if another soft 
gluon is emitted after a time interval $\simge 1/\varepsilon_p$,
it will ``see'' a different configuration of $\rho$, 
without quantum interference between the different configurations. 
This can be any of the configurations allowed by the dynamics
of the fast partons. We are thus led to treat $\rho_a(x^-,x_\perp)$ as
a {\it  classical random} variable (here, a  {\it field} variable),
with some probability density, or {\it weight function},
$W_{k^+}[\rho]$, which is a functional of $\rho$. 

As suggested by its notation, the weight function 
depends upon the soft scale $k^+$ at which we measure correlations. 
Indeed, as we shall see in Sect. \ref{RGE-CGC}, $W_{k^+}[\rho]$ 
is obtained by integrating out degrees of freedom with 
longitudinal momenta larger than $k^+$. It turns out
that it is more convenient to use the {\em rapidity}\footnote{Strictly
speaking, this is the rapidity {\it difference}
between the small-x gluon and the hadron, as defined previously
in eq.~(\ref{tau-Intro}).
But this difference is the relevant quantity for what follows,
so from now on it
will be simply referred to as ``the  rapidity'', for brevity.}
\be
\tau\equiv\ln(P^+/k^+) = \ln(1/\x)\,\ee
to indicate this
dependence, and thus write $W_\tau[\rho]\equiv W_{k^+}[\rho]$.

To deal with field variables and functionals of them,
it is convenient to consider a discretized (or lattice) version
of the 3-dimensional configuration space, with 
lattice points $(x^-,x_\perp)$. (We use the same notations for 
discrete and continuous coordinates, to avoid a proliferation of
symbols.) A configuration of the colour source is specified by
giving its values $\rho^a(x^-,x_\perp)$ at the $N$ lattice 
points.
The functional $W_\tau[\rho]$ is a (real) function of these $N$ values. 
To have a meaningful probabilistic interpretation, this
function must be positive 
semi-definite ($W_\tau[\rho]\ge 0$ for any $\rho$), and normalized 
to unity:
\be\label{norm}
\int { D}[\rho]\, \,W_\tau[\rho]\,=\,1\,,\ee
with the following functional measure:
\be\label{measure} D[\rho]\,\equiv\,
\prod_{a}\prod_{x^-} \prod_{x_\perp}\,d\rho^a(x^-,x_\perp)\,.\ee

Gluon correlation functions at the soft scale 
$k^+=\x P^+= P^+{\rm e}^{-\tau}$ are obtained by first
solving the classical equations of motion (\ref{cleq0})
and then averaging the solution over $\rho$ with the weight function 
$W_\tau[\rho]$ (below ${\vec x}\equiv (x^-,{x}_{\perp})$) :
\be\label{clascorr}
\langle A^i_a(x^+,\vec x)A^j_b(x^+,\vec y)
\cdots\rangle_\tau\,=\,
\int { D}[\rho]\,\,W_\tau[\rho]\,{\cal A}_a^i({\vec x})
{\cal A}_b^j({\vec y})\cdots\,,\ee
where ${\cal A}_a^i\equiv {\cal A}_a^i[\rho]$ is the 
solution to the classical Yang-Mills equations with static source
$\rho_a$, and is itself independent of time (cf. Sect. \ref{YMSOL}
below). Note that only equal-time correlators can be computed in this
way; but these are precisely the correlators that are
measured by a small-x external probe, which
is absorbed almost instantaneously by the hadron
(cf. eq.~(\ref{lifetime})).

The formula (\ref{clascorr}) is readily extended to any
operator which can be related to $\rho$. To guarantee that
only the physical,  gauge-invariant, operators acquire
a non-vanishing expectation value, we shall require $W_\tau[\rho]$ 
to be gauge-invariant. In practical calculations, 
one generally has to fix a gauge, so the gauge symmetry of $W_\tau[\rho]$ 
may not be always manifest.

To summarize, the effective theory is defined by 
eqs.~(\ref{cleq0}) and (\ref{clascorr}) together with the (so far,
unspecified) weight function $W_{\tau}[\rho]$. In what follows,
we shall devote much effort to derive this theory from QCD,
and construct the weight function $W_{\tau}[\rho]$ in the process
(in Sects. \ref{RGE-CGC}--\ref{S-SOL}).
But before doing that, let us gain more 
experience with the classical theory by solving the equations of motion
(\ref{cleq0}) (in Sect. \ref{YMSOL}), and then using the result to compute 
the gluon distribution of a large nucleus (in Sect. \ref{MVmodel}).
In performing these calculations, we shall need a more precise definition
of the gluon distribution function and, more generally,
of the relevant physical observables, so we start by discussing that.

\subsection{Some useful observables}
\label{OBS-DEF}

In subsequent applications of the effective theory, we shall
mainly focus on two observables which, because of their physical
content and of the specific structure of the effective theory,
are particularly suggestive for studies of non-linear phenomena
like saturation. These observables, that we introduce now,
are the gluon distribution 
function and the cross-section for the scattering of a ``colour dipole''
off the hadron.

\subsubsection{The gluon distribution function}

We denote by $G(\x,Q^2) d\x$ the number of gluons in the
hadron wavefunction having longitudinal momenta between 
$\x P^+$ and $(\x+d\x)P^+$, and a transverse size $\Delta x_\perp \sim 1/Q\,$.
In other terms, the {\em gluon distribution}
$\x G(\x,Q^2)$ is the number of gluons with transverse momenta  
$k_\perp \simle Q$ per unit rapidity : 
\be\labe{GDFdef}
\x G(\x,Q^2)&=&\int^{Q^2} {d^2k_\perp}\,
k^+\frac{dN}{dk^+d^2k_\perp}\bigg|_{k^+=xP^+}\nn&=&
\int {d^3k}\,\Theta(Q^2-k_\perp^2)\,
\x\delta(\x-k^+/P^+)\,\frac{dN}{d^3 k}\,,\ee
where $\vec k \equiv (k^+,{\bf k}_\perp)$ and 
\be
\frac{dN}{d^3 k}\,=\,\frac{dN}{dk^+d^2k_\perp}\,,\ee
is the Fock space gluon density, i.e., the
number of gluons per unit of volume in momentum space. 
The difficulty is, however, that this number depends upon the
gauge, so in general it is not a physical observables. Still, as
we shall shortly argue, this quantity can be given a gauge-invariant
meaning when computed in the light-cone (LC) gauge 
\be\label{LCG}
        A^+_a \,=\, 0\,.
\ee
(We define the light-cone components of $A^\mu_a$
in the standard way, as $A^\pm_a = (A^0_a \pm A^3_a)/\sqrt{2}$.)
In this gauge, the equations of motion\footnote{For the purposes
of LC quantization we use the equations of motion without sources;
that is, we consider real QCD, and not the effective theory (\ref{cleq0}).}
\be
        D_\mu F^{\mu \nu} = 0,
\ee
imply for the $+$ component
\be
        D_iF^{i+} + D^+F^{-+} = 0,
\ee
which allows one to compute $A^-$ in terms of $A^i$ as
\be
        A^- = {1 \over \partial^{+2}} D^i \partial^+ A^i\,.
\ee
This equation says that we can express the longitudinal field
in terms of the transverse degrees of freedom which are specified by
the transverse fields entirely and explicitly.  These degrees of freedom
correspond to the two polarization states of the gluons. The quantization
of these degrees of freedom  proceeds by writing \cite{KS70}:
\be\label{Fock}
        A^i_c (x^+,\vec x) = \int_{k^+ > 0} {{d^3k} \over {(2\pi)^3 2k^+}} 
\left(
e^{i{\vec k\cdot \vec x}} a^i_c(x^+,\vec k) + e^{-i{\vec k\cdot \vec x}} 
a^{i\dagger}_c
 (x^+,\vec k)\right)\,\,
\ee
($\vec x\cdot \vec k = x^-k^+-{\bf x}_\perp\cdot{\bf k}_\perp$)
with the creation and annihilation operators satisfying the
following commutation relation at equal LC time $x^+$ :
\be
        [a^i_b (x^+,\vec k), a^{j\dagger}_c(x^+,\vec q)]\, = \,\delta^{ij}
 \delta_{bc}\,  2k^+(2\pi)^3
\delta^{(3)} (k - q).
\ee
In terms of these Fock space operators, the gluon density is computed as:
\be
\frac{dN}{d^3 k}= \langle   
a^{i\dagger}_{c}(x^+, \vec k)\,a_{ c}^i(x^+, \vec k)  \rangle
=  \frac{2k^+}{(2 \pi)^3}\, \langle    A^i_c(x^+,\vec k)
A^i_c(x^+,-\vec k)\rangle\, ,\ee
where the average is over the hadron wavefunction.
By homogeneity in time, this equal-time average is independent
of the coordinate $x^+$, which will be therefore
omitted in what follows.
By inserting this into eq.~(\ref{GDFdef}) and using the fact that,
in the LC-gauge, $F^{i+}_a(k)=ik^+A^i_a(k)$, one obtains
(with $k^+=\x P^+$):
\be\labe{GDF}
\x G(\x,Q^2)=\frac{1}{\pi}\int {d^2k_\perp \over (2 \pi)^2}\,\Theta(Q^2-
k_\perp^2)\Bigl\langle F^{i+}_a(\vec k)
F^{i+}_a(-\vec k)\Bigr\rangle.\ee
As anticipated, this does not look gauge invariant. In coordinate space:
\be\labe{FF}
F^{i+}_a(\vec k)F^{i+}_a(-\vec k)=
\int d^3x \int d^3y\,{\rm e}^{i(\vec x- \vec y)\cdot \vec k}\,
F^{i+}_a(\vec x)F^{i+}_a(\vec y)\ee
involves the electric fields\footnote{The component
$F^{i+}_a=-\partial^+ A^i_a$ is usually referred to as the
(LC) ``electric field'' by analogy with the standard electric field
$E^i_a=F^{i0}_a=-\partial^0 A^i_a$ (in the temporal gauge $A^0_a=0$).}
at different spatial points
$\vec x$ and $\vec y$. A manifestly gauge invariant operator
can be constructed by appropriately inserting Wilson lines.
Specifically, in some arbitrary gauge, we define
\be\labe{GIFF}
{\cal O}_\gamma(\vec x,\vec y)\equiv\,
{\rm Tr}\,\left\{F^{i+}(\vec x)\,U_\gamma(\vec x,\vec y)\,
F^{i+}(\vec y)\,U_\gamma(\vec y,\vec x)\right\},\ee
where (with $\vec A_a\equiv (A^+_a,{\bf A}_\perp^a)$, 
$\vec A\equiv \vec A_a T^a$)
\be\labe{UGEN}
U_\gamma(\vec x,\vec y)\,=\,{\rm P}\,{\rm exp}\left\{ig\int_\gamma d\vec z
\cdot \vec A(\vec z)\right\},\ee
and $\gamma$ is an arbitrary oriented path from $\vec y$
to $\vec x$. The (omitted) temporal coordinates $x^+$ are
 the same for all fields. For any path 
$\gamma$, the operator in eq.~(\ref{GIFF}) is gauge-invariant,
since the chain of operators there makes a closed loop.

\begin{figure} 
\begin{center} 
\includegraphics[width=0.8\textwidth]
{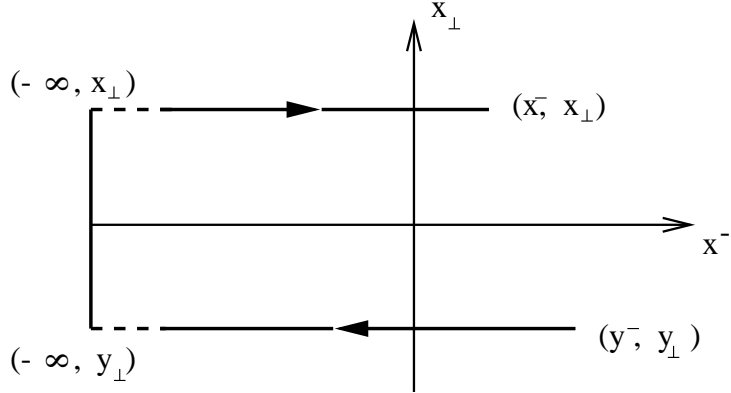} 
\caption{The path $\gamma$ used for the evaluation of the
gauge-invariant operator (\ref{GIFF}).}
\label{PATH} 
\end{center}
\end{figure}

We now show that, by appropriately chosing the path, the gauge,
and the boundary conditions, the
gauge-invariant operator (\ref{GIFF}) can be made to coincide with the
simple 2-point function (\ref{FF}). Specifically, consider the path 
shown in Fig. \ref{PATH}, with the
the following three elements: two ``horizontal'' pieces
going along the $x^-$ axis from $(y^-,y_\perp)$ to $(-\infty,y_\perp)$,
and, respectively, from $(-\infty,x_\perp)$ to $(x^-,x_\perp)$,
and a ``vertical'' piece from $(-\infty,y_\perp)$ to $(-\infty,x_\perp)$.
Along the horizontal pieces, $d\vec z \cdot \vec A =
dz^- A^+$, so these pieces do not matter in the LC gauge. 
Along the vertical piece, $d\vec z \cdot \vec A = d{\bf z}_\perp 
\cdot {\bf A}_{\perp}(-\infty, z_\perp)$, and the path  $\gamma$
between $y_\perp$ and $x_\perp$ is still arbitrary.
But the contribution of any such a path to the Wilson line
vanishes once we impose the following, ``retarded'', boundary condition:
\be\label{retAi}
A^i_a(x)\,\to\,0 \quad {\rm as} \quad x^-\,\to\,-\infty.\ee
(Note that the ``retardation'' 
property refers here to $x^-$, and not to time.)

To summarize, for the particular class of paths mentioned above, 
in the LC gauge $A^+=0$, and with the boundary condition
(\ref{retAi}), $U_\gamma(\vec x,\vec y)\to 1$, and
the manifestly gauge-invariant operator in eq.~(\ref{GIFF})
reduces to the simpler operator (\ref{FF}) which defines the number
of gluons in this gauge. Converserly, the latter quantity
has a  gauge-invariant meaning, as the expression of a
gauge-invariant operator in a specific gauge.

We shall need later also the gluon distribution function in the transverse 
phase-space (in short, the ``gluon density''), 
i.e., the number of gluons per unit rapidity
per unit transverse momentum per unit transverse area:
\be\label{TPS0}
{\cal N}_\tau(k_\perp,b_\perp)\,\equiv\,
\frac{d^5 N}{d\tau d^2k_\perp d^2 b_\perp}\,=\,
\frac{d^2 \,\x G(\x,k_\perp^2)}{d^2k_\perp d^2 b_\perp}\,,
\ee
where $\tau=\ln(1/x)=\ln(P^+/k^+)$ and $b_\perp$ is the impact
parameter in the transverse plane (i.e., the central coordinate
$b_\perp=(x_\perp+y_\perp)/2$ in eq.~(\ref{FF})).
This phase-space distribution 
is a meaningful quantity since the typical transverse
momenta we consider are relatively large,
\be 
k_\perp^2\,\gg\,\Lambda_{QCD}^2\,\sim 1/R^2,\ee
so that the transverse de Broglie wavelengths $\sim 1/k_\perp$
of the partons under consideration are much shorter than the
typical scale of transverse variation in the hadron,
$1/\Lambda_{QCD}$. (In particular, this explains why
we can consider the hadron to have a well defined transverse
size $R$.) 

In fact, for simplicity, we shall mostly consider a hadron which is
homogeneous in the transverse plane, with a sharp boundary at radial
distance $R$. Then, the density (\ref{TPS0})
is independent of $b_\perp$ (within the disk $b_\perp < R$), and reads
(cf. eq.~(\ref{GDF})) :
\be\label{TPS}
{\cal N}_\tau(k_\perp)\,=\,\frac{1}{\pi R^2}\,
\frac{d^3 N}{d\tau d^2k_\perp}\,=\,\frac{1}{4\pi^4 R^2}\,
\langle F^{i+}_a(\vec k)F^{i+}_a(-\vec k) \rangle.\ee

\subsubsection{The dipole-hadron cross-section}
\label{sect-dipole}

Consider high energy deep inelastic scattering (DIS) in a special 
frame --- the ``dipole frame'' ---
in which the virtual photon $\gamma^*$ is
moving very fast, say, in the negative $z$
direction, but most of the total energy is still carried by the 
hadron, which moves nearly at the speed of light in the positive $z$ direction.
Thus, the rapidity gap between the hadron and the virtual photon
is 
\be\label{tau-dipole}
\tau=\y_{hadron} - \y_{\gamma^*}\,,\qquad
{\rm  with}\qquad  |\y_{\gamma^*}|\ll \y_{hadron}.\ee
(As in Sect. \ref{SECT-DIS}, $\tau=\ln(1/\x) \approx \ln(s/Q^2)$, where
$Q^2$ is the virtuality of $\gamma^*$
and $s$ is the invariant energy squared. Note also that
$\y_{\gamma^*}< 0$, since  $\gamma^*$ is a left mover.)

The dipole frame is special in two respects \cite{AMCARGESE} 
(and references therein):

i) The DIS looks like a two step process, in which $\gamma^*$
fluctuates first into a quark--antiquark pair, which then
scatters off the hadron. The $q\bar q$ pair is in a colour singlet
state, so it forms a {\em colour dipole}.

ii) The essential of the quantum evolution is put
in the hadron wavefunction, which carries most of the energy.
The dipole wavefunction, on the other hand, is simple and given
by lowest order perturbation theory. More precisely, 
if $\alpha_s |\y_{\gamma^*}|\ll 1$, then the dipole is just a
quark--antiquark pair, without additional gluons.

Thus, in this frame,
all the non-trivial dynamics is in the dipole-hadron scattering.
Because of the high energy of the $q\bar q$ pair, this scattering
can be treated in the eikonal approximation 
\cite{AM0,NZ91,BH,B} : the quark (and the antiquark) follows a
straight line trajectory with $z=-t$ (or $x^+=0$),
and the effect of its interactions with the colour field of the hadronic
target is contained in the Wilson line:
\be\label{Wilson0}
V^\dagger(x_{\perp})\,=\,{\rm P}\,{\rm exp}\left(ig\int_{-\infty}
^\infty dx^- A^+_a(x^-,x_{\perp}) t^a\right),\ee
where $x_{\perp}$ is the transverse coordinate of the quark,
$t^a$'s are the generators of the colour group in the 
fundamental representation, and the symbol P denotes the ordering of
the colour matrices $A^+(\vec x)=A^+_a(\vec x) t^a$ in the
exponent from  right to left in increasing order of their $x^-$ arguments.
Note that $A^+$ is the
projection of $A^\mu$ along the trajectory of the fermion.
For an antiquark with transverse coordinate $y_{\perp}$
the corresponding gauge factor is $V(y_{\perp})$.
Clearly, we adopt here a gauge where $A^+_a\ne 0$ (e.g.,
the covariant gauge to be discussed at length in Sect. \ref{YMSOL}).

It can then be shown that the $S$-matrix element for the 
dipole-hadron scattering is obtained by averaging the total gauge factor
${\rm tr}(V^\dagger(x_{\perp}) V(y_{\perp}))$ (the colour
trace occurs since we consider a colourless  $q\bar q$ state)
over all the colour field configurations in the hadron wavefunction:
\be\label{Stau}
S_\tau(x_{\perp},y_{\perp})\,\equiv\,\frac{1}{N_c}\,
\Big\langle {\rm tr}\big(V^\dagger(x_{\perp}) V(y_{\perp})\big)
\Big\rangle_\tau.\ee
The dipole frame is like the hadron infinite momentum frame
in that $\y_{hadron}\approx \tau$, cf. eq.~(\ref{tau-dipole}),
so the average in eq.~(\ref{Stau})
can be computed within the effective theory of Sect.
\ref{EFT-YM}, that is, like in eq.~(\ref{clascorr}).

The dipole--hadron cross section for a dipole
of size $r_\perp=x_{\perp}-y_{\perp}$ is obtained by integrating
$2(1-S_\tau(x_{\perp},y_{\perp}))$ over all
the impact parameters $b_\perp=(x_{\perp}+y_{\perp})/2\,$:
\be\label{sigmadipole}
\sigma_{dipole}(\tau,r_\perp)\,=\,2\int d^2b_\perp\,\frac{1}{N_c}
\Big\langle {\rm tr}\Big(1- V^\dagger(x_{\perp}) V(y_{\perp})\Big)
\Big\rangle_\tau.\ee
Finally, the $\gamma^*$--hadron cross-section is obtained by 
convoluting the dipole cross-section (\ref{sigmadipole}) with the
probability that the incoming photon splits into
a  $q\bar q$ pair:
\be
\sigma_{\gamma^* h}(\tau,Q^2)\,=\,
\int_0^1 dz \int d^2r_\perp\,|\Psi(z,r_\perp;Q^2)|^2\,
\sigma_{dipole}(\tau,r_\perp).\ee
Here, $\Psi(z,r_\perp;Q^2)$ is the light-cone wavefunction for a photon
splitting into a $q\bar q$ pair with transverse size $r_\perp$ and
a fraction $z$ of the photon's longitudinal momentum carried
by the quark \cite{AM0,NZ91}.

\subsection{The classical colour field}
\label{YMSOL}

From the point of view of the effective theory, the high
density regime at small x is characterized by strong classical
colour fields, whose non-linear dynamics must be treated exactly.
Indeed, we shall soon discover that, at saturation,
$\x G(\x,Q^2)\sim 1/\alpha_s$,
which via eqs.~(\ref{GDF}) and (\ref{clascorr}) implies classical
fields with amplitudes ${\cal A}^i\sim 1/g$. Such strong fields
cannot be expanded out from the covariant derivative
$D^i=\partial^i-igA^i$. Thus, we need the exact solution to the classical
equations of motion (\ref{cleq0}), that we shall now construct.

We note first that, for a large class of gauges,
it is consistent to look for solutions having the following 
properties:
\be\label{YMprop}
F^{ij}_a=0,\qquad A^-_a=0,\qquad A^+_a,\,A^i_a\,:\,\,{\rm static}\,,
\ee
where ``static'' means independent of $x^+$.
(In fact, once such a static solution is found in a given gauge,
then the properties (\ref{YMprop}) will be preserved by any
time-independent gauge transformation.) This follows from
the specific structure of the colour source which has just a ``$+$''
component, and is static. For instance, the  component $\mu=i$
of eq.~(\ref{cleq0}) reads:
\be\label{YMi}
0\,=\,D_\nu F^{\nu i}\,=\,D_jF^{ji}+D_+F^{+i}+D_-F^{-i}.\ee
But $D_+=D^-=\partial^--igA^-$ vanishes by eq.~(\ref{YMprop}),
and so does $F^{-i}$. Thus eq.~(\ref{YMi}) reduces to $D_jF^{ji}=0$,
which implies $F^{ij}=0$, as indicated in eq.~(\ref{YMprop}).
This further implies that the transverse fields $A^i$
form a two-dimensional pure gauge. That is, there exists a gauge rotation
$U(x^-,x_\perp)\in {\rm SU}(N)$ such that
(in matrix notations appropriate for the adjoint representation:
${A}^i={ A}^i_a T^a$, etc) :
\be
{A}^i(x^-,x_{\perp})\,=\,{i \over g}\,
U(x^-,x_{\perp})\,\partial^i U^{\dagger}(x^-,x_{\perp})\,.
\labe{tpg} 
\ee
Thus, the requirements (\ref{YMprop}) leave just two independent
field degrees of freedom, $A^+(\vec x)$ and $U(\vec x)$, which are
further reduced to one (either $A^+$ or $U$)
by imposing a gauge-fixing condition.

We consider first the covariant gauge (COV-gauge)
$\partial_\mu {A}^\mu =0$. By eqs.~(\ref{YMprop}) and (\ref{tpg}),
this implies $\partial_i A^i=0$, or $U=0$. Thus, in this gauge:
\be\label{YMCOV}
\tilde {\cal A}^\mu_a(x)\,=\,\delta^{\mu +}\alpha_a(x^-,x_{\perp}),\ee
with $\alpha_a(\vec x)$ linearly related to the colour source
$\tilde\rho_a$ in the COV-gauge :
\be\labe{EQTA}
- \nabla^2_\perp \alpha_a({\vec x})\,=\,{\tilde \rho}_a(\vec x)\,.
\ee
Note that we use curly letters
to denote solutions to the classical field
equations (as we did already in eq.~(\ref{clascorr})).
Besides, we generally use a tilde to indicate quantities
in the COV-gauge, although we keep the simple notation
$\alpha_a(\vec x)$ for the classical field in this gauge,
since this  quantity will be frequently used. 

Eq.~(\ref{EQTA}) has the solution :
\be\labe{alpha}
\alpha_a (x^-,{ x}_\perp)&=&\int d^2y_\perp\,
\langle x_\perp|\,\frac{1}{-\grad^2_\perp}\,|y_\perp\rangle\,
\tilde\rho_a  (x^-,{ y}_\perp)\nn&=&\int \frac{d^2y_\perp}{4\pi}\,
\ln\frac{1}{({x}_\perp - {y}_\perp)^2\mu^2}\,
\tilde\rho_a  (x^-,{ y}_\perp),\ee
where the infrared cutoff $\mu$ is necessary to invert the
Laplacean operator in two dimensions, but it will eventually
disappear from (or get replaced by the confinement scale
$\Lambda_{QCD}$ in) our subsequent formulae.

The only non-trivial field strength is the electric field:
\be\label{F+i} 
\tilde {\cal F}^{+i}_a\,=\,-\partial^i\alpha_a.\ee
In terms of the usual electric (${\bf E}$) and magnetic
(${\bf B}$) fields, this solution is characterized
by purely transverse fields,
${\bf E}_\perp=(E^1,E^2)$ and ${\bf B}_\perp=(B^1,B^2)$,
which are orthogonal to each other: ${\bf E}_\perp \cdot
{\bf B}_\perp =0$ (since $B^1=-E^2$ and $B^2=E^1$).

To compute the gluon distribution (\ref{GDF}), one needs
the classical solution in the LC-gauge $A^+=0$. This is of
the form ${\cal A}^\mu_a=\delta^{\mu i}{\cal A}^i_a$ with
${\cal A}^i_a(x^-,{ x}_\perp)$ a ``pure gauge'', cf. eq.~(\ref{tpg}).
The gauge rotation $U(\vec x)$
can be obtained by inserting the Ansatz (\ref{tpg})
in eq.~(\ref{cleq0}) with $\mu=+$ to deduce an equation for
$U$. Alternatively, and simpler,
the LC-gauge solution can be obtained by a gauge rotation
of the solution (\ref{YMCOV}) in the COV-gauge:
\be\label{COVLC}
{\cal A}^\mu\,=\,U\Bigl(\tilde {\cal A}^\mu+\frac{i}{g}\partial^\mu
\Bigr)U^\dagger,\ee
where the gauge rotation $U(\vec x)$ is chosen such that ${\cal A}^+=0$,
i.e.,
\be
\alpha\, =\, {i \over g}\, U^{\dagger} \left( \partial^+ U \right ).
\labe{ta+}
\ee
Eq.~(\ref{ta+}) is easily inverted to give
\be
U^{\dagger}(x^-,x_{\perp})=
 {\rm P} \exp
 \left \{ig \int_{-\infty}^{x^-} dz^-\,\alpha (z^-,x_{\perp})
 \right \}.\labe{UTA}
\ee
From eq.~(\ref{COVLC}), ${\cal A}^i$ is obtained indeed
in the form (\ref{tpg}), with $U$ given in eq.~(\ref{UTA}).
The lower limit $x^-_0\to -\infty$ in the integral over $x^-$ in 
eq.~(\ref{UTA}) has been chosen such as to impose
the ``retarded'' boundary condition (\ref{retAi}). Furthermore:
\be\label{Fi+}
{\cal F}^{+i}(\vec x)\equiv \partial^+
{\cal A}^{i}(\vec x)\,=\,
U(\vec x)\tilde {\cal F}^{+i}(\vec x)U^\dagger(\vec x).\ee
Together, eqs.~(\ref{tpg}), (\ref{alpha}) and (\ref{UTA}) provide 
an explicit expression for the  LC-gauge solution ${\cal A}^i$ in 
terms of the colour source ${\tilde \rho}\,$ in the { COV}-gauge.
The corresponding expression in terms of the colour source 
in the { LC}-gauge $\rho$ cannot be easily obtained: Eq.~(\ref{EQTA})
implies indeed
\be\labe{EQTA-LC}
- \nabla^2_\perp \alpha\,=\,U^\dagger\,\rho\, U,
\ee
which implicitly determines $\alpha$ (and thus $U$) in terms
of $\rho$, but which we don't know how to solve explicitly.
But this is not a difficulty, as we argue now:

Recall indeed that the classical source is just a ``dummy'' variable
which is integrated out in computing correlations according to
eq.~(\ref{clascorr}). 
Both the  measure and the weight function in eq.~(\ref{clascorr})
are gauge invariant. Thus, one can compute correlation functions
in the LC-gauge by performing a change
of variables $\rho \to \tilde \rho$, and thus replacing the
a priori unknown functionals ${\cal A}^i[\rho]$ by the
functionals ${\cal A}^i[\tilde\rho]$, which are known explicitly.
In other terms, one can replace eq.~(\ref{clascorr}) by
\be\labe{COVclascorr}
\langle A^\mu(x^+,\vec x)A^\nu(x^+,\vec y)
\cdots\rangle_\tau\,=\,
\int {\cal D}[\tilde\rho]\,\,W_\tau[\tilde\rho]\,{\cal A}^\mu_{x}
[\tilde \rho]\,{\cal A}^\nu_{y}[\tilde \rho]\cdots\,,
\ee
where ${\cal A}^\mu[\tilde \rho]$ is the classical solution in some
{\it generic} gauge (e.g., the LC-gauge), 
but expressed as a functional  of the colour source 
${\tilde \rho}$ in the {\it COV}-gauge.

Moreover, the gauge-invariant observables can be expressed
directly in terms of the gauge fields in the COV-gauge, although
the corresponding expressions may look more complicated than
in the LC-gauge. For instance, the operator which enters the
gluon distribution can be written as (cf. eq.~(\ref{Fi+}))
\be\label{FF-gauges}
{\rm Tr}\left\{{\cal F}^{+i}(\vec x)
{\cal F}^{+i}(\vec y)\!\right\} =
{\rm Tr}\left\{U(\vec x)\tilde {\cal F}^{+i}(\vec x)U^\dagger(\vec x)\,
U(\vec y)\tilde {\cal F}^{+i}(\vec y)\,U^\dagger(\vec y)\right\},\,\,
\ee
where the classical fields are in the LC-gauge in the l.h.s. and in
the COV-gauge in the r.h.s, and $U$ and $U^\dagger$ are given by
eq.~(\ref{UTA}). Both writings express the
gauge-invariant operator (\ref{GIFF}) 
(with the path $\gamma$ in Fig. \ref{PATH})
in the indicated gauges. (Indeed,
$U_\gamma(\vec x,\vec y) = U^\dagger(\vec x)U(\vec y)$
for the COV-gauge field $\tilde {\cal A}^\mu=\delta^{\mu +}\alpha$.)
Note that, while in the LC-gauge the non-linear effects
are encoded in the electric fields ${\cal F}^{+i}$, 
in the COV-gauge they are rather encoded in the Wilson lines
$U$ and $U^\dagger$ (the corresponding field  
$\tilde {\cal F}^{+i}_a=-\partial^i\alpha_a$ being linear in
$\tilde\rho_a$).

\begin{figure} 
\begin{center} 
\includegraphics[width=0.9\textwidth]
{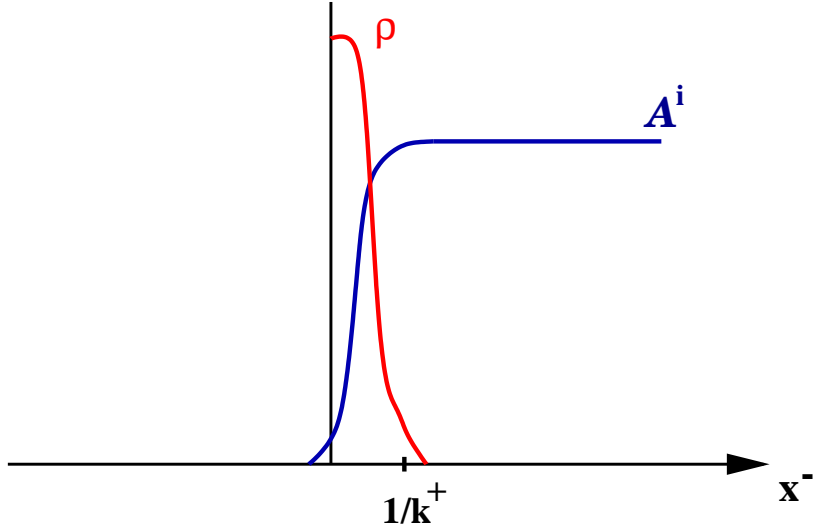} 
\caption{The longitudinal structure of the colour source $\rho$
and of the classical field solution ${\cal A}^i$ for the effective
theory at the scale $k^+\,$. As functions of $x^-$,
$\alpha$ and ${\cal F}^{+i}$ are
as localized as $\rho$.}
\label{YM-PLOT} 
\end{center}
\end{figure}

Up to this point, the longitudinal structure of the source 
has been arbitrary: the solutions written
above hold for any function $\rho^a(x^-)$. 
For what follows, however, it is useful to recall,
from Sect. \ref{EFT-YM}, that $\rho$ has is localized near $x^-=0$.
More precisely, the quantum analysis in Sect. \ref{SRGE}
will demonstrate that the classical source at the longitudinal
scale $k^+$ has support at
positive $x^-$, with $0\le x^-\le 1/k^+$. From 
eqs.~(\ref{EQTA})--(\ref{alpha}), it is clear that
this is also the longitudinal support of the ``Coulomb field'' 
$\alpha(\vec x)$.
Thus, integrals over $x^-$ as that in eq.~(\ref{UTA}) receive
contributions only from $x^-$ in this limited range. 
The resulting longitudinal structure for the
classical solution is
illustrated in Fig.~\ref{YM-PLOT}, and can be approximated as 
follows:
\be\labe{APM}
{\cal A}^i(x^-,x_\perp)&\approx&\theta(x^-)\,
\frac{i}{g}\,V(\del^i V^\dagger)
\,\equiv\,\theta(x^-){\cal A}^i_\infty(x_\perp),\\
\label{FASYMP}
{\cal F}^{+i}(\vec x) &\equiv&
\partial^+{\cal A}^i\,\approx\,\delta(x^-)\,
{\cal A}^{i}_\infty(x_\perp).\ee
It is here understood that the $\delta--$ and $\theta--$functions
of $x^-$ are smeared over a distance $\Delta x^-\sim 1/k^+$.
In the equations above, $V$ and $V^\dagger$ are
the asymptotic values of the respective
gauge rotations as $x^-\to\infty$ :
\be\labe{v}
V^\dagger(x_{\perp})\,\equiv\,{\rm P} \exp
 \left \{
ig \int_{-\infty}^{\infty} dz^-\,\alpha (z^-,x_{\perp})
 \right \}.\ee
In practice, $U(x^-,x_{\perp})=V(x_{\perp})$ for any
$x^-\gg 1/k^+$.
Note that (\ref{v}) is
the same Wilson line as in the discussion of the eikonal approximation 
in Sect. \ref{sect-dipole} (compare to eq.~(\ref{Wilson0}) there).
In the present context, the eikonal approximation is implicit in
the special geometry of the colour source in eq.~(\ref{cleq0}),
which is created by fast moving particles.

\subsection{The gluon distribution of the valence quarks}
\label{MVmodel}

To compute observables in the effective theory, one still
needs an expression for the weight function $W_\tau[\rho]$.
Before discussing the general construction of  $W_\tau[\rho]$ in
Sect. \ref{RGE-CGC}, let us present a simple model for it, due to
McLerran-Venugopalan (MV) \cite{MV94}, which takes into account 
the colour charge of the valence quarks alone. That is,
it ignores the quantum evolution of the colour sources with $\tau$.
This model is expected to work better for a large nucleus, with atomic
number $A\gg 1\,$; indeed, this has many valence quarks ($A\times N_c$), 
and therefore as many colour sources,
which can create a strong colour field already at moderate
values of x, where the quantum evolution can be still neglected.
In this model, $\tau$ is fixed, but one can study the strong field
effects (in particular, gluon saturation) in the limit where $A$ is large.
Besides, the MV model provides a reasonable initial condition for the
quantum evolution towards small x, to be described later.

The main assumption of the MV model is that the $A\times N_c$
valence quarks can be treated as {\it independent} colour sources.
This relies on confinement. Note first that confinement plays
no role for the dynamics in the transverse plane: Indeed, we probe
the nucleus with large transerse momenta $Q^2\gg\Lambda_{QCD}^2$,
that is, over distance scales much shorter than those where
confinement sets in. On the other hand, even at moderate values
of x, we are still probing an integrated version of the hadron 
in the longitudinal direction, i.e., we measure all the ``partons''
(here, valence quarks) in a tube of transverse area 
$\Delta S_\perp \sim 1/Q^2$ and longitudinal extent $\Delta x^-
\sim 1/\x P^+ > 1/P^+$. The number of valence quarks which are
crossed by this tube,
\be
\Delta N\approx n\,\Delta S_\perp\,=\,\Delta S_\perp\,
\frac{AN_c}{\pi R^2_A}\sim A^{1/3}\,,\ee
(with $n=$ the number of quarks per unit transverse area,
$R$ the radius of a single nucleon, and $R_A=A^{1/3}R$
the radius of the nucleus) increases with $A$, but these
quarks are confined within different nucleons, so they are uncorrelated.
When the number of partons $\Delta N$ is large enough, the external
probe ``sees'' them as a classical colour source with a random 
distribution over the transverse area. The total colour charge
${\cal Q}^a$ in the tube is the incoherent sum of the colour charges
of the individual partons. Thus,
\be\label{MV-Q}
\langle {\cal Q}^a\rangle\,=\,0, \qquad \langle {\cal Q}^a {\cal Q}^a
\rangle\,=\,g^2C_f\Delta N\,=\,\Delta S_\perp\,
\frac{g^2C_f N_c A}{\pi R^2_A},\ee
where we have used the fact that the colour charge squared 
of a single quark is $g^2t^at^a=g^2C_f$. One can treat this charge
as classical since, when $\Delta N$ is large enough, we
can ignore commutators of charges:
\be
    \mid [{\cal Q}^a, {\cal Q}^b] \mid \, = \,\mid i f^{abc} 
{\cal Q}^c \mid \,\ll\, {\cal Q}^2\,.
\ee
In order to take the continuum limit (i.e., the limit where
the transverse area $\Delta S_\perp$ of the tube is 
small\footnote{This amounts to increasing $Q^2$, so, strictly
speaking,  at this step one should also include
the DGLAP quantum evolution (i.e., the fact that,
with increasing transverse resolution, the original ``quark''
is resolved into a set of smaller constituents). The quantum analysis
to be discussed later will include that in the ``double-log
approximation''; see Sect. \ref{MFA-HIGH}.}), it is convenient to introduce
the colour charge densities $\rho^a(x^-,x_\perp)$ (with the
same meaning as in Sect. \ref{EFT-YM}) and 
\be
\rho^a(x_\perp)\equiv \int dx^- \rho^a(x^-,x_\perp)\ee
(the colour charge per unit area in the transverse plane).
Then,
\be
{\cal Q}^a\,=\,\int_{\Delta S_\perp} d^2x_\perp\,\rho^a(x_\perp)\,=\,
\int_{\Delta S_\perp} d^2x_\perp\int dx^-\,\rho^a(x^-,x_\perp),\ee
and eqs.~(\ref{MV-Q}) imply (recall that $C_f=(N_c^2-1)/2N_c$) :
\be\label{MV-corr}
\langle \rho_a(x_\perp)\rho_b(y_\perp)\rangle_A&=&
\delta_{ab}\delta^{(2)}(x_\perp-y_\perp)\,\mu_A,\qquad
\mu_A\equiv \frac{g^2A}{2\pi R^2_A}\,,\nn
\langle \rho_a(x^-,x_\perp)\rho_b(y^-,y_\perp)\rangle_A&=&
\delta_{ab}\delta^{(2)}(x_\perp-y_\perp)\delta(x^- -y^-)\,
\lambda_A(x^-),\nn  \int dx^-\,\lambda_A(x^-)&=&\mu_A\,.\ee
Here, $\mu_A\sim A^{1/3}$ is the average colour charge squared
of the valence quarks per unit transverse area and per
colour, and $\lambda_A(x^-)$ is the corresponding
density per unit volume. The latter has some dependence upon
$x^-$, whose precise form is, however, not important since
the final formulae will involve only the integrated density $\mu_A$.
There is no explicit dependence upon $x_\perp$ in $\mu_A$ or
$\lambda_A(x^-)$ since we assume transverse homogeneity within
the nuclear disk of radius $R_A$. Finally, the correlations are local
in $x^-$ since, as argued before, colour sources at different values
of  $x^-$ belong to different nucleons, so they are uncorrelated.
All the higher-point, connected, correlation functions of
$\rho_a(\vec x)$ are assumed to vanish. 
The non-zero correlators (\ref{MV-corr}) are generated
by the following weight function \cite{MV94} :
\be\label{MV-W}
W_A[\rho]\,=\,{\cal N}{\rm exp}\left\{-\,{1\over 2}\, \int d^3x
\frac{\rho_a(\vec x)\rho_a(\vec x)}{\lambda_A(x^-)}\right\},\ee
which is a Gaussian in $\rho_a$, with a local kernel. 
This is gauge-invariant, so the variable $\rho_a$ in this expression
can be the colour source in any gauge.
The integral over $x_\perp$ in eq.~(\ref{MV-W})
is effectively cutoff at $R_A$. By using this weight function,
we shall now compute the observables introduced
in Sect. ~\ref{OBS-DEF}.

Consider first the gluon distribution in the low density regime,
i.e., when the atomic number $A$ is not too high, 
so that the corresponding classical field is weak and
can be computed in the linear approximation.
By expanding the general solution (\ref{tpg}) to linear order in 
$\rho$, or, equivalently, by directly solving the linearized version
of eq.~(\ref{cleq0}), one easily obtains:
\be
{\cal A}^i_a(k)\,\simeq\, -{k^i 
\over k^+ +i \varepsilon} {\rho_a(k^+,k_{\perp}) 
\over k_{\perp}^2}\,,\qquad {\cal F}^{+i}_a(k)
\,\simeq\,i{k^i \over k_{\perp}^2}\,
{\rho_a(\vec k)}\,,
\labe{aaimom}
\ee
which together with  eq.~(\ref{MV-corr}) implies:
\be\label{linphi}
\langle{\cal F}^{i+}_a(\vec k){\cal F}^{i+}_a(-\vec k)\rangle_A
\simeq\frac{1}{ k_{\perp}^2}\,\langle\rho_a(\vec k)\rho_a(-\vec k)\rangle_A
\,=\,\pi R^2_A(N_c^2-1)\frac{\mu_A}{ k_{\perp}^2}\,.\,\,\,\,\ee
By inserting this approximation in eqs.~(\ref{TPS}) and (\ref{GDF}),
one obtains the following estimates for the gluon density and
distribution function:
\be\label{MV-LOW}
{\cal N}_A(k_\perp)&\simeq&
\frac{N_c^2-1}{4\pi^3}\,\frac{\mu_A}{ k_{\perp}^2}\,,
\\
\x G(\x,Q^2)&\simeq&\frac{(N_c^2-1)R_A^2}{4\pi}\,\mu_A\int^{Q^2}_{\Lambda_{QCD}^2}
\frac{dk_\perp^2}{ k_{\perp}^2}\,=\,\frac{\alpha_s AN_cC_f}{\pi}\,
\ln\frac{Q^2}{\Lambda_{QCD}^2}\,,\nonumber\ee
(with $\alpha_s=g^2/4\pi$). The integral over $ k_{\perp}$
in the second line has a logarithmic infrared divergence which
has been cut by hand at the scale $\Lambda_{QCD}$ since we know that,
because of confinement, there cannot be gluon modes with transverse
wavelengths larger than $1/\Lambda_{QCD}$ (see also Ref. \cite{LM00}).

We recognize in eq.~(\ref{MV-LOW}) the standard bremsstrahlung
spectrum of soft ``photons'' radiated by fast moving charges
\cite{TB-DIS}. In deriving this result, we have however neglected the
non-Abelian nature of the radiated fields, i.e., the fact that
they represent gluons, and not photons. This will be corrected
in the next subsection.

\subsection{Gluon saturation in a large nucleus}
\label{MV-SAT}

According to eq.~(\ref{MV-LOW}), the gluon density in the
transverse phase-space is proportional to $A^{1/3}$, and becomes
arbitrarily large when $A$ increases. This is however an artifact
of our previous approximations which have neglected the interactions
among the radiated gluons, i.e., the non-linear effects in the
classical field equations. To see this, one needs to recompute
the gluon distribution by using the exact, non-linear solution
for the classical field, as obtained in Sect. \ref{YMSOL}.
This involves the following LC-gauge field-field correlator:
\be\label{FF0}
\langle {\cal F}^{+i}_a(\vec x) {\cal F}^{+i}_a
(\vec y) \rangle_A &=&
\Bigl\langle\Bigl (U_{ab}^\dagger\partial^i\alpha^b\Bigr)_{\vec x}\,
\Bigl(U_{ac}^\dagger\partial^i\alpha^c\Bigr)_{\vec y}\Bigr\rangle_A,\ee
which, in view of the non-linear calculation,
has been rewritten in terms of the classical field in the
COV-gauge (cf. eq.~(\ref{FF-gauges})),
where $\tilde {\cal F}^{+i}_a=-\partial^i\alpha_a$.
To evaluate  (\ref{FF0}), one expands the Wilson lines in powers
of $\alpha$ and then contracts the $\alpha$ fields in all the possible
ways with the following propagator:
\be\label{MV-aa}
\langle \alpha_a(\vec x)\alpha_b(\vec y)
\rangle_A&=&
\delta_{ab}\delta(x^- -y^-)\,\gamma_A(x^-,x_\perp-y_\perp),\nn
\gamma_A(x^-,k_\perp)&\equiv&\frac{1}{ k_{\perp}^4}\,\lambda_A(x^-).\ee
We have used here
$\tilde\rho^a(x^-, k_\perp)=k_{\perp}^2\alpha^a(x^-, k_\perp)$, 
cf. eq.~(\ref{alpha}), together with eq.~(\ref{MV-corr})
which holds in any gauge and, in particular, in the COV-gauge.
The propagator (\ref{MV-aa}) is very
singular as $k_\perp\to 0$, but this turns out to be (almost)
harmless for the considerations to follow.

The fact that the fields $\alpha$ are uncorrelated
in $x^-$ greatly
simplifies the calculation of the correlator (\ref{FF0}).
Indeed, this implies that the two COV-gauge electric fields
$\partial^i\alpha_b(\vec x)$ and $\partial^i\alpha_c(\vec y)$
can be contracted only together, and not with the other fields 
$\alpha$ generated when expanding the Wilson lines. That is:
\be\label{CONTRACTION}
\Bigl\langle\Bigl (U_{ab}^\dagger\partial^i\alpha^b\Bigr)_{\vec x}\,
\Bigl(U_{ac}^\dagger\partial^i\alpha^c\Bigr)_{\vec y}\Bigr\rangle
=\left\langle \partial^i\alpha^b(\vec x)
\partial^i\alpha^c(\vec y)\right\rangle
\left\langle U_{ab}^\dagger 
(\vec x) U_{ca}(\vec y)\right\rangle\nn
=\delta(x^- -y^-)\langle{\rm Tr}\,U^\dagger(\vec x) U(\vec y)
\rangle\big(-\grad^2_\perp\gamma_A(x^-,x_\perp-y_\perp)\big),\quad\ee
where we have used $U_{ac}^\dagger=U_{ca}$ in the adjoint
representation. Eq.~(\ref{CONTRACTION})
can be proven as follows: i) By rotational
symmetry, $\partial^i\alpha(\vec x)$ cannot be contracted with a
field $\alpha(z^-,x_\perp)$ resulting from the expansion of
$U^\dagger(\vec x)$; indeed:
\be
\langle\alpha(z^-,x_\perp)\partial^i\alpha(x^-,x_\perp)\rangle
\propto \partial^i\gamma_A(x^-,r_\perp)\Big|_{r_\perp=0}
\propto \int {d^2k_\perp \over (2 \pi)^2}\,\frac{-ik^i}{k_\perp^2}
\,=\,0\,.\nonumber\ee
ii) Contractions of the type
\be\label{2contr}
\langle\alpha(z^-,y_\perp)\partial^i\alpha(x^-,x_\perp)\rangle\,
\langle\alpha(u^-,x_\perp)\partial^i\alpha(y^-,y_\perp)\rangle 
\propto\delta(x^- -z^-)\delta(u^- -y^-)\nonumber\ee
are not allowed by the ordering of the Wilson lines in $x^-\,$:
$\alpha(z^-,y_\perp)$ has been generated by expanding $U^\dagger(\vec y)$,
which requires $z^-<y^-$ (and similarly $u^- < x^-$). Then, the
first contraction in (\ref{2contr}) implies $x^-=z^-<y^-$,
while the second one leads to the contradictory requirement
$y^-=u^- < x^-$. 

The allowed contractions in eq.~(\ref{CONTRACTION}) involve:
\be\label{SAdef}
S_A(x^-, x_\perp-y_\perp)\,\equiv\,\frac{1}{N_c^2-1}\,
\langle{\rm Tr}\,U^\dagger(x^-,x_\perp) U(x^-,y_\perp)\rangle_A,\ee
which is like the $S$-matrix element (\ref{Stau})
for the dipole-hadron scattering, but now for a colour dipole
in the adjoint representation (i.e., a dipole made of two gluons).
This can be computed by expanding the Wilson lines, performing
contractions with the help of eq.~(\ref{MV-aa}),
and recognizing the result as the expansion of an ordinary
exponential. One thus finds (see
also Sect. \ref{MFA} for a more rapid derivation):
\be\label{SA}
S_A(x^-, r_\perp)&=&{\rm exp}\Big\{-g^2N_c[\xi_A(x^-,0_\perp)
-\xi_A(x^-,r_\perp)]\Big\},\nn
\xi_A(x^-,r_\perp)&\equiv& \int_{-\infty}^{x^-}dz^- 
\,\gamma_A(z^-, r_\perp),\ee
where the exponent can be easily understood: It arises as
\be\label{Vdipole}
\Big\langle gT^a(\alpha_a(\vec x)-\alpha_a(\vec y))\,
gT^b (\alpha_b(\vec x)-\alpha_b(\vec y)) \Big\rangle
\ee
where $igT^a(\alpha_a(\vec x)-\alpha_a(\vec y))$ is the
amplitude for the dipole scattering
off the ``Coulomb'' field $\alpha_a$, to lowest order in this 
field (i.e., the amplitude for a single scattering). Then, (\ref{Vdipole})
is the amplitude times the complex conjugate amplitude, that is,
the {\em cross section} for such a single scattering. This appears as an
exponent in eq.~(\ref{SA}) since this equation resums
multiple scatterings to all 
orders, and, in the eikonal approximation, the all-order
result is simply the exponential of the lowest order result. 
Since, moreover, $\alpha_a$ is the field created by the colour sources 
in the hadron (here, the valence quarks), we deduce that eq.~(\ref{SA})
describes the multiple scattering of the colour dipole off
these colour sources. 

If the field $\alpha_a$ is slowly varying over the transverse size
$r_\perp=x_\perp-y_\perp$ of the dipole (``small dipole''), one can expand
\be\label{dipole-F}
gT^a(\alpha_a(\vec x)-\alpha_a(\vec y))\approx-gT^a(x^i-y^i)
\partial^i\alpha_a(\vec x)=gT^a(x^i-y^i)\tilde {\cal F}^{+i}_a(\vec x),
\quad\ee
and then eq.~(\ref{Vdipole}) involves the correlator of two
(COV-gauge) electric fields. This is indeed the case, at it
can be seen by an analysis of the exponent in eq.~(\ref{SA}) :
\be\label{mux}
\xi_A(x^-,0_\perp)-\xi_A(x^-,r_\perp)&=&\mu_A(x^-)\int 
{d^2k_\perp\over (2\pi)^2}\,
\frac{1}{k_\perp^4}\,\Bigl[1-
{\rm e}^{ik_\perp\cdot r_\perp}\Bigr]\,,\nn
\mu_A(x^-)&\equiv&\int_{-\infty}^{x^-}dz^-\lambda_A(z^-).
\label{xiox}\ee
The above integral over $k_\perp$ is dominated by soft momenta,
and has even a logarithmic
divergence which reflects the lack of confinement in our model
(see also \cite{LM00}).
Note, however, that the dominant, quadratic, infrared divergence 
$\sim \int (d^2k_\perp/k_\perp^4)$, which would characterize the
scattering of a {\em coloured} particle (a single gluon) off the hadronic
field\footnote{Such a divergence would occur in
$\langle U^\dagger(x)\rangle_A$, which describes the scattering
of a single gluon.},
has cancelled between the two components of the 
{\em colourless} dipole. The remaining, logarithmic, divergence 
can be cut off by hand, 
by introducing an infrared cutoff $\Lambda_{QCD}$. Then one can expand:
\be\label{xi1}
\int {d^2k_\perp\over (2\pi)^2}
\frac{1-
{\rm e}^{ik_\perp\cdot r_\perp}}{k_\perp^4}\simeq
\int\limits^{1/r_\perp^2} {d^2k_\perp\over (2\pi)^2}
\frac{1}{k_\perp^4}{(k_\perp\cdot r_\perp)^2\over 2}
\simeq{r_\perp^2\over 16\pi}
\ln{1\over r_\perp^2\Lambda^2_{QCD}}\,.\,\,\,\,\,\,\,\,\ee
(This is valid to leading logarithmic accuracy, 
since the terms neglected in this way are not enhanced by a large 
transverse logarithm.)
We thus obtain:
\be\label{SMV}
S_A(x^-, r_\perp)\,\simeq\,
{\rm exp}\left\{-\,\frac{\alpha_s N_c}{4}\, r_\perp^2 \,\mu_A(x^-)\,
\ln{1\over r_\perp^2\Lambda^2_{QCD}}\right\},\ee
which together with eq.~(\ref{CONTRACTION}) can be used to finally
evaluate the gluon density (\ref{TPS}). This requires a double Fourier
transform (to $k^+$ and $k_\perp$), as shown in eq.~(\ref{FF}).
The presence of the $\delta$-function in eq.~(\ref{CONTRACTION}) makes
the Fourier transform to $k^+$ trivial, and one gets:
\be\label{NMFA}
{\cal N}_A(k_\perp)=\frac{N_c^2-1}{4\pi^3}
\int d^2r_\perp {\rm e}^{-ik_\perp\cdot r_\perp}
\int dx^- S_A(x^-, r_\perp)(-\grad^2_\perp\gamma_A(x^-,r_\perp)),
\nn\ee
where (cf. eqs.~(\ref{MV-aa}) and (\ref{mux})) :
\be\label{gradgamma}
-\grad^2_\perp\gamma_A(x^-,r_\perp)=\lambda_A(x^-)
\int {d^2p_\perp\over (2\pi)^2}\,
\frac{{\rm e}^{ip_\perp\cdot x_\perp}}{p_\perp^2}
=\frac{1}{4\pi}\ln{1\over r_\perp^2\Lambda^2_{QCD}}\,\frac
{\partial\mu_A(x^-) }{\partial x^-}.\nn\ee

The non-linear effects in eq.~(\ref{NMFA}) are encoded in the
quantity $S_A(x^-, r_\perp)$, which finds its origin in the gauge rotations
 in the r.h.s. of eq.~(\ref{FF0}). In fact, by replacing
$S_A(x^-, r_\perp)\to 1$ in eq.~(\ref{NMFA}),
one would recover the  linear approximation of eq.~(\ref{MV-LOW}). 
To  perform the integral over $x^-$ in eq.~(\ref{NMFA}), we note
that the quantity (\ref{gradgamma}) is essentially the derivative
w.r.t. $x^-$ of the exponent in $S_A(x^-, r_\perp)$, 
eq.~(\ref{SMV}). Therefore:
\be\label{NMV}
{\cal N}_A(k_\perp)=\frac{N_c^2-1}{4\pi^4}
\!\int \! d^2r_\perp {\rm e}^{-ik_\perp\cdot r_\perp}\,
\frac{1-{\rm exp}\Big\{-\frac{1}{4}\, r_\perp^2 Q_A^2
\ln{1\over r_\perp^2\Lambda^2_{QCD}}\Big\}}{\alpha_s N_cr_\perp^2 }\,,\,\,\ee
where 
\be\label{QAMV}
Q_A^2\,\equiv\,\alpha_s N_c \mu_A\,=\,\alpha_s N_c 
\int dx^-\lambda_A(x^-)\,\sim\,A^{1/3}.\ee
Eq.~(\ref{NMV}) is the complet result for the gluon density
of a large nucleus in the MV model \cite{JKMW97,KM98}. 
To study its dependence
upon $k_\perp$, one must still perform the Fourier transform,
but the result can be easily anticipated: 

i) At high momenta
$k_\perp\!\gg\! Q_A$, the integral is dominated by small
distances $r_\perp\!\ll \!1/ Q_A$, and can be evaluated by
expanding out the exponential. To lowest non-trivial order (which
corresponds to the linear approximation), one
obtains the bremsstrahlung spectrum of eq.~(\ref{MV-LOW}):
\be\label{High-N}
{\cal N}_A(k_\perp)\,\propto\,
\frac{ 1}{\alpha_s N_c }\,\frac{Q_A^2}{k_\perp^2}
\,=\,\frac{\mu_A}{k_\perp^2}\qquad {\rm for}\quad k_\perp\gg Q_A.\ee
ii) At small momenta, $k_\perp\!\ll \! Q_A$, the dominant contribution
comes from large distances $r_\perp\!\gg\! 1/ Q_A$, where one
can simply neglect the exponential in the numerator and recognize
$1/r_\perp^2$ as the Fourier transform\footnote{The saturation scale
provides the ultraviolet cutoff for the logarithm in eq.~(\ref{Low-N})
since the short distances $r_\perp\ll 1/ Q_A$ are cut off
by the exponential in eq.~(\ref{NMV}).} of $\ln k_\perp^2\,$:
\be\label{Low-N}
{\cal N}_A(k_\perp)\,\approx\,\frac{N_c^2-1}{4\pi^3} \,
\frac{ 1}{\alpha_s N_c }\ln\frac{Q_A^2}{k_\perp^2}
\qquad {\rm for}\quad k_\perp\ll Q_A.\ee
There are two fundamental differences between eqs.~(\ref{High-N})
and (\ref{Low-N}), which refer both to a {\it saturation}
of the increase of the gluon density: either with $1/k_\perp^2$
(at fixed atomic number $A$), or with $A$
(at fixed transverse momentum $k_\perp$). In both cases,
this saturation is only {\it marginal} : in the low--$k_\perp$ regime,
eq.~(\ref{Low-N}), the gluon density keeps increasing with 
$1/k_\perp^2$, and also with $A$, but this increase is only {\it logarithmic}, 
in contrast to the strong, power-like, increase $\propto (A^{1/3}/k_\perp^2)$ 
in the high--$k_\perp$ regime, eq.~(\ref{High-N}).

\begin{figure} 
\begin{center} 
\includegraphics[width=0.9\textwidth]
{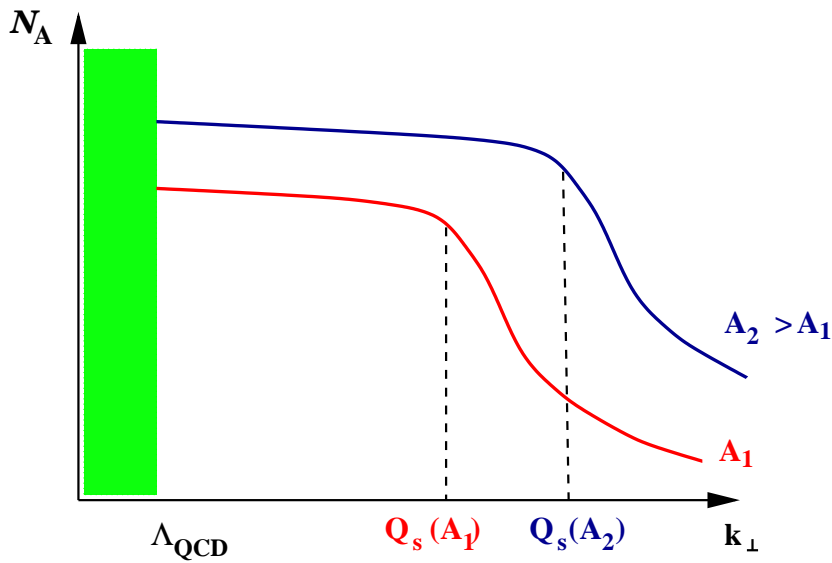} 
\caption{The gluon phase-space density ${\cal N}_A(k_\perp)$
of a large nucleus
(as described by the MV model) plotted as a function of $k_\perp$.}
\label{SATURATION-MV} 
\end{center}
\end{figure}

Moreover, the gluon density at low $k_\perp$ is of order $1/{\alpha_s}\,$,
which is the maximum density allowed by the repulsive interactions
between the strong colour fields $\bar A^i=\sqrt{\langle A^iA^i\rangle}\sim
1/g$. When increasing the atomic number $A$, the new gluons are
produced preponderently at large transverse momenta $\simge Q_{A}$,
where this repulsion is less important. This is illustrated in Fig. 
\ref{SATURATION-MV}.

To be more precise, the true scale which separates between the two
regimes (\ref{High-N}) and (\ref{Low-N}) is not $Q_A$, but rather
the {\it saturation momentum} $Q_s(A)$ which is the reciprocal
of the distance $1/r_\perp$ where the exponent
in eq.~(\ref{NMV}) becomes of order one. Thus, this is defined
as the solution to the following equation:
\be\label{QSMV}
Q_s^2(A)\,=\,{1\over 4}\,
\alpha_s N_c \mu_A\,\ln{Q_s^2(A)\over \Lambda^2_{QCD}}\,.
\ee
To clarify its physical interpretation,
note that, at short-distances $r_\perp\!\ll \!1/ Q_A$,
\be\label{QSMV1}
\mu_A\,
\ln{1\over r_\perp^2\Lambda^2_{QCD}}\,\propto\,
\frac{\x G(\x, 1/r_\perp^2)}{(N^2_c-1)\pi R_A^2}\,\ee
is the number of gluons (of each colour) having tranverse
size $r_\perp$ per unit of transverse area (cf. eq.~(\ref{MV-LOW})).
Since each such a gluon carries a colour charge squared $(gT^a)(gT^a)
=g^2N_c$, we deduce that
\be\label{chargeQs}
\alpha_s N_c\,\mu_A\, \ln{1\over r_\perp^2\Lambda^2_{QCD}}\ee
is the average colour charge squared of the
gluons having tranverse size $r_\perp$ per unit area and per colour. 
Then, eq.~(\ref{QSMV}) is the condition that the total colour
charge squared within the area occupied by each gluon is of
order one. This is the original criterion of saturation by
Gribov, Levin and Ryskin \cite{GLR}, 
for which the MV model offers an explicit realization.

To conclude this discussion of the MV model, note that,
in the previous computation, we have
also obtained the $S$-matrix element $S_A(r_\perp)$ for the
dipole-hadron scattering (cf. Sect. \ref{sect-dipole}).
This is given by eq.~(\ref{SMV}) with $\mu_A(x^-)\to \mu_A$
and $N_c=T^aT^a$ replaced in general
by the colour Casimir $t^at^a$ for the representation of interest
(e.g., $C_f=(N^2_c-1)/2N_c$ for the fundamental representation).
As discussed after eq.~(\ref{Vdipole}), this describes the
multiple scattering of the colour dipole on the colour field in the 
hadron (here, the field of the valence quarks). 
According to  eq.~(\ref{SMV}), one
can distinguish, here too, between a short-distance and
a large-distance regime, which moreover are separated by the
same ``saturation scale'' as for the gluon distribution:

i) A small-size dipole $r_\perp\!\ll \!1/ Q_s$ is only
weakly interacting with the hadron:
\be
1-S_A(r_\perp)\,\approx\,\frac{1}{4}\, r_\perp^2 Q_A^2
\ln{1\over r_\perp^2\Lambda^2_{QCD}}\qquad {\rm for}\quad
r_\perp \ll 1/ Q_s(A),\ee
a phenomenon usually referred to as ``colour transparency''.

ii) A relatively large dipole, with  $r_\perp\! \gg \!1/ Q_s$, 
is strongly absorbed: 
\be\label{BDL}
S_A(r_\perp)\,\approx\,0\qquad {\rm for}\quad
r_\perp \gg 1/ Q_s(A),\ee
a situation commonly referred to as the ``black disk'', or ``unitarity'',
 limit.

The remarkable fact that the critical dipole size is set by the
saturation scale $Q_s$ can be understood as follows: A small dipole
--- small as compared to the typical variation scale of the external 
Coulomb field --- couples to the associated electric field 
$\tilde {\cal F}^{+i}$ (cf. eq.~(\ref{dipole-F})), 
so its cross-section for one scattering, eq.~(\ref{Vdipole}),
is proportional to the number of 
gluons $\langle\tilde {\cal F}^{+i}\tilde {\cal F}^{+i}\rangle$
within the transverse area $r_\perp^2$ explored by the dipole.
This is manifest on eq.~(\ref{SMV}), whose exponent
is precisely the colour charge squared of the gluons
within that area (cf. the remark after eq.~(\ref{chargeQs})).
At saturation, this charge becomes of order one,
and the dipole is strongly interacting. The important lesson 
is that the unitarity limit (\ref{BDL}) for the scattering 
of a small dipole on a high energy hadron is equivalent to gluon saturation
in the hadron wavefunction \cite{AM0,AM2,SAT,K,AMCARGESE}. 

\section{Quantum evolution and\\ the Colour Glass Condensate}
\label{RGE-CGC}
\setcounter{equation}{0}

In this section, we show that the classical Yang-Mills theory
described in Sect. \ref{EFT} can be actually derived from QCD as
an effective theory at small x. This requires integrating out
quantum fluctuations in layers of $p^+$, which can be done
with the help of a renormalization group equation (RGE) for
the weight function $W_\tau[\rho]$. We shall not present all
the calculations leading to this RGE; this would require heavy
technical developments going far beyond the purpose of these lectures.
(See Ref. \cite{PI} for more details.) Rather, we shall emphasize the
general strategy of this construction and the physical picture 
behind it (that of the colour glass), together with those elements
of the calculation which are important to understand the structure
of the final equation.

\subsection{The BFKL cascade}
\label{SBFKL}

In Sect. \ref{EFT-YM}, we have argued that the radiation
of a soft gluon by a fast parton via the tree-level graph
shown in Fig. \ref{cascade}.a can be described as a classical
process with a colour source whose structure is largerly fixed
by the kinematics. Our main goal in this section will be to show
that this picture is not spoilt by quantum corrections.
We start by showing that the dominant quantum corrections,
those which will be resummed in what follows, preserve indeed
the separation of scales which lies at the basis of the effective
theory developed in Sect. \ref{EFT}.

\begin{figure} 
\begin{center} 
\includegraphics[width=0.9\textwidth]
{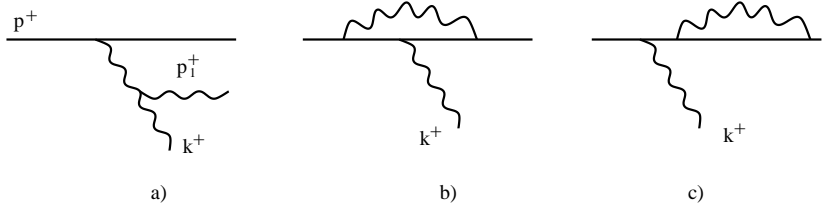} 
\caption{Lowest-order quantum corrections to the
emission of a soft gluon by a fast parton: a) a real-gluon emission;
b) a vertex correction; c) a  self-energy correction. } 
\label{one-gluon-fig} 
\end{center}
\end{figure}

Consider first the lowest-order radiative correction
to the tree-level graph in Fig. \ref{cascade}.a, namely,
the emission of one additional (quantum) gluon, as
shown in Fig. \ref{one-gluon-fig}.a. At the same level of accuracy,
one should include also the vertex and self-energy
corrections illustrated in
Fig. \ref{one-gluon-fig}.b, c. This will be done in the complete calculation
presented in Sect. \ref{SRGE}. But in order to 
get a simple order-of-magnitude estimate for the quantum 
corrections --- which is our purpose in this subsection ---
it is enough to consider the radiative process in Fig. \ref{one-gluon-fig}.a.

The probability for the emission of a quantum
gluon with longitudinal momentum $p_1^+$ in the
range $p^+ > p_1^+ > k^+$ is
\beq\label{one-gluon}
\Delta P\,\propto\,\frac{\alpha_s N_c}{\pi}
\int_{k^+}^{p^+}\,\frac{d p_1^+}{p_1^+}
\,=\,\frac{\alpha_s N_c}{\pi}
\,\ln\,\frac{p^+}{k^+}\,\sim \,\alpha_s\ln\frac{1}{\rm x}\,.
\eeq
This becomes large when the available interval of rapidity
$\Delta \tau= \ln(1/{\rm x})$ is large.
This is the typical kind of quantum correction that
we would like to resum here. A calculation which
includes effects of order $(\alpha_s\ln(1/{\rm x}))^n$ to all orders
in $n$ is said to be valid to ``leading logarithmic accuracy'' (LLA).

\begin{figure}[htb]
\centering
\resizebox{.9\textwidth}{!}{%
\includegraphics*{{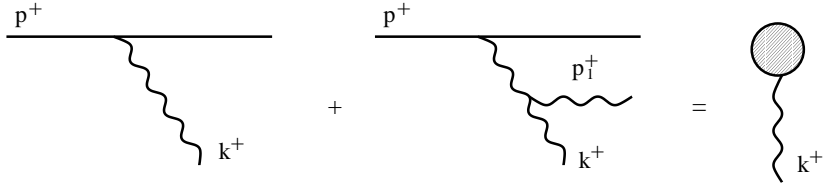}}}
   \caption{Effective colour source after including
the lowest-order radiative correction.}
\label{SOURCE}
\end{figure}

The typical contributions to the
logarithmic integration in eq. (\ref{one-gluon}) come from modes
with momenta $p_1^+$ {\em deeply} inside the strip:
$p^+ \gg p_1^+ \gg k^+$. Thus, in Fig. \ref{one-gluon-fig}.a, the
soft final gluon with momentum $k^+$ is emitted typically
from a relatively fast gluon, with momentum $p_1^+ \gg k^+$. This
latter gluon can therefore
be seen as a component of the {\it effective} colour 
source at the soft scale $k^+$. In other terms,
one can visualise the combined effect of the tree-level
process, Fig. \ref{cascade}.a, and the first-order radiative correction,
Fig. \ref{one-gluon-fig}.a, as the generation of a modified
colour source at the scale $k^+$, 
which receives contributions {\it only} 
from the modes with longitudinal momenta much larger than $k^+$.
This is illustrated in Fig. \ref{SOURCE}.

Clearly, when x is small enough, $\ln(1/{\rm x}) \sim 1/\alpha_s$,
the ``correction'' (\ref{one-gluon}) becomes of ${\cal O}(1)$,
and it is highly probable that more gluons will be emitted along the way.
This gives birth to the gluon cascade depicted in Fig.  \ref{cascade}.b,
whose dominant contribution, for a fixed number of ``rungs''
$N$, is of order $(\alpha_s\ln(1/x))^N$,
and comes from the kinematical domain where
the longitudinal momenta are strongly ordered:
\beq\label{HL}
p^+\equiv p_0^+\,\gg\,p_1^+\,\gg\,p_2^+\,\gg\,\cdots\,\gg\,
p_N^+\equiv k^+.\eeq
(Other momentum orderings give contributions which are suppressed by,
at least, one factor of $1/\ln(1/{\rm x})$, 
and thus can be neglected to LLA.) With this ordering,
this is the famous BFKL cascade, that we would like to include in our 
effective source. This should be possible since the
hierarchy of scales in eq.~(\ref{HL}) is indeed consistent
with the  kinematical assumptions in Sect. \ref{EFT}.

Note first that, the strong ordering (\ref{HL}) in longitudinal
momenta implies a corresponding ordering in the lifetimes
of the emitted  gluons (cf. eq.~(\ref{lifetime})):
\be
\Delta x^+_0\,\gg\,\Delta x^+_1\gg\,\Delta x^+_2\,
\gg\,\cdots\,\gg\,\Delta x^+_N.\ee
Because of this, any newly emitted gluon lives too shortly
to notice the dynamics of the gluons above it. 
This is true in particular for the last emitted gluon,
with momentum $k^+$, which ``sees'' the $N$ previous gluons 
in the cascade as
a frozen colour charge distribution, with an average colour charge
${\cal Q}\equiv \sqrt{\langle {\cal Q}_a{\cal Q}_a\rangle} \sim N\/$. Thus,
this $(N+1)$th gluon is emitted {\em coherently} off the colour
charge fluctuations of the $N$ previous ones, 
with a differential probability (compare to eq.~(\ref{one-gluon})) :
\beq\label{PN}
d P_N\,\propto \,\frac{\alpha_s N_c}{\pi}\,N(\tau)\,d\tau.\eeq
When increasing the rapidity by one more step,
$\tau \to \tau+d\tau$, the number of radiated gluons
changes according to
\be
N(\tau+d\tau)=(1+N(\tau)) d P_N+N(\tau)(1-d P_N),\ee
which together with eq.~(\ref{PN}) implies  (with
$\bar\alpha_s \equiv \alpha_s N_c/\pi$)
\beq\label{expgr}
\x G(\x,Q^2)\,\equiv \,{dN\over d\tau}\,\sim\,C\alpha_s
{\rm e}^{\kappa \bar\alpha_s\tau}.\eeq
Thus, the gluon distribution 
grows exponentially with $\tau=\ln(1/{\rm x})$. A more refined 
treatment, using the BFKL equation, gives $\kappa=4\ln 2$,
and shows that the prefactor $C$ in the r.h.s. of
eq.~(\ref{expgr}) has actually a weak dependence on $\tau\,$: $C\propto
(\alpha_s\tau)^{-1/2}$ \cite{BFKL,TB-BFKL}.

Thus, the BFKL picture is that of an unstable growth of the
colour charge fluctuations as x  becomes smaller and smaller.
However, this evolution assumes the radiated gluons to behave as free 
particles, so it ceases to be valid at very low ${\rm x}$, 
where the gluon density becomes so large that their
mutual interactions cannot be neglected anylonger.
This happens, typically, when the interaction probability for the radiated
gluons becomes of order one, cf. eq.~(\ref{QSGLR}), which is also the
criterion for the saturation effects to be important
(compare in this respect
eq.~(\ref{QSGLR}) and eqs.~(\ref{QSMV})--(\ref{QSMV1})).
Thus one cannot study saturation consistently
without including non-linear effects in the quantum evolution.
It is our main objective in what follows to explain
how to do that.

\subsection{The quantum effective theory}
\label{QEFT}

To the accuracy of interest, 
quantum corrections can be incorporated in the effective
theory by renormalizing the source $\rho_a$
and its correlation functions (i.e., the weight function $W_\tau[\rho]$).
The argument proceeds by induction: We assume 
the effective theory to exist at some scale $\Lambda^+$ and show that
it can be extended at the lower scale $b\Lambda^+\ll \Lambda^+ $.
Specifically:

\bigskip
I) We assume that a {\em quantum}
effective theory exists at some original scale $\Lambda^+$
with $\Lambda^+ \ll P^+$. That is,  we assume that the {\em fast}
quantum modes with momenta $p^+\gg \Lambda^+$ can be replaced, 
as far as their effects on the correlation 
functions at the scale $\Lambda^+$ are concerned, by a 
classical random source $\rho_a$ with weight function 
$W_{\Lambda^+}[\rho]$. (We shall eventually convert $\Lambda^+$
into the rapidity $\tau$ by using $\tau=\ln(P^+/\Lambda^+)$.)
On the other hand, the {\em soft} gluons,  with momenta $p^+<\Lambda^+$,
are still explicitely present in the theory, as quantum gauge fields.
Thus, this effective theory
includes both the classical field ${\cal A}^i[\rho]$ generated by $\rho$,
and the { soft} quantum gluons.

Within this theory, the correlation functions of the
soft ($k^+\le \Lambda^+ $)
fields are obtained as (e.g., for the 2-point function)
\be\labe{2point} 
\langle {\rm T}A^\mu(x)A^\nu(y)\rangle
=\int {\cal D}\rho\,W_{\Lambda^+}[\rho]\left\{
\frac{\int^{\Lambda^+} {\cal D}A\,\delta(A^+)
\,A^\mu(x)A^\nu(y)\,{\rm e}^{\,iS[A,\,\rho]}}
{\int^{\Lambda^+} {\cal D}A\,\delta(A^+)\,\,{\rm e}^{\,iS[A,\,\rho]}}
\right\},\nonumber\\\ee
where T stays for time ordering (i.e. ordering in $x^+$).
This is written in the LC-gauge $A^+_a=0$, and
involves two functional integrals:

a) a quantum path integral
over the soft gluon fields $A^\mu$ at fixed $\rho$:
\be\labe{2point-Q} 
\langle {\rm T} A^\mu(x)A^\nu(y)\rangle_\rho
\,=\,\frac{\int^{\Lambda^+} {\cal D}A\,\delta(A^+)
\,A^\mu(x)A^\nu(y)\,{\rm e}^{\,iS[A,\,\rho]}}
{\int^{\Lambda^+} {\cal D}A\,\delta(A^+)\,\,{\rm e}^{\,iS[A,\,\rho]}}
,\ee
b) a classical average over $\rho$, like in 
eq.~(\ref{clascorr}) :
\be\labe{2point-CL} 
\langle {\rm T}A^\mu(x)A^\nu(y)\rangle
\,=\,\int {\cal D}\rho\,W_{\Lambda^+}[\rho]\,
\langle {\rm T} A^\mu(x)A^\nu(y)\rangle_\rho\,.\ee
The upper script ``$\Lambda^+$'' on the quantum path integral
is to recall the
restriction to soft ($|p^+| < \Lambda^+$) longitudinal 
momenta\footnote{The separation between fast and soft degrees of freedom 
according to their longitudinal momenta has a gauge-invariant
meaning (within the LC-gauge) since the residual gauge 
transformations, being independent of  $x^-$,
cannot change the $p^+$ momenta.}.
The action $S[A,\,\rho]$ is chosen such as
to generate the classical field equations (\ref{cleq0})
in the saddle point approximation $\delta S/\delta A^\mu=0$.
This requirement, together with gauge symmetry and the
eikonal approximation, single out the following action \cite{JKLW97} :
\be\label{ACTION}
S[A,\rho]=- \int d^4x \,{1 \over 4} \,F_{\mu\nu}^a F^{\mu\nu}_a
+{i \over {gN_c}} \int d^3 \vec x\, {\rm Tr}\,\Bigl\{ \rho(\vec x)
\,W[A^-](\vec x)\Bigr\},\,\,\,
\ee
where $W[A^-]$ is a Wilson line in the temporal direction:
\be\label{WLINE}
W[A^-](\vec x)\, =\,{\rm T}\, \exp\left\{\,ig\int dx^+ A^-(x) \right\}.
\ee
With this action, the condition $\delta S/\delta A^\mu
=0$ implies indeed eq.~(\ref{cleq0}) for field
configurations having $A^-_a=0$. Thus, the classical
solution ${\cal A}^\mu_a=\delta^{\mu i}{\cal A}^i_a[\rho]$
found in Sect. \ref{YMSOL} is the tree-level field
in the present quantum theory.

As long as we are interested in correlation functions at the scale
$\Lambda^+$, or slightly below it, we can satisfy ourselves with
this classical (or saddle point) approximation.
That is,  to the accuracy
to which holds the effective theory in eq.~(\ref{2point}), the gluon
correlations at the scale $\Lambda^+$ can be computed
from the classical field solution, as in eq.~(\ref{clascorr}).
But quantum corrections 
become important when we consider
correlations at a much softer scale $k^+\ll \Lambda^+$,
such that $\alpha_s\ln(\Lambda^+/k^+)\sim 1$.


\bigskip
II) Within the quantum effective theory, we integrate
out the {\em semi-fast} quantum fluctuations, i.e., the fields
with longitudinal momenta inside the strip:
\be\labe{strip}\,\,
 b\Lambda^+ \,\,\ll\,\, |p^+|\,\, \ll\,\,\Lambda^+\,,\quad
{\rm with}\quad b \ll 1\quad {\rm and} \quad\alpha_s\ln(1/b)< 1.\ee
This generates quantum corrections to the correlation functions
at the softer scale  $b\Lambda^+$,  which can be computed
by decomposing the total gluon field as follows:
\be\label{A-decomp}
A^\mu_c\,=\,{\cal A}^\mu_c[\rho]+a^\mu_c+\delta A^\mu_c.\ee
Here, ${\cal A}^\mu_c$ is the tree-level field,
$a^\mu_c$ are the semi-fast fluctuations to be integrated out, 
and $\delta A^\mu_c$ are the {\em soft} modes with momenta
$|p^+| \le b\Lambda^+$ whose correlations receive quantum
corrections from the semi-fast gluons.
 
These {\it induced} correlations  must be computed to leading
order in $\alpha_s\ln(1/b)$ (LLA), but to {\em all} orders in the 
classical fields ${\cal A}^i[\rho]$ (since we expect
${\cal A}^i\sim 1/g$ at saturation).
This amounts to an one-loop calculation, but with the exact
background field propagator $\langle a^\mu(x)a^\nu(y) \rangle_\rho$
of the semi-fast gluons. 
For instance, the quantum corrections to the 2-point function read
schematically:
\be\label{induced-2p}
\big\langle({\cal A}^i[\rho]+\delta A^i)
({\cal A}^j[\rho]+\delta A^j)\big\rangle_\rho\,-\,
{\cal A}^i[\rho]{\cal A}^j[\rho]\,=\,\nn
=\,{\cal A}^i[\rho]\langle \delta A^j\rangle_\rho+
\langle \delta A^i\rangle_\rho\,{\cal A}^j[\rho]+
\langle \delta A^i \delta A^j\rangle_\rho
\ee
where the brackets $\langle\cdots\rangle_\rho$ 
stand for the quantum average over the semi-fast fields
in the background of $\rho\,$; this average is defined 
as in eq.~(\ref{2point-Q}), but with the functional integral
now restricted to the fields $a^\mu_c$. The purpose of the quantum
calculation is to provide explicit expressions for the
1-point function $\langle \delta A^i\rangle_\rho$ and the
2-point function $\langle \delta A^i \delta A^j\rangle_\rho$
as functionals of $\rho$ (to the indicated accuracy). Once these
expressions are known, the 2-point function 
$\langle A^i(x) A^j(y)\rangle$ at the scale $b\Lambda^+$
can be finally computed as:
\beq\label{evolaa}
\langle A^i A^j\rangle\,=\,
\Big\langle\big\langle({\cal A}^i[\rho]+\delta A^i)
({\cal A}^j[\rho]+\delta A^j)\big\rangle_\rho\,\Big\rangle_{W_{\Lambda}},
\ee
where the external brackets $\langle\cdots\rangle_{W_{\Lambda}}$ denote
the classical average over $\rho$ with weight function $W_\Lambda[\rho]$,
as in eq.~(\ref{2point-CL}).

\bigskip
III) We finally show that the induced correlations
can be absorbed into a functional change $W_{\Lambda^+}[\rho]
\to W_{b\Lambda^+}[\rho]$ in the weight function for $\rho$.
That is, the result (\ref{evolaa}) of the classical+quantum
calculation in the effective theory at the scale $\Lambda^+$
can be reproduced by a purely classical calculation, but with
a modified weight function $W_{b\Lambda^+}[\rho]$, corresponding
to a new effective theory:
\beq
\Big\langle\langle({\cal A}^i[\rho]+\delta A^i)
({\cal A}^j[\rho]+\delta A^j)\rangle_\rho\,\Big\rangle_{W_{\Lambda}}
\,=\, \big\langle{\cal A}^i[\rho]
{\cal A}^i[\rho]\big\rangle_{W_{b\Lambda}},
\ee
where the average in the r.h.s. is defined as
in eq.~(\ref{clascorr}), or (\ref{2point-CL}),
but with weight function $W_{b\Lambda^+}[\rho]$.
This demonstrates the existence of
the effective theory at the softer scale $b\Lambda^+$.

Since $\Delta W\equiv  W_{b\Lambda^+}-W_{\Lambda^+}
\propto \alpha_s\ln(1/b)$, the evolution of the weight function
is best written in terms of rapidity:
$ W_{\tau+\Delta\tau} - W_\tau = -\Delta\tau H W_\tau $, where
$\tau=\ln(P^+/\Lambda^+)$, $\Delta\tau=\ln(1/b)$, and
$H\equiv H[\rho, {\delta\over \delta\rho}]$ is a functional
differential operator acting on $ W_\tau $ (generally,
a non-linear functional of $\rho$). 
In the limit $\Delta\tau\to 0$, this gives a  
{\em renormalization group equation} (RGE) describing
the flow of the weight function with $\tau$ \cite{JKMW97,JKLW97} :
\be\labe{RGE0}
{\del W_\tau[\rho] \over {\del \tau}}\,=\,- 
H\Big[\rho\,, {\delta\over \delta\rho}\Big]W_\tau[\rho] \,.\ee
By integrating this equation with initial conditions at
$\tau \ll 1$ (i.e., at $\Lambda^+\sim P^+$), one can
obtain the weight function at the rapidity $\tau$ of interest.
The initial conditions are not really perturbative, 
but one can rely on some non-perturbative model, like the MV model 
discussed in Sects. \ref{MVmodel}--\ref{MV-SAT}.

\begin{figure}[htb]
\centering
\resizebox{.9\textwidth}{!}{%
\includegraphics*{{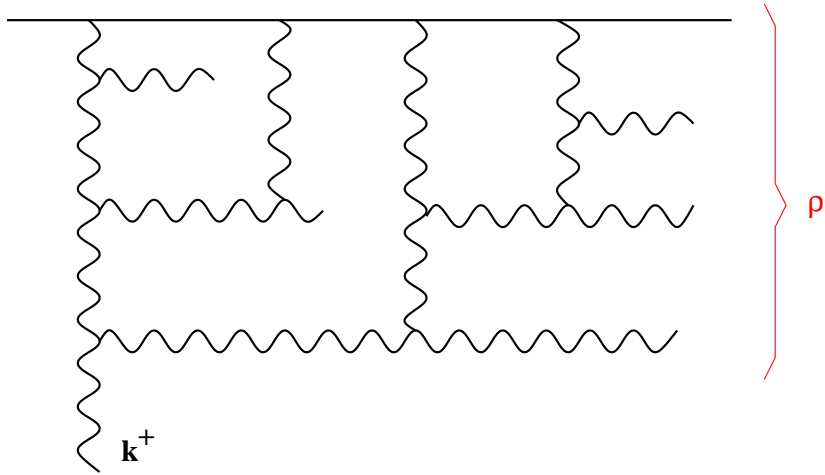}}}
   \caption{A typical Feynman diagram that is implicitely resummed
in the quantum evolution of the effective theory.}
\label{CAR_RHO}
\end{figure}

A key ingredient in this approach, which makes the difference
w.r.t. the BFKL equation, are the non-linear effects 
encoded in the background field calculation. Recall that
$\rho$, and therefore the classical fields ${\cal A}^i[\rho]$,
are random variables whose correlators (\ref{clascorr}) reproduce  
the gluon density and, more generally, the $n$-point correlation 
functions of the gluon fields at the scale $\Lambda^+$. 
Thus by computing
quantum corrections in the presence of these background fields,
and then averaging over the latter, one is effectively studying
quantum evolution in a medium with high gluon density.
After each step in this evolution, the properties
of the medium (i.e., the correlators of $\rho$) are updated,
by including the latest quantum corrections. In terms
of Feynman graphs of the ordinary perturbation theory,
this corresponds to a complicated resummation of diagrams
describing the interactions between the gluons radiated
in different parton cascades and at different rapidities.
A typical such a diagram is shown in Fig. \ref{CAR_RHO}. 
At low density, where the non-linear effects
can be neglected, eq.~(\ref{RGE0}) correctly reproduces the
BFKL equation \cite{JKLW97},  as it should (see Sect. \ref{S-BFKL} below).

\subsection{The Colour Glass Condensate}
\label{CGC}

Note the special form of the average in eq.~(\ref{2point}).
This is {\em not} the same as :
\be\label{AV-ORD}
\frac{\int {\cal D}\rho\,\,W_\Lambda[\rho]\,
\int^\Lambda {\cal D}A
\,A^\mu(x)A^\nu(y)\,\,{\rm e}^{\,iS[A,\,\rho]}}
{\int {\cal D}\rho\,\,W_\Lambda[\rho]\,
\int^\Lambda {\cal D}A\,\,{\rm e}^{\,iS[A,\,\rho]}}\,.\ee
In eq.~(\ref{AV-ORD}), both the colour source $\rho_a$ and
the gauge fields  $A^\mu_a$
are dynamical variables that are summed over on the same
footing. They are free to take on values which extremize
the total ``effective action'' :
\be
S_{eff}[A,\,\rho]\,=\,S[A,\,\rho]\,-\,i\ln W_\Lambda[\rho].\ee
By contrast, in eq.~(\ref{2point}), the average over
$A^\mu$ is taken at fixed $\rho$ : the gauge fields
can vary in response to $\rho$, but $\rho$ cannot vary in
response to the gauge fields. That is, $\rho$ is not
a dynamical variable, but rather an ``external'' source.
Giving a colour charge distribution $\rho_a(\vec x)$ 
specifies a medium in which propagate
the quantum gluons. But this medium is,  by itself, random,
so after performing the quantum analysis at fixed $\rho$,
one must also perform an average over $\rho$.
The reason for treating $\rho$ and $A^\mu$  differently lies
is the separation of scales in the problem: the changes
in $\rho$ happens on time scales much larger than the 
lifetime of the soft gluons.
This situation is typical for amorphous materials called
``glasses''.

The prototype of such systems is a ``spin glass''
\cite{SG-TB}, that is,
a collection of magnetic impurities 
(the ``spins'') which are randomly distributed in a 
non-magnetic metal host. For instance, one can take the spins
to sit on a regular lattice with lattice sites $i,j,\dots$,
and interaction Hamiltonian
\be
H_J[S]\,=\,-\sum_{< i,j>} J_{ij} S_i S_j,\ee
(the sum runs over all pairs $< i,j>$, and the spins
$S_i$ are allowed to take two values, $+1$ or $-1$),
but let their interactions (the ``link variables''
$J_{ij}$) to be random, with a Gaussian probability distribution,
for simplicity:
\be\label{PROB-J}
dP[J]\,=\,\prod_{< i,j>} dJ_{ij} P(J_{ij}),\qquad
P(J_{ij})\,=\,\frac{1}{\sqrt{2\pi\Delta_{ij}}}\,
{\rm e}^{-\frac{J_{ij}^2}{2\Delta_{ij}}}.\ee
Physically, this corresponds to the fact that
the modifications in $J_{ij}$ occur
on time scales much larger than the time scales 
characterizing the dynamics of the spins (e.g.,
their thermalization when the system is brought in
contact with a thermal bath). In practice, the $J_{ij}$'s are
frozen into their fixed values by rapid cooling when the
sample is prepared. This kind or rapid cooling is called
``quenching'', and one says that the $J_{ij}$'s are
``quenched variables'', as opposed to the ``dynamical
variables'', the spins $S_i$. This procedure selects 
random values for the $J_{ij}$'s,
with the probability distribution (\ref{PROB-J}).

Thus, the spins thermalize for a given set of
``quenched variables'', and for each such a set one can
compute the thermal partition function and the free energy:
\be\label{SG-FJ}
Z[J]\,=\,\sum_{\{S\}}\,{\rm e}^{-\beta H_J[S]},\qquad
F[J]\,=\,-T\ln Z[J].\ee
But the $J_{ij}$'s are themselves random, so the experimentally relevant 
quantity is the following average
\be \label{SG-F}
F\,=\,\langle F[J] \rangle\,\equiv\,\int dP[J]\,F[J]\,=\,-T
\int dP[J]\,\ln Z[J].\ee
Note that it is $\ln Z[J]$, not $ Z[J]$ itself, which should be 
averaged (``quenched average''). 
Similarly, (connected) correlation functions are 
generated by the free energy in the presence of a site-dependent
external magnetic field:
\be\label{SG-2P}
\langle S_iS_j \rangle\,-\,\langle S_i\rangle
\langle S_j\rangle\,=\,T^2\int dP[J]\,\frac{\partial ^2 \ln Z[J,h]}
{\partial h_i\partial h_j}\,,\ee
with $\ln Z[J,h]$ defined as in eq.~(\ref{SG-FJ}), but with
$H_J[S]\to H_J[S]-\sum_i h_i S_i$.

Eqs.~(\ref{SG-F})--(\ref{SG-2P}) are the analogs of
eq.~(\ref{2point}) for the problem at hand: the colour source $\rho_a$
is our ``quenched variable'', and the  quantum average over the fields
$A^\mu$ at fixed $\rho$, eq.~(\ref{2point-Q}), corresponds to the
thermal average at fixed $J_{ij}$'s, eq.~(\ref{SG-FJ}).
As in eq.~(\ref{SG-F}), it is $\ln Z$, and not
$Z$, which is effectively averaged in eq.~(\ref{2point}) 
(the average of
$Z$ would rather corresponds to eq.~(\ref{AV-ORD})).
In fact, the {\em connected} correlation functions of the soft
gluons in the effective theory are obtained from 
the following generating functional:
\be\labe{PART}
{F}[j^\mu_a]\,=\,\int {\cal D}\rho\,\,W_\Lambda[\rho]
\,\,\ln\left(\int^\Lambda {\cal D}A\,
\delta(A^+)\,\,{\rm e}^{\,iS[A,\,\rho]-i\int j\cdot A}\right),\ee
which is the analog of eq.~(\ref{SG-F}) with $\ln Z[J]
\to \ln Z[J,h]$. (The external current $j^\mu_a$ in (\ref{PART})
is just a device to generate
Green's functions via differentiations,
and should not be confused with the physical source $\rho_a$.)
 
We are thus naturally led to interprete the small-x component
of the hadron wavefunction as a {\em glass}, with the colour
charge density playing the role of the spin for spin glasses. 
Thus, this is a {\em colour} glass. 
Unlike what happens for spin glasses, which may have 
a non-zero value for the average magnetization $\langle S_i\rangle$
(at least locally, i.e., at a given site), the
{\em average} colour charge must be zero, 
\be\label{no-rho}
\langle \rho_a(\vec x) \rangle\,=\,0 \quad{\rm at\,\,any\,\,}
\vec x,\ee
by gauge symmetry. In practice, this is insured by the fact that 
we sum over all the possible configurations of 
$\rho_a(\vec x)$ with a gauge-invariant weight function.
Let us however examine a particular configuration 
$\rho_a(\vec x)$ from this ensemble. We now argue that,
at sufficiently small x (or large atomic number $A$), this
configuration describes typically a {\em Bose condensate}.

This applies to the {\em saturated} modes, i.e.,
the modes with transverse momenta
$\Lambda_{QCD}\ll k_\perp\ll Q_s(\tau)$ and longitudinal momenta
$k^+={\rm x}P^+\ll P^+$. As argued in Sect. \ref{MV-SAT}, 
these modes are characterized by a high gluon number density
in the transverse phase-space,
${\cal N}_\tau( k_\perp) \sim 1/\alpha_s$. (This prediction
of the classical MV model remains valid after including the
quantum evolution, as we shall see in Sect. 5.4 below.)
Microscopically, these modes correspond to bosonic
states with large occupation numbers $\sim 1/\alpha_s$. 
Each such a state is a Bose condensate.

More precisely, the general definition of a Bose condensate is
that of a 
quantum state in which the Fock space annihilation operator
$a^i_c(\vec k)$ (cf. eq.~(\ref{Fock})), or, equivalently,
the field operator $A^i_c (x)$, takes on a non-zero
expectation value. This situation may be characterized  as
the spontaneous generation of a classical field.
Of course, this cannot happen for gluons in the vacuum, 
as it would violate gauge symmetry.
And, in an absolute sense, this cannot happen in a hadron
neither, since the average colour charge vanishes there too
(cf. eq.~(\ref{no-rho})),
and therefore so does the associated classical field:
$\langle{\cal A}^i_c[\rho]\rangle=0$. But in the hadron there
{\em are} colour sources, and, as argued before, they can be even
treated as a classical charge distribution which is frozen
during the short lifetime of the small-x gluons. Thus, over such
a short time scale (short as compared to the typical time scale
for changes in the colour distribution), one effectively has a
non-trivial classical field ${\cal A}^i_c[\rho]$. At saturation,
this field is typically strong (cf. eqs.~(\ref{NMV}) and (\ref{FASYMP})) :
\be\Bigl\langle  {\cal A}^{ia}_\infty(x_\perp)\,
 {\cal A}^{ia}_\infty(y_\perp)\Bigr\rangle
 &\approx &\frac{N_c^2-1}{\pi N_c}
\frac{1-{\rm e}^{-r_\perp^2 Q_s^2}}{\alpha_s r_\perp^2}\,,\\
\bar {\cal A}^i \,\sim \, \sqrt{{\langle  {\cal A}^{i}_\infty
{\cal A}^{i}_\infty\rangle}} &\sim &\frac{1}{ \sqrt
{\alpha_s r_\perp^2 }}\qquad
{\rm for}\,\,\,\,r_\perp\gg 1/Q_s\,,\labe{GCLNON}
\ee
and its typical amplitude (\ref{GCLNON}) at large $r_\perp$
is even independent of the actual strength 
$\bar\rho \sim\sqrt{\langle\rho_a \rho_a \rangle}\sim Q_s$
of the colour source. This can be thus
characterized as a Bose condensate.

We thus see that it is the same fundamental separation
in time scales which allows us to speak about both the 
{\em colour glass} and the {\em Bose condensate}, although
these two concepts seem at a first sight contradictory:
the notion of a {``glass''} makes explicit reference to the
{\em average} over $\rho$, while the ``condensate'' rather
refers to a specific realization of $\rho$, before averaging.

\subsection{The renormalization group equation}
\label{SRGE}

As explained in Sect. \ref{QEFT}, the quantum evolution of the
effective theory is obtained by matching correlations computed
in two ways: (a) via a classical+quantum
calculation in the effective theory at the scale $\Lambda^+$, and
(b) via a purely classical calculation within 
 the effective theory at the scale  $b\Lambda^+$. The quantum
corrections that are included in this way are those generated
by the coupling between the ``semi-fast'' gluons with $p^+$ momenta
in the strip (\ref{strip}) and the ``soft'' gluons $\delta A^\mu_c$ 
with momenta $|p^+| \le b\Lambda^+$. To the accuracy of interest,
it is sufficient to consider the eikonal coupling 
$\delta A^-_a\delta\hat\rho_a$ to the plus component
$\delta\hat J^+_a\equiv \delta\hat\rho_a$ of the colour current
of the semi-fast gluons. Indeed, these gluons are relatively
fast moving in the $x^+$ direction, so $\delta\hat\rho_a$ is the
large component of their current.

The results of the matching can be summarized as follows:

i) To ${\cal O}(\alpha_s\ln(1/b))$, the induced correlations
of the transverse fields $A^i_a$ (see eq.~(\ref{induced-2p})
for an example) can be all related to the following 1-point and
2-point functions of $\delta\hat\rho$ (with $\Delta\tau=\ln(1/b)$) :
\be\label{sigperp}
\sigma_a ({x}_\perp)&\equiv &\frac{1}{\Delta\tau}
\int dx^- \,\langle\delta \hat\rho_a(x)\rangle_\rho\,,\\
\chi_{ab}(x_\perp, y_\perp)&\equiv &\frac{1}{\Delta\tau}
\int dx^- \int dy^-\,
\langle\delta \hat\rho_a(x^+,\vec x)\,
\delta \hat\rho_b(x^+,\vec y)\rangle_\rho\,,
\label{chiperp}\ee
where, as in eq.~(\ref{induced-2p}), $\langle\cdots\rangle_\rho$ 
denotes the average over semi-fast quantum fluctuations in the
background of the tree-level source $\rho$.

Thus the quantum evolution consists in adding new correlations
$\sigma$ and $\chi$ to $\rho$.

ii) These new correlations can be included in the weight function
$W_\tau[\rho]$ by allowing this to evolve with 
$\tau$ according to the following RGE
\cite{JKLW97,PI} :
\be\label{RGE}
{\del W_\tau[\rho] \over {\del \tau}}\,=\,
 {1 \over 2} {\delta^2 \over {\delta
\rho_\tau^a(x_\perp) \delta \rho_\tau^b(y_\perp)}} 
[W_\tau\chi_{xy}^{ab}] - 
{\delta \over {\delta \rho_\tau^a(x_\perp)}}
[W_\tau\sigma_{x}^a]\,.
\ee 
We use here compact notations
where $\sigma_{x}^a\equiv \sigma_a(x_\perp)$,
$\chi_{xy}^{ab}\equiv\chi_{ab}(x_\perp,y_\perp)$, and
repeated colour indices 
(and coordinates) are understood to be summed (integrated) over. 
The notation $\rho_\tau^a(x_\perp)$ will be explained later
(see eq.~(\ref{rhotau})).

A complete proof of the statements above would require the lengthy analysis
of Refs. \cite{PI}. But assuming them to be true, it is easy to
understand the general structure of the RGE (\ref{RGE}). Indeed,
according to eqs.~(\ref{sigperp})--(\ref{chiperp}), the induced
correlations that we need to take into account are
(with the notations of eq.~(\ref{evolaa})):
\be\label{rho-2p}
\Big\langle\big\langle(\rho_a+\delta\hat\rho_a)_{x_\perp}
(\rho_b+\delta\hat\rho_b)_{y_\perp}\big\rangle_\rho\,\Big\rangle_{W_\tau}
- \Big\langle \rho_a(x_\perp)\rho_b(y_\perp)\Big\rangle_{W_\tau}\quad\qquad\\
= \int { D}[\rho]\,W_\tau[\rho]\,\,\Delta\tau\left\{
\sigma_a ({x}_\perp)\rho_b(y_\perp) + \rho_a(x_\perp)\sigma_b ({y}_\perp)
+ \chi_{ab}(x_\perp, y_\perp)\right\}\,=\nn
= \int { D}[\rho]\,W_\tau[\rho]\,\Delta\tau\bigg\{
\sigma_{z}^c{\delta \over {\delta \rho^c(z_\perp)}}+
{1 \over 2} \chi_{zu}^{cd}{\delta^2 \over {\delta
\rho^c(z_\perp) \delta \rho^d(u_\perp)}} \bigg\}
\rho_a(x_\perp)\rho_b(y_\perp),\nonumber\ee
where the colour indices $c,d$ (the transverse coordinates
$z_\perp,u_\perp$) in the last line are to be summed (integrated) over.
After a few integrations by parts w.r.t. $\rho$, the last expression
can be recast into the form:
\be\label{rho-2p1}
\int { D}[\rho]\,\,\rho_a(x_\perp)\rho_b(y_\perp)\,\Delta W_\tau[\rho],
\ee
with $\Delta W_\tau[\rho]$ given by the finite-difference 
version of  eq.~(\ref{RGE}). 

In eqs.~(\ref{rho-2p})--(\ref{rho-2p1}), we have considered only correlators
of {\em two}--dimensional (or ``integrated'') charge densities, like
\be\label{rho-2}
\rho_a(x_\perp)\,\equiv\,\int dx^- \,\rho_a(x^-,x_\perp),\ee
and similarly $\delta\hat\rho_a(x_\perp)$. This is in agreement with
eqs.~(\ref{sigperp})--(\ref{chiperp}), which show that only such {\em
integrated} (over $x^-$) quantum corrections are relevant to the order
of interest, and is moreover physically intuitive:
The soft gluons ($k^+\simle b\Lambda^+$) to which applies the effective
theory are unable to discriminate the internal longitudinal structure
of their sources, which are localized in $x^-$ over relatively short
distances $\ll 1/b\Lambda^+$, because of their large $p^+$ momenta.
Although essentially correct, this argument is a little too simplistic
as shown by the fact that some of the quantities encountered before
{\em are} in fact sensitive to the longitudinal structure of $\rho$ 
(i.e., they are not
simply functionals of the integrated charge density (\ref{rho-2})). 
A generic example is the background field ${\cal A}^i[\rho]$,
or any other quantity built with the Wilson lines (\ref{UTA})
or (\ref{v}). Such quantities are sensitive to the $x^-$ dependence of 
$\rho$ because of the path-ordering of the Wilson lines in $x^-$. 
The ordering is important since
colour matrices $\rho(x^-)=\rho_a(x^-)T^a$ at different values of
$x^-$ do not commute with each other. This suggests that the correct
way to think of an ``integrated'' version of the hadron (over $x^-$)
is in terms of Wilson lines --- which take into account
the colour precession in the colour field of the hadron, with the
proper ordering of colour matrices ---, 
and not of 2-dimensional 
charge densities like (\ref{rho-2}). This will be confirmed by the
subsequent analysis of the quantum corrections.

\subsubsection{The quantum colour source}

For the purposes of the quantum calculation, it is useful to
expand the action $S[A, \rho]
\equiv S[{\cal A}+a+\delta A, \rho]$ to quadratic order in the
small fluctuations $a^\mu$, and retain only their eikonal coupling
to the component $\delta A^-_a$ of the {soft} fields:
\be\labe{SEXP}
S[{\cal A}+\delta A+a, \rho]\,\approx\,S[{\cal A}+\delta A, \rho]
+{1\over 2} a^\mu(x) G_{\mu\nu}^{-1}({x,y})[\rho]a^\nu(y)
- \delta A^-_a \delta\hat\rho_a,\nn\ee
with
\be\labe{invG}
G_{\mu\nu}^{-1\,ab}(x,y)[\rho]\,\equiv\,
\frac{\delta^2 S [A,\,\rho]}
{\delta A^\mu_a(x)\delta A^\nu_b(y)}\bigg|_{{\cal A}},\ee
\be\label{rho-q}
\delta\hat\rho_a(x)\equiv
-\frac{\delta^2 S}{\delta A^-_a(x)\delta A^\nu_b(y)}\bigg|_{
{\cal A}}a^\nu_b(y)-\frac{1}{2}
\frac{\delta^3 S}{\delta A^-_a(x)\delta A^\nu_b(y)
\delta A^\lambda_c(z)}\bigg|_{
{\cal A}}a^\nu_b(y)a^\lambda_c(z),\nn \ee
where it is understood that only
the soft modes with $k^+\simle b\Lambda^+$ are kept in the
products of fields.

The expansion (\ref{SEXP}) corresponds to a one-loop approximation
for the soft correlation functions like 
$\langle \delta A^i\rangle_\rho$ and
$\langle \delta A^i \delta A^j\rangle_\rho$
(cf. eq.~(\ref{induced-2p})), but where the propagator
$iG^{\mu\nu}(x,y)=\langle {\rm T}a^\mu(x)a^\nu(y) \rangle_\rho$ 
of the semi-fast gluons running along the loop is computed in the
background of the tree-level field ${\cal A}^i[\rho]$,
by inverting the differential operator in eq.~(\ref{invG}).
\begin{figure}[htb]
\centering
\resizebox{.8\textwidth}{!}{%
\includegraphics*{{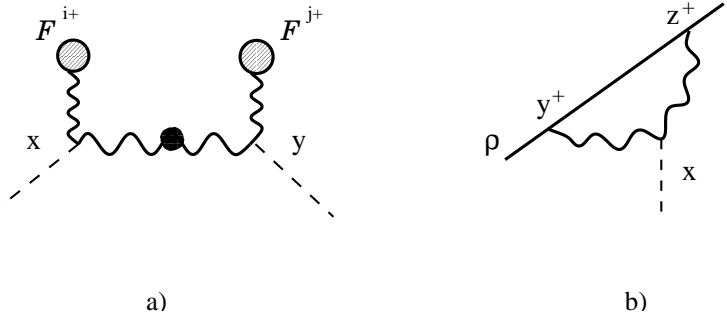}}}
\caption{Some typical Feynman diagrams for $\chi$ and $\sigma$.
The internal wavy lines are propagators of
the semi-fast gluons; the external dotted lines
carry soft momenta, and couple to the fields $\delta A^-$.
(a) A contribution to $\chi$. The external blobs
denote insertions of the electric field ${\cal F}^{+i}\,$;
the internal line with a blob denotes the
background field propagator. (b) A contribution to $\sigma$
to linear order in $\rho$. The continuous line represents the source $\rho$.}
\label{CHISIG-LIN}
\end{figure}

To gain some more intuition, we use as an example
the contributions to $\delta \hat\rho_a$ coming from the Yang-Mills
piece of the action, $S_{YM}=\int d^4x(-F_{\mu\nu}^2/4)$ :
\be\label{drhoYM}
\delta \hat\rho_a (x)\big|_{YM}\,=\,
2 gf^{abc}{\cal F}^{+i}_b (\vec x) a^{i}_c (x)
+g f^{abc} (\partial^+ a_{b}^{i}(x) )a_{c}^{i}(x).\ee
The first term in the r.h.s., which is linear in $a^i$,
is the only one to contribute to $\chi$, eq.~(\ref{chiperp}), 
to leading order in $\alpha_s$. It generates the
tree-like diagram in Fig. \ref{CHISIG-LIN}.a, where the
internal line with a blob represents the  background field
propagator $G^{ij}(x,y)$ of the semi-fast gluons. 
Physically, Fig. \ref{CHISIG-LIN}.a describes 
the emission of an on-shell (or ``real'') semi-fast gluon 
by the classical source.

Since $\langle a^i\rangle=0$, it is only the second, quadratic term in
the r.h.s. of eq.~(\ref{drhoYM}) which
contributes to $\sigma$, eq.~(\ref{sigperp}).
In Fig. \ref{CHISIG-LIN}.b we show such a contribution of
lowest order in $\rho\,$. (This involves also vertices from the
Wilson line piece of the action, eqs.~(\ref{ACTION})--(\ref{WLINE}).)
Obviously, this represents a vertex
correction to the tree-level emission in Fig. \ref{cascade}.a.

The structures illustrated by Figs. \ref{CHISIG-LIN}.a and b
are generic:  $\chi$ is the
``real'' correction, whose iteration generates the gluon cascades;
 $\sigma$ is the ``virtual'' correction, which provides one-loop 
corrections to the emission vertices in these cascades. Both  $\chi$ 
and $\sigma$ include terms non-linear in $\rho$ which describe
interactions among gluons at different rapidities in different
cascades. In general, real and virtual
corrections are related by gauge symmetry, and this is also
the case for $\chi$ and $\sigma$, as we shall discuss later.

The diagrams contributing to $\sigma$ and $\chi$
in the general case, together with their explicit evaluation, can
be found in Ref. \cite{PI}. Here, we shall present only the final
results of this calculation.

\subsubsection{The induced colour source and field}
\label{sec:sigma}

For the reasons explained in Sect. \ref{YMSOL}, it is more
convenient to work with the colour source $\tilde\rho_a$ 
in the {\em covariant} gauge. 
The corresponding weight function $W_\tau[\tilde\rho]$
obeys an evolution equation similar to (\ref{RGE}), but with
modified coefficients $\tilde\sigma$ and $\tilde\chi$, which are
obtained from the LC-gauge coefficients $\sigma$ and $\chi$ via the
gauge rotation (\ref{COVLC}). 
In what follows, we shall give directly the final results for
these COV-gauge quantities. 

Consider first the {\em induced source} $\delta\rho^a
= \langle\delta \hat\rho^a\rangle_\rho$, that is, the correction to
the average colour charge density generated by the polarization 
of the semi-fast gluons. After rotation to the COV-gauge, 
this reads \cite{PI} :
\be\label{drho}
\delta\tilde\rho_a(\vec x) \,=\,F_\Lambda(x^-)(-\grad_\perp^2\nu_a (x_\perp)),
\ee
where the ``form factor'' 
\be\label{FormF}
F_\Lambda(x^-)\,\equiv\,\theta(x^-)\,\frac{e^{-ib\Lambda x^-}-
e^{-i\Lambda x^-}}{x^-}\,,
\ee
specifies the longitudinal profile of $\delta\tilde\rho_a$,  while
($ V^\dagger_x\equiv V^\dagger(x_\perp)$, cf. eq.~(\ref{v}))
\be
\nu^a (x_\perp)&=&{ig\over 2\pi}
\int {d^2z_\perp\over (2\pi)^2}\,\frac{1}{(x_\perp-z_\perp)^2}
\,{\rm Tr}\Bigl(T^a V^\dagger_x V_z\Bigr)
\label{nu}\ee
contains the dependence upon the background field $\alpha_a$ (via the
Wilson lines $V$ and $V^\dagger$), together with the
transverse and colour structure of $\delta\tilde\rho_a$. By comparing
eqs.~(\ref{drho}) and (\ref{EQTA}), we deduce that $F(x^-)\nu^a (x_\perp)$
is the {\em induced field} in the COV-gauge, i.e., the quantum
correction to the tree-level field $\alpha_a$. Since:
\be 
\int dx^- F_\Lambda(x^-) \,=\,\ln {1 \over b}\,=\,\Delta\tau,\ee
eqs.~(\ref{sigperp}) and (\ref{drho}) immediately imply :
\be\label{sig-nu}
\tilde\sigma_a(x_\perp)&=&-\grad_\perp^2\nu_a (x_\perp).\ee
This is the coefficient of the virtual term in the RGE for
$W_\tau[\tilde\rho]$.

But the longitudinal structure of $\delta\tilde\rho_a$ is also interesting.
Eq.~(\ref{drho}) shows that the induced source and field
have typically support 
at\footnote{Indeed, $F_\Lambda(x^-)\approx 0$ both for small $x^-\ll
1/\Lambda^+$ (since in this case the two exponentials 
mutually cancel), and for large $x^-\gg 1/b\Lambda^+\/$
(where the two exponentials are individually small).} 
 \be\label{stripx-}
1/\Lambda^+\,\,\simle\,\,x^-\,\,\simle\,\,1/(b\Lambda^+)\,.\ee
Recall that  $\delta\tilde\rho_a$ has been generated by integrating
out quantum fluctuations in the strip $b\Lambda^+ \ll |p^+|
\ll \Lambda^+$.
Thus, when integrating out quantum gluons in layers of $p^+$,
one builds the classical source $\rho$ (or field $\alpha$)
in layers of $x^-$, with a one-to-one correspondence between 
the $x^-$ coordinate of a given layer and the $p^+$ momenta of the modes
that have been integrated out to generate that layer. 
By induction, we deduce that $\rho_a(\vec x)$ 
($\equiv$ the colour source generated by the quantum evolution down to 
$\Lambda^+$)
has support at $0\le x^-\simle 1/\Lambda^+$, as anticipated
in Sect. \ref{EFT}. This allows us to consider only
positive values for $x^-$ in what follows.

To exploit this tight correspondence between $p^+$ and $x^-$, it is
convenient to use the {\em space-time rapidity} y,
\be {\rm y}\,\equiv\, \ln(x^-/ x^-_0),\quad 
x^-_0\equiv 1/P^+\,,\quad -\infty < {\rm y} < \infty\,,\ee
to indicate the longitudinal
coordinate of a field. We shall set, e.g.,
\be\label{rhotau}
\rho_{{\rm y}}^a(x_\perp)&\equiv&
 x^-\rho^a(x^-,x_\perp)\qquad {\rm for}\quad
x^-=x^-_{\rm y}\equiv x^-_0{\rm e}^{{\rm y}},\nn
\int d{\rm y}\,\rho_{{\rm y}}^a(x_\perp)&=&\int dx^-\,\rho^a(x^-,x_\perp),
\ee
and similarly for the other fields ($\tilde\rho$, $\alpha$, etc.).
 The previous discussion on the
longitudinal structure can then be summarized as follows:

The source $\rho_{{\rm y}}^a(x_\perp)$ generated by the quantum evolution
from $\tau' =0$ up to $\tau$ has support at y in the interval
$0\le {\rm y} \le \tau$. 
When new quantum modes, with rapidities $\tau'$ in the interval
$\tau <\tau' <\tau+\Delta\tau$, are integrated out, 
the preexisting colour source at ${\rm y} \le \tau$ is 
not changed, but some new contribution is added to it, in
the rapidity bin $\tau <{\rm y} <\tau+\Delta\tau$.
Because of that, $\Delta W \equiv W_{\tau+\Delta\tau} - W_\tau$ 
involves only the change in $\rho_{\rm y}$ within that last bin. In the
continuum limit $\Delta\tau\to 0$, this generates the functional 
derivatives of $W_\tau$ with respect to $\rho_{\rm y}^a$ at ${\rm y}=\tau$,
as shown in eq.~(\ref{RGE}). This clarifies the longitudinal
structure of the RGE.

Consider also the transverse and colour structure of 
the induced field (\ref{nu}). This can be understood by reference
to Fig. \ref{CHISIG-LIN}.b. The transverse kernel in eq.~(\ref{nu})
has been generated as:
\be
{1 \over (2\pi)^2}\,\frac{1}{(x_\perp-z_\perp)^2}\,=\,
{1 \over 2\pi} \frac{x^i-z^i}{(x_\perp-z_\perp)^2}\,
{1 \over 2\pi} \frac{x^i-z^i}{(x_\perp-z_\perp)^2}\,,\ee
where (compare with eqs.~(\ref{alpha}) and (\ref{aaimom}))
\be\label{id0}
{1 \over 2\pi} \frac{x^i-z^i}{(x_\perp-z_\perp)^2}=
\partial^i_x\langle
x_\perp|\,\frac{1}{-\grad^2_\perp}\,|z_\perp\rangle\,
=\int {d^2p_\perp \over (2 \pi)^2}\, {-i\,p^i \over p_{\perp}^2}\,
{\rm e}^{ip_{\perp}\cdot(x_{\perp}-z_{\perp})}
\ee
is the propagator of the semi-fast gluon 
emitted by the source $\rho$ (recall that ${\cal F}^{+j}\approx
(ip^j/p_\perp^2)\rho$ to linear order).
The two Wilson lines in eq.~(\ref{nu}) account for the
scattering of this semi-fast gluon off the background field
at $z_{\perp}$ (this brings in a factor $V^\dagger (z_{\perp})$ 
in the eikonal approximation), 
and for its gauge rotation by the classical field ${\cal A}^i(\vec x)$ at
$x^- >1/\Lambda^+$ (cf. eq.~(\ref{stripx-})), which is a pure gauge
(cf. eq.~(\ref{APM})).

\subsubsection{The RGE in the $\alpha$--representation}

Eq.~(\ref{sig-nu}) suggests that it may be technically simpler
and physically more transparent to work directly with
 the classical field $\alpha_a$ and the quantum corrections to it
(like $\nu_a$), rather than with the colour source $\tilde\rho_a$ 
and the corresponding corrections (like  $\tilde\sigma_a$).
This point of view is also
supported by the fact that the LC-gauge field and the 
related observables are primarily related to $\alpha_a$ 
(cf. Sects. \ref{OBS-DEF} and \ref{YMSOL}), and reexpressing them
in terms of $\tilde\rho$ --- with the help of eq.~(\ref{alpha}) --- would
introduce a dependence upon the unphysical infrared cutoff $\mu$.

For these reasons, we prefer to work in the
$\alpha$--{\em representation}, in which observables are expressed
in terms of $\alpha$, and the average is performed with the
weight function $W_\tau[\alpha]\equiv W_\tau[\tilde\rho=
-\grad_\perp^2\alpha]$. This satisfies the following RGE,
which is obtained after a change of variables in eq.~(\ref{RGE}) :
\be\labe{RGEA}
{\del W_\tau[\alpha] \over {\del \tau}}\,=\,
 {1 \over 2} {\delta^2 \over {\delta
\alpha_\tau^a(x_\perp) \delta \alpha_\tau^b(y_\perp)}}\, 
[W_\tau\eta_{xy}^{ab}] - 
{\delta \over {\delta \alpha_\tau^a(x_\perp)}}\,
[W_\tau\nu_{x}^a] \,,
\ee
where $\nu_{x}^a\equiv \nu^a(x_\perp)$, cf.
eq.~(\ref{nu}), and $\eta_{xy}^{ab}\equiv \eta^{ab}(x_\perp,y_\perp)$,
with
\be
\eta^{ab}(x_\perp,y_\perp)\equiv \int d^2z_\perp d^2u_\perp
\langle x_\perp|\frac{1}{-\grad^2_\perp}|z_\perp\rangle\,
\tilde\chi^{ab}(z_\perp,u_\perp)\,
\langle u_\perp|\frac{1}{-\grad^2_\perp}|y_\perp\rangle.\,\,\,\,\,\,
\label{etadef}\ee
It is thus sufficient 
to give the result for the ``real correction''
$\langle\delta \hat\rho_a \delta \hat\rho_b\rangle_\rho$ 
directly in the $\alpha$-representation (cf. eqs.~(\ref{chiperp})
and (\ref{etadef})). This reads \cite{PI}:
\be\label{eta}
\eta^{ab}(x_\perp,y_\perp)
&=&{1\over \pi}\!\int 
{d^2z_\perp\over (2\pi)^2}\,
\frac{(x^i-z^i)(y^i-z^i)}{(x_\perp-z_\perp)^2(y_\perp-z_\perp)^2 }\nn
&{}&\qquad\quad\times\,\,
\Bigl\{1
+ V^\dagger_x V_y- V^\dagger_x V_z -  V^\dagger_z V_y\Bigr\}^{ab}.
\ee
The transverse and colour structure of $\eta$ have the same pattern
as discussed after eq.~(\ref{DIFFU}) in connection with $\nu$.

The r.h.s. of eq.~(\ref{RGEA}) involves functional derivatives
w.r.t. the colour field $\alpha_\tau^a(x_\perp)$ at the end point
$\y=\tau$. When applied to the coefficients $\eta$ and $\nu$, this
requires the corresponding  derivatives of the Wilson lines
$V$ and $V^\dagger$, that we compute now. Note first
that, since $\alpha_{{\rm y}}=0$ for $ {\rm y} > \tau$, we can
rewrite
\be\label{vtau} V^\dagger(x_{\perp})\,\equiv\,{\rm P}\, {\rm e}^
{ig \int_{-\infty}^{\infty} d{\rm y}\,\alpha_{{\rm y}} (x_{\perp})}
\,=\,{\rm P}{\,\rm e}^{
ig \int_{-\infty}^{\tau} d{\rm y}\,\alpha_{{\rm y}} (x_{\perp})}
.\ee Therefore (with $\delta_{xy}\equiv \delta^{(2)}(x_{\perp}-y_\perp)$):
\be\label{DIFFU}
{\delta V^\dagger (x_{\perp})\over \delta \alpha^a_\tau(y_\perp)}=
ig\delta_{xy}T^a V^\dagger (x_{\perp}),\quad
{\delta V(x_{\perp})\over \delta \alpha^a_\tau(y_\perp)}=
-ig\delta_{xy}V(x_{\perp})T^a\,.\ee

A simple interpretation of the four terms in
eq.~(\ref{eta}) follows from the dual picture of the dipole-hadron
scattering, in which the quantum evolution is put in the dipole
wavefunction, and, more generally, in the Wilson line operators through which
a generic external projectile scatters off the hadronic target
\cite{B,K,AM3,W,AM01,KMW00,B00,BB01}. 
(See also the lectures notes by Al Mueller
in this volume \cite{AMCARGESE}.) Recent analyses of the
high energy scattering from this dual perspective 
have led to a set of coupled evolution
equations for the correlation functions of Wilson lines, originally
derived by Balitsky \cite{B} (see also \cite{K,B00}), 
and subsequently reformulated
by Weigert \cite{W} in a compact way, as a functional evolution equation 
for the generating functional of these correlation functions.
It turns out that
Weigert's equation is equivalent to the RGE (\ref{RGEA}) \cite{PI,BIW},
which demonstrates the
equivalence between the two descriptions --- the
 target picture and the projectile picture ---
of the nonlinear evolution in QCD at small x. We shall say more on
Balitsky's equations in Sect.  \ref{S-BK}.

\subsection{Recovering the BFKL equation}
\label{S-BFKL}

Before studying more general properties and consequences of the RGE
in the next section, let us rapidly show that, in the weak field
(or low density) limit, this equation reproduces the BFKL equation,
as expected \cite{JKLW97}.
Eq.~(\ref{rho-2p}) implies the following evolution equation for
the 2-point function $\langle\rho\rho\rangle_\tau\,$:
\be\labe{RGE2p}
{d\over {d\tau}}
\langle\rho_a(x_\perp)\rho_b(y_\perp)\rangle_\tau\! =\! 
\langle\sigma_a(x_\perp)\rho_b(y_\perp)
+\rho_a(x_\perp)\sigma_b(y_\perp)+
\chi_{ab}(x_\perp,y_\perp)\rangle_\tau\,\,\,\,\,\,\,\,\ee
For a generic, strong, source $\rho$, the coefficients
$\sigma$ and $\chi$ are non-linear in
$\rho$ to all orders, so the r.h.s. of eq.~(\ref{RGE2p})
involves $n$-point correlators 
$\langle\rho(1) \rho(2)\cdots\rho(n)\rangle_\tau$ of arbitrarily high
order\footnote{Incidentally, this shows that the $n$-point functions
of $\rho$ do not form a convenient basis to study the non-linearities
in the evolution. By contrast, the correlators of the Wilson lines form a
more convenient such a
 basis \cite{B,BB01}, as we shall discuss in Sect. \ref{S-BK}.}
 $n$. But in the weak field limit, where $\sigma$ is linear in
$\rho$ and $\chi$ is quadratic, this becomes a closed equation for
the 2-point function, which coincides with the BFKL equation, as we 
show now.

Specifically, consider the evolution equation for the following 
2-point function:
\be\label{varphiWF}
\mu_\tau (k^2_{\perp})\,\equiv\,
\langle\rho_a(k_\perp)\rho_a(-k_\perp)\rangle_\tau,\ee
 ($\rho_a(k_\perp)$
is the Fourier transform of $\rho_a(x_\perp)$),
which according to eqs.~(\ref{GDF}) and  (\ref{linphi})
represents the ``unintegrated gluon distribution'' :
\be
{\partial\, \x G(\x,k^2_{\perp}) \over \partial \ln k^2_{\perp}}
\,\propto\, 
\mu_\tau  (k^2_{\perp}),\ee
in the weak field regime. In this regime, one can also expand 
$V^\dagger_x\approx 1+ig\alpha(x_\perp)$, and therefore
$${\rm Tr}\bigl(T^a V^\dagger_x V_z\bigr)\approx
igN_c(\alpha^a(x_\perp)
-\alpha^a(z_\perp)).$$
Then, eq.~(\ref{nu}) reduces to (with  $\alpha_s=g^2/4\pi$
and $k^2_{\perp}\alpha_a(k_\perp)=\rho_a(k_\perp)$) :
\be\label{sigmaq}
\sigma^{(0)}_a (k_\perp)
\,=\,-\alpha_s N_c \,\rho_a(k_\perp)
\int \frac{d^2 p_\perp}{(2\pi)^2}\,{k^2_{\perp} \over p_{\perp}^2 
(p_{\perp}- k_{\perp})^2}\,.\ee
For $\chi$ one obtains similarly 
\be\label{chiBFKL}
\chi^{(0)}_{aa}(k_\perp,-k_\perp)\,=\,4\alpha_sN_ck_\perp^2
\int \frac{d^2 p_\perp}{(2\pi)^2}\,\frac{\rho_a(p_\perp)\rho_a(-p_\perp)}
{p^2_{\perp} (k_{\perp}-p_{\perp})^2}\,.\ee
By inserting eqs.~(\ref{sigmaq}) and (\ref{chiBFKL}) into 
the evolution equation (\ref{RGE2p}), and using (\ref{varphiWF}),
one finally obtains:
\begin{eqnarray}
 {\partial \mu_\tau (k^2_{\perp}) \over \partial \tau} & = & \,\,\,
{\alpha_s N_c \over \pi^2}\,
\int d^2 p_{\perp}
 {k^2_{\perp} \over p^2_{\perp} (k_{\perp}-p_{\perp})^2}\,
 \mu_\tau(p^2_{\perp}) \nonumber \\
& & -\,
{\alpha_s N_c \over2 \pi^2}\,
\int d^2 p_{\perp}
 {k^2_{\perp} \over p^2_{\perp} (k_{\perp}-p_{\perp})^2}\,
 \mu_\tau(k^2_{\perp})\,,
\label{BFKL}
\end{eqnarray}
which coincides, as anticipated, with the BFKL equation \cite{BFKL,TB-BFKL}.
The first term in the r.h.s., which here is generated by 
$\chi^{(0)}$, is the {\it real} BFKL kernel, 
while the second term, coming from $\sigma^{(0)}$,
is the corresponding {\it virtual} kernel.

Note finally that the BFKL approximation has been obtained 
by expanding the Wilson lines to linear order in $g\alpha^a$; thus, this
is formally the same as the lowest order perturbative expansion
of the RGE.

\section{A functional Fokker-Planck equation}
\setcounter{equation}{0}

We now dispose of a powerful tool --- the functional
RGE (\ref{RGEA}) --- to construct the effective theory by integrating
out quantum fluctuations in perturbation theory. Eq.~(\ref{RGEA}) has
a rich and elegant mathematical structure, to be described 
in Sects. \ref{S-FP} and \ref{sec:FP}. Then, in Sects. \ref{S-BK}
and \ref{S-SOL}, we shall indicate two strategies to make use
of this equation:

i) One can use it to derive ordinary (i.e., non-functional) evolution 
equations for the correlation functions of interest, like we did for
the 2-point function $\langle\rho\rho\rangle_\tau\,$ in  
Sect. \ref{S-BFKL}. When specialized to correlation functions
of the Wilson lines, this strategy leads to a system of equations
originally derived by Balitsky \cite{B}. This will be discussed
in Sect. \ref{S-BK}.

A difficulty with this approach is that it generally leads to {\em coupled}
equations (the 2-point function is coupled to the 4-point one, 
etc.), so that one has to follow simultaneously the evolution of
infinitely many correlators. Still, some progress has been done,
by using functional techniques \cite{Balitsky2001}, and, especially,
by recognizing that, in the large $N_c$ limit, a closed equation can
be written for the 2-point function: this is the Kovchegov equation \cite{K}.

ii) One can try and solve directly the functional RGE, with
appropriate initial conditions. An exact but formal solution can be
written in the form of a path integral \cite{BIW}. 
This is well suited for lattice simulations in 2+1 dimensions. 
But approximate analytic solutions, which allow for a 
more direct physical insight, have been found as well \cite{SAT,IIM}.
These solutions will be described in Sect. \ref{S-SOL}.

\subsection{General properties and consequences of the RGE}
\label{S-FP}

We start with a summary of the most important properties of the RGE
(\ref{RGEA}).

{\bf i)} {\bf The coefficients $\eta$ and $\nu$ are real quantities.
Moreover, $\eta$ is symmetric: 
$\eta_{ab}(x_\perp,y_\perp)=\eta_{ba}(y_\perp,x_\perp)$,
and positive semi-definite.} 

{\bf ii)} {\bf The RGE preserves the
normalization of the weight function:} 
\be\label{norm1}
\int {\cal D}\alpha\, \,W_\tau[\alpha]\,=\,1\qquad
{\rm at\,\, any}\,\, \tau.\ee
Indeed, the r.h.s. of eq.~(\ref{RGEA})
is a total derivative with respect to $\alpha$. Thus, if
eq.~(\ref{norm1}) is satisfied by the initial condition at
$\tau_0$,  it remains true at any $\tau>\tau_0$.

Properties {\sf (i)} and {\sf (ii)} guarantee that the solution
$W_\tau[\alpha]$ to the RGE has a meaningful probabilistic interpretation
(cf. the discussion prior to  eq.~(\ref{norm})).

{\bf iii)} {\bf The momentum rapidity $\tau$ and the space-time rapidity 
${\rm y}$ are identified by the quantum evolution.} That is,
the field $\alpha_{\rm y}$ in the rapidity bin 
$({\rm y}, \,{\rm y}+d{\rm y})$ is generated by the quantum evolution
from $\tau={\rm y}$ up to $\tau={\rm y}+d{\rm y}$.
This follows from the discussion in Sect. \ref{sec:sigma}, and implies
that the two rapidities  can be treated as only one variable, the
``evolution time''.

With this interpretation, the
function $\{\alpha_{{\rm y}}^a(x_\perp)\,|\,-\infty < {\rm y}<\infty \}$ 
--- which physically represents the longitudinal profile of the
3-dimensional  field $\alpha^a(x^-,x_\perp)$ in units of rapidity
(cf. eq.~(\ref{rhotau}))  ---
is viewed as a {\it trajectory} in the functional space spanned by
the 2-dimensional fields $\alpha^a(x_\perp)$. 
Quantum evolution then appears as the progression of the ``point'' 
$\alpha^a(x_\perp)$ along this trajectory.
Thus, eq.~(\ref{RGEA}) describes effectively a
field theory in 2+1 dimensions (the transverse coordinates and the
``evolution time''), which is however {\it non-local} in
both $x_\perp$ and ${\rm y}$
(since the coefficients (\ref{nu}) and (\ref{eta}) of the RGE involve
$\alpha_{\rm y}$ at all the ``times'' ${\rm y}\le\tau$, via the
Wilson lines (\ref{vtau})).

{\bf iv) The initial condition.} Let the quantum evolution
proceed from some original ``time'' $\tau_0$ up to the actual
``time'' $\tau$. The ``trajectory'' 
$\{\alpha_{{\rm y}}^a(x_\perp)\,|\,-\infty < {\rm y}<\infty \}$ 
can be decomposed into three pieces: a) The field
$\alpha_{\rm y}$ at ${\rm y}\le \tau_0$ belongs to the initial
conditions. b) The field $\alpha_{\rm y}$
at $\tau_0 < {\rm y}\le \tau$ is generated
by the quantum evolution. c) There is no field at all at larger
$ {\rm y}$: $\alpha_{{\rm y}}=0$ for any $ {\rm y} > \tau$. Thus:
\be\label{longitW}
W_\tau[\alpha]\,=\,\delta[\alpha^>]\,
{\cal W}_\tau[\alpha^<],\ee
where $\alpha^<_{{\rm y}}$ ($\alpha^>_{{\rm y}}$) 
is the function $\alpha_{{\rm y}}$
for ${\rm y}< \tau$ (respectively, ${\rm y}> \tau$) :
\be
\alpha_{{\rm y}}(x_\perp)\,\equiv\,
\theta(\tau-{\rm y})\alpha^<_{{\rm y}}(x_\perp)\,+\,
\theta({\rm y}-\tau)\alpha^>_{{\rm y}}(x_\perp),\ee
and the $\delta$--functional $\delta[\alpha^>]$
should be understood with a discretization of the configuration space,
as in eq.~(\ref{measure}):
\be\label{fdelta}
\delta[\alpha^>]\,\equiv\,\prod_{{\rm y}> \tau}
\prod_{a}\prod_{x_\perp}\,
\delta(\alpha_{{\rm y}}^a(x_\perp)).\ee
Moreover, it can be shown \cite{SAT,BIW} that ${\cal W}_\tau[\alpha^<]$
has the factorized structure:
\be \label{calW}
{\cal W}_\tau[\alpha^<]\,=\,
{\cal W}_{\tau,\,\tau_0}[\alpha|V_0]\,{\cal W}_{\tau_0}[\alpha]\,,\ee
where ${\cal W}_{\tau_0}[\alpha]$ is the initial weight function
at $\tau_0\,$, and
\be\label{V0}
V_0^\dagger(x_\perp) \,\equiv\,{\rm P}\, {\rm e}^
{ig \int_{-\infty}^{\tau_0} d{\rm y}\,\alpha_{{\rm y}} (x_{\perp})}\ee
is the Wilson line built with the initial field.
In eq.~(\ref{calW}), it is
understood that, in ${\cal W}_{\tau_0}$, the field argument 
$\alpha_{\rm y}$ has support at ${\rm y}\le \tau_0$,
while in ${\cal W}_{\tau,\,\tau_0}$ it has support
at $\tau_0 < {\rm y}\le \tau$. The ``propagator'' ${\cal W}_{\tau,\,\tau_0}$
from $\tau_0$ to $\tau$ depends also upon the initial field at 
${\rm y}\le \tau_0$, but only in an integrated way,
via the Wilson lines $V_0$ and $V_0^\dagger$. 
From eq.~(\ref{calW}) we deduce
that ${\cal W}_{\tau,\,\tau_0}[\alpha|V_0] \to 1$ when
$\tau\to\tau_0$.

The initial weight function ${\cal W}_{\tau_0}$
cannot be obtained within the present formalism, but rather
requires some model for the hadron wavefunction at 
rapidity $\tau_0$. It is convenient to choose a moderate value
for $\tau_0=\ln(1/\x_0)$, e.g., $\x_0\simeq 10^{-2}$. This
$\x_0$ is small enough for the LLA to apply, but still large enough for 
the non-linear effects to remain negligible. Then one can use initial
conditions which are consistent with the standard, linear, evolution 
equations (cf. Sect. \ref{MFA-HIGH} below).
Once a convenient value for
$\x_0$ has been chosen, one can always redefine
$\tau\equiv\ln(\x_0/\x)$ so that the initial condition is formulated
at $\tau_0=0$. With this choice, the field 
$\alpha_{{\rm y}}$ at positive rapidities $ {\rm y} > 0$ is generated
by the quantum evolution, while the field at negative rapidities
$ {\rm y} < 0$ must be specified by the initial condition.

{\bf v)} {\bf The Hamiltonian structure of the RGE.}
Eq.~(\ref{RGEA}) can be rewritten as:
\be\labe{RGEAnew}
{\del W_\tau[\alpha] \over {\del \tau}}\,=\,
{\delta \over {\delta \alpha_\tau^a(x_\perp)}}
\left\{ {1 \over 2} \eta_{xy}^{ab} \,{\delta W_\tau\over 
{\delta \alpha_\tau^b(y_\perp)}}  + \left({1 \over 2} \,
{\delta \eta_{xy}^{ab}\over {\delta \alpha_\tau^b(y_\perp)}}
 -\nu_{x}^a\right)W_\tau \right\}\,,
\ee
A crucial property, with many consequences, is that the second
term within the braces is actually zero.
Indeed, the following relation holds between
the coefficients of the RGE \cite{W,PI}:
 \be\label{sigchi}
{1 \over 2} \int d^2y_\perp 
{\delta \eta^{ab}(x_\perp,y_\perp)\over {\delta \alpha_\tau^b(y_\perp)}}
\,=\,\nu^a(x_\perp)\,.\ee
It is easy to prove this relation by using eq.~(\ref{DIFFU})
to act with $\delta/\delta \alpha_\tau^b(y_\perp)$ on 
$\eta_{ab}(x_\perp,y_\perp)$, eq.~(\ref{eta}). This yields, e.g.,
\be\label{anti}
{\delta\over {\delta \alpha_\tau^b(y_\perp)}}\,(V^\dagger_x V_y)^{ab}
&=& {\delta V^{\dagger\,ac}_x\over {\delta \alpha_\tau^b(y_\perp)}}\,V_y^{cb}
\,+\,V^{\dagger\,ac}_x{\delta V_y^{cb}\over 
{\delta \alpha_\tau^b(y_\perp)}}\\
&=& ig\delta_{xy}\Bigl(T^bV^\dagger_y\Bigr)_{ac}V_y^{cb}
-ig\delta^{(2)}(0_{\perp})V^{\dagger\,ac}_x\Bigl(V_yT^b\Bigr)_{cb}\,=\,0,
\nonumber \ee
where both terms in the second line
vanish because of the antisymmetry of the
colour group generators in the adjoint representation (e.g.,
$(T^b)_{ab}=0$). The only nonvanishing contribution is
\be
-{\delta\over {\delta \alpha_\tau^b(y_\perp)}}\,(V^\dagger_x V_z)^{ab}
\,=\,-ig\delta_{xy}\Bigl(T^b V^\dagger_x V_z\Bigr)_{ab}
\,=\,ig\delta_{xy}
{\rm Tr}\Bigl(T^a V^\dagger_x V_z\Bigr),\ee
which reproduces indeed eq.~(\ref{nu}) after integration over $y_\perp$,
since:
\be\label{Kxyz}
{\cal K}(x_\perp,y_\perp,z_\perp)\equiv
\frac{(x^i\!-\!z^i)(y^i\!-\!z^i)}{(x_\perp\!-\!z_\perp)^2
(y_\perp\!-\!z_\perp)^2} \rightarrow \frac{1}{(x_\perp\!-\!z_\perp)^2}
\,\,\,\,{\rm for}\,\,\,\,y_\perp\! \to\! x_\perp.\,\,\,\,\ee
With eqs.~(\ref{RGEAnew}) and (\ref{sigchi}), the RGE
can be brought into a Hamiltonian form:
\be\label{RGEH}
{\del W_\tau[\alpha] \over {\del \tau}}\,=\,-\,H W_\tau[\alpha], 
\ee
with the following Hamiltonian:
\be\labe{H}
H&\equiv&
{1 \over 2}
\int \!d^2x_\perp\!\int d^2y_\perp
{i\delta \over {\delta
\alpha_\tau^a(x_{\perp})} }\,\eta_{xy}^{ab}\,
{i\delta \over {\delta \alpha_\tau^b(y_{\perp})}}\,=\!
\int \!{d^2 z_\perp\over 2\pi }\,J^i_a(z_\perp)J^i_a(z_\perp),\nn
J^i_a(z_\perp)&\equiv& \int {d^2 x_\perp\over 2\pi }\,
\frac{z^i-x^i}{(z_\perp-x_\perp)^2}\,(1 - V^\dagger_zV_x)_{ab}\,
{i \delta \over {\delta
\alpha_\tau^b(x_\perp)} }\,,\ee
which is Hermitian (since $\eta_{xy}^{ab}$ is real and symmetric)
and positive semi-definite (since the ``current'' $J^i_a(z_\perp)$
is itself Hermitian). 

{\bf vi)} {\bf  The infrared and ultraviolet behaviours of the RGE.} 
These are determined by the kernel $\eta^{ab}(x_\perp,y_\perp)$ in the
Hamiltonian. 
In the infrared limit, where $z_\perp$ is much larger than both 
$x_\perp$ and $y_\perp$ (see eq.~(\ref{eta})),
${\cal K}(x_\perp,y_\perp,z_\perp)\approx
1/z_\perp^2 $, and the ensuing integral 
$(d^2z_\perp/z_\perp^2)$ has a logarithmic infrared 
divergence\footnote{This infrared behaviour is not modified by
the $z_\perp$ dependence of the Wilson lines since, e.g., 
$\langle V^\dagger_x V_z \rangle \to 0$ 
as $|z_\perp-x_\perp|\to \infty$; cf. Sect. 5.2 below.}.
Thus, there is potentially an IR problem in the RGE. This is
not necessarily a real difficulty, since IR problems are expected
to be absent only for the {\it gauge-invariant} observables. We shall
see indeed, on specific  examples, that the IR divergences cancel 
when the RGE is used to derive evolution equations for 
gauge-invariant quantities. This cancellation relies in a crucial 
way on the property (\ref{sigchi}).

Coming now to the ultraviolet, or short-distance,
behaviour, it is easy to see on eq.~(\ref{eta}) 
that no UV problem is to be anticipated.
For instance, the would-be linear pole of ${\cal K}(x_\perp,y_\perp,z_\perp)$
at $|z_\perp-x_\perp|\to 0$ is actually cancelled by the
factor $1- V^\dagger_z V_x$ which vanishes in the same limit.

\subsection{Quantum  evolution as Brownian motion}
\label{sec:FP}

To clarify the probabilistic interpretation of the RGE (\ref{RGEA}),
we start by recalling the
simplest example of a stochastic process, namely the Brownian
motion of a small particle in a viscous liquid 
and in the presence of some external force, like gravitation
\cite{ZJ}. The particle is
so small that it can feel the collisions with the molecules in
the liquid; after each such a collision, the velocity of the
particle changes randomly.
And the liquid is so viscous that, after each collision, the
particle enters immediately a constant velocity regime in which
the friction force $\propto v^i$ (with $v^i$ the velocity of the
particle) is equilibrated by the random force due to collisions 
together with the external force $F^i(x)$.
In these conditions, the particle executes a random walk whose
description is necessary statistical. The relevant quantity is
the  probability density $P(x,t)$ to find the particle at point 
$x$ at time $t$. This is normalized as:
\be\label{normP}
\int d^3x\,P(x,t)\,=\,1,\ee
and obeys an evolution equation of the diffusion
type, known as the Fokker-Planck equation \cite{ZJ} :
\be\label{FPBM}
{\del P(x,t)\over {\del t}}\,=\,D{\del^2\over \del x^i\del x^i}\,P(x,t)\,-\,
{\del\over \del x^i}\Bigl(F^i(x) P(x,t)\Bigr).\ee
Here, $D$ is the diffusion coefficient, which is a measure of the
strength of the random force; for simplicity, we 
assume this to be a constant, i.e., independent of $x$ or $t$.
The solution to eq.~(\ref{FPBM}) corresponding to some arbitrary
initial condition $P(x,t_0)$ can be written as
\be
P(x,t)\,=\,\int d^3x_0\,P(x,t|x_0,t_0)\,P(x_0,t_0),\ee
where $P(x,t|x_0,t_0)$ is the 
solution to (\ref{FPBM}) with the initial condition:
\be
P(x,t_0|x_0,t_0)\,=\,\delta^{(3)}(x-x_0).\ee
Physically, this is the  probability density to find the particle at point 
$x$ at time $t$ knowing that it was at $x_0$ at time $t_0$.

If $F^i=0$, this solution is immediately obtained by going to
momentum space: The Fourier transform $\tilde P(k,t)$ of 
$P(x,t_0|x_0,0)\equiv P(x-x_0,t)$ obeys to:
\be\label{FPBMk}
{\del\tilde P(k,t)\over {\del t}}\,=\,-Dk^2\,\tilde P(k,t),\qquad
\tilde P(k,t=0)\,=\,1,\ee
with the obvious solution $\tilde P(k,t)={\rm e}^{-Dk^2 t}$, or,
finally, 
\be\label{Diff0} P(x-x_0,t)\,=\,{1\over (4\pi Dt)^{3/2}}\,\,
{\rm e}^{-\frac{({ x-x_0})^2}{4Dt}}.\ee
This shows a purely diffusive behaviour: the probability to find
the particle within a fixed volume centered at some point $x$
goes smoothly to zero as $t\to \infty$ for any $x$
(runaway solution). The correlations of $x$ reflect this behaviour
too; for instance:
\be \label{r2}
\overline{r^2}(t)\,\equiv\,\langle ({ x-x_0})^2\rangle (t)\,\equiv\int d^3x\,
({ x-x_0})^2\,P(x-x_0,t)\,=\,6Dt,\ee
showing that, on the average, the particle gets further and further
away from the original point $x_0$, but along a non-differentiable
trajectory: $\bar r(t)\propto \sqrt{t}$, so the
average velocity $\bar v=\bar r(\Delta t)/\Delta t$ has no
well-defined limit when $\Delta t\to 0$.

This situation may change, however, if the motion of the particle
is biased by an external force. Assume this force to be derived from
a potential: $F^i=-{\del V/ \del x^i}$.
Then one can check that the time-independent distribution $P_0(x)\sim
{\rm exp}[-\beta V(x)]$ is a stationary solution to eq.~(\ref{FPBM})
provided $\beta D=1$. Of course, this solution is acceptable as a probability
density only if it is normalizable, which puts some constraints on the form 
of the potential. But assuming this to be the case, then $P_0(x)\sim
{\rm e}^{-\beta V}$ represents an equilibrium distribution which is
(asymptotically) reached by the system at large times \cite{ZJ}. 
Once this is done, all the correlations become independent of time (unlike
(\ref{r2})). This solution
is a ``fixed point'' in the functional space of 
all (acceptable) distributions.

Returning to our RGE (\ref{RGEA}), it should be clear by now
that this is a functional Fokker-Planck equation
which describes a random walk in the functional 
space of the colour fields $\alpha^a(x_\perp)$. In this equation,
$\eta$ plays the role of the ``diffusion coefficient'', 
while $\nu$ is like a ``force term'', although this identification
is somehow ambiguous since $\eta$ is itself a functional of $\alpha$, 
so its derivatives can generate other contributions to the
force term, as shown in eq.~(\ref{RGEAnew}).
(In the analogous problem of the Brownian motion, this would
correspond to a diffusion coefficient which depends on
$x$ and has a tensorial structure: $D\to D_{ij}(x)$.
This situation occurs, e.g., in the 
description of a random walk on a curved manifold \cite{ZJ}.)
In fact, it is more correct to identify the combination
${1 \over 2} ({\delta \eta/{\delta \alpha_\tau}}) -\nu\,$
as the effective  ``force term'', since the remaining second-order 
differential operator in eq.~(\ref{RGEAnew}) --- which describes
diffusion --- is then Hermitian and positive semi-definite.

A fixed point of the quantum evolution would be
a solution $W[\alpha]$ to eq.~(\ref{RGEA})
which is normalizable and independent of ``time'' $\tau$.
If such a solution existed, then the high energy limit of QCD
scattering would be trivial (at least, within the present approximations): 
At sufficiently high energies,
all the cross sections would become independent of energy
(recall that $\tau\sim \ln s$).
The relation (\ref{sigchi}) between the coefficients in the RGE
guarantees, however, that such a ``fixed point''
does not exist: The effective force 
in eq.~(\ref{RGEAnew}) vanishes, and the corresponding evolution
Hamiltonian (\ref{H}) is just a kinetic operator, which
describes pure diffusion. We thus expect gluon
correlations to keep growing with
$\tau\sim \ln s$ even at asymptotically large energies. In Sect. \ref{S-SOL},
we shall find approximate solutions to eq.~(\ref{RGEH}) which show
indeed such a behaviour \cite{SAT}.

\subsection{The Balitsky-Kovchegov equation}
\label{S-BK}

If $\langle O[\alpha] \,\rangle_\tau$ is any observable which
can be computed as an average over $\alpha\,$:
\be\label{OBSERV}
\langle O[\alpha] \,\rangle_\tau&=&
\int\,{\cal D}[\alpha]\,O[\alpha] \,W_\tau[\alpha],\ee
(cf. eq.~(\ref{COVclascorr})), then its evolution with $\tau$ 
is governed by the following equation:
\be\labe{evolO}
{\del \over {\del \tau}}\langle O[\alpha] \,\rangle_\tau&=&\int
{\cal D}\alpha\,O[\alpha] \,{\del W_\tau[\alpha] \over {\del \tau}}\nn
&=&\left\langle {1 \over 2}\,{\delta \over {\delta
\alpha_\tau^a(x_{\perp})} }\,\eta_{xy}^{ab}\,
{\delta \over {\delta \alpha_\tau^b(y_{\perp})}}\,O[\alpha]\right
\rangle_\tau\,,\ee
where, in writing the second line,
we have used eq.~(\ref{RGEH}) for $\del W_\tau/\del \tau$ and
then integrated twice by parts within the
functional integral over $\alpha$. 

Let us apply this to the 2-point function (\ref{Stau}) of the Wilson
lines in the fundamental representation.
We recall that, physically, this is the $S$-matrix element 
for dipole-hadron scattering (cf. Sect. \ref{sect-dipole}).
A straightforward calculation yields (see \cite{PI} for details):
\be\labe{evolV}
{\del \over {\del \tau}}\langle {\rm tr}(V^\dagger_x V_y)
\rangle_\tau&=&-{\alpha_s\over 2 \pi^2}\int d^2z_\perp
\frac{(x_\perp-y_\perp)^2}{(x_\perp-z_\perp)^2(y_\perp-z_\perp)^2 }\nn
&{}&\quad \times \left\langle N_c {\rm tr}(V^\dagger_x V_y)
- {\rm tr}(V^\dagger_x V_z){\rm tr}(V^\dagger_z V_y)\right\rangle_\tau.\ee
This is the equation originally obtained by Balitsky \cite{B},
within a quite different formalism : by an analysis of the quantum
evolution of the dipole itself.

Not that the above equation is not closed: It relates
the 2-point function to the 4-point function $\langle
{\rm tr}(V^\dagger_x V_z){\rm tr}(V^\dagger_z V_y)\rangle$.
One can similarly derive an evolution equation for the latter \cite{B}, but 
this will 
in turn couple the 4-point function  to a 6-point function, and so on.
That is, eq.~(\ref{evolV}) is just the first in an 
infinite hierarchy of coupled equations \cite{B}. 

A closed equation can still be obtained in
the large $N_c$ limit, in which the 4-point function 
in eq.~(\ref{evolV}) factorizes:
\be
\left\langle {\rm tr}(V^\dagger_x V_z)\,
{\rm tr}(V^\dagger_z V_y)\right\rangle_\tau
\longrightarrow 
\left\langle {\rm tr}(V^\dagger_x V_z)\right\rangle_\tau\,
\left\langle{\rm tr}
(V^\dagger_z V_y)\right\rangle_\tau\quad {\rm for}\,\,N_c\to\infty.
\nonumber\ee
Then eq.~(\ref{evolV}) reduces to a closed equation
for $S_\tau(x_{\perp},y_{\perp})=
\langle {\rm tr}(V^\dagger_x V_y)\rangle_\tau/N_c$ :
\be\labe{evolN}
{\del \over {\del \tau}} S_\tau(x_{\perp},y_{\perp})
&=&- {\alpha_s N_c\over 2 \pi^2}\int d^2z_\perp
\frac{(x_\perp-y_\perp)^2}{(x_\perp-z_\perp)^2(y_\perp-z_\perp)^2 } \nn
&{}&\qquad \times
\left\{S_\tau(x_{\perp},y_{\perp})-
S_\tau(x_{\perp},z_{\perp})S_\tau(z_{\perp},y_{\perp})\right\}.\ee
The same equation has been independently obtained by Kovchegov \cite{K} within
Mueller's dipole model \cite{AM3,AMCARGESE}. 
(See also Ref. \cite{B00} for another derivation.)

An important observation refers to the transverse kernel
in eqs.~(\ref{evolV}) or (\ref{evolN}): This is not the same
as the original kernel
${\cal K}(x_\perp,y_\perp,z_\perp)$, eq.~(\ref{Kxyz}), of the RGE.
Rather, this has been generated as
\be\label{decay}
{\cal K}(x_\perp,x_\perp,z_\perp)+
{\cal K}(y_\perp,y_\perp,z_\perp)- 2
{\cal K}(x_\perp,y_\perp,z_\perp)=
\frac{(x_\perp-y_\perp)^2}{(x_\perp-z_\perp)^2(y_\perp-z_\perp)^2 },
\nonumber\ee
and has the remarkable feature to 
show a better infrared behaviour than eq.~(\ref{Kxyz}):
When $z_\perp \!\gg\! x_\perp,\,y_\perp$, the kernel above
decreases like $(x_\perp-y_\perp)^2/z_\perp^4$, so its integral
over $z_\perp$ is actually finite.

There is currently a large interest in the solutions to eq.~(\ref{evolN}),
and significant progress has been achieved by combining
analytic and numerical methods \cite{LT99,K,B00,LL,LL01,GB01}.
The conclusions reached in 
this way are equivalent to those obtained from direct investigations
of the RGE (\ref{RGEA}) \cite{SAT,IIM} that we shall review in what follows.

\section{Approximate solutions to \\the Renormalization Group Equation}
\label{S-SOL}

\setcounter{equation}{0}

We shall now construct approximate solutions to the RGE (\ref{RGEH})
and study their physical implications \cite{SAT,IIM}.

\subsection{The  mean field approximation}
\label{MFA}

As compared to the standard diffusion equation (\ref{FPBM}), 
the main complication with the RGE (\ref{RGEH}) comes
from the fact that its kernel $\eta$ is itself dependent on $\alpha$. 
In this respect, eq.~(\ref{RGEH}) is similar
to the following diffusion equation:
\be\label{FPDIFF}
{\del P(x,t)\over {\del t}}\,=\,{\del\over \del x^i} D_{ij}(x)
{\del\over \del x^j}\,P(x,t),\ee
in which the diffusivities $D_{ij}(x)$ are allowed to depend
upon the position $x$ of the particle. This dependence makes
eq.~(\ref{FPDIFF}) difficult to solve in general (i.e., for some
arbitrary tensor field $D_{ij}(x)$). But since
$x$ is a random variable, with probability density $P(x,t)$, 
 a reasonable approximation is obtained by replacing
$D_{ij}(x)$ in eq.~(\ref{FPDIFF}) by its expectation value:
\be\label{barD}
D_{ij}(x)\longrightarrow \langle D_{ij}(x)\rangle(t)\equiv
\int d^3x\,\bar P(x,t)\,D_{ij}(x)\,\equiv\,\delta_{ij}\bar D(t),\ee
which is independent of $x$, but a function of time. We denote with a bar
quantities evaluated in this ``mean field approximation'' (MFA).
In particular, $\bar P(x,t)$ is itself related to $\bar D(t)$,
as the solution to the following approximate equation:
\be\label{FPMFA}
{\del\bar P(x,t)\over {\del t}}\,=\,\bar D(t)
{\del^2\over \del x^i\del x^i}\,\bar P(x,t).\ee
Thus, eq.~(\ref{barD}) is actually a {\em self-consistent} equation
for  $\bar D(t)$. Being homogeneous in $x$, eq.~(\ref{FPMFA}) is easily
solved by Fourier transform, as in eqs.~(\ref{FPBMk})--(\ref{Diff0}).
For the initial condition $\bar P(x,t=0)=\delta^{(3)}(x)$, one thus obtains:
\be\label{DiffBar}
\bar P(x,t)\,=\,{1\over (4\pi \xi(t))^{3/2}}\,\,
{\rm e}^{-\frac{x^2}{4\xi(t)}}\,,\qquad
\xi(t)\,\equiv\,\int_0^t dt'\, \bar D(t').\ee
By inserting this solution in eq.~(\ref{barD}), one can compute the
average there (as a functional of $\bar D(t)$), and then solve 
the self-consistent equation for $\bar D(t)$, thus completely
specifying the approximate solution (\ref{DiffBar}).

This is the strategy that we shall use to obtain approximate
solutions to the functional diffusion equation (\ref{RGEH}).
The corresponding MFA reads:
\be\label{RGE-MFA}
{\del \bar W_\tau[\alpha] \over {\del \tau}}\,=\,{1 \over 2}
\int_{x_\perp,y_\perp}\!\gamma_\tau(x_{\perp},y_\perp)\,
{\delta^2 \bar W_\tau[\alpha]\over {\delta
\alpha_\tau^a(x_{\perp})\delta \alpha_\tau^a(y_{\perp})}}\,, 
\ee
with
\be\label{gammatau}
\delta^{ab}\gamma_\tau(x_{\perp},y_\perp)\,\equiv\,
\langle \eta^{ab}(x_\perp,y_\perp)\rangle_\tau\,\equiv\,
\int D[\alpha]\,\eta^{ab}(x_\perp,y_\perp)\,\bar W_\tau[\alpha]\,
,\ee
where the trivial colour structure in the l.h.s.
follows from gauge symmetry. By the same argument, $\langle \nu^a(x_\perp)
\rangle_\tau=0$, which is indeed consistent with the MFA
(\ref{gammatau}) for $\eta$ and the condition (\ref{sigchi}).

Eq.~(\ref{RGE-MFA}) is homogeneous in the functional variable
$\alpha_{\rm y}^a(x_{\perp})$ (since its kernel 
$\gamma_\tau$ is independent of $\alpha$),
so it can be solved by functional Fourier analysis. This is
the straightforward extension of the corresponding
analysis for ordinary functions, and can be more rigourously introduced
by using a discretized version of the 3-dimensional configuration
space $({\rm y},x_\perp)$, as in eqs.~(\ref{measure}) or 
(\ref{fdelta}). We write:
\be\label{barW-F}
\delta[\alpha]&=&\int D[\pi]\,\,{\rm e}^{-i\int d{\rm y}
\int d^2x_\perp\,\pi_{\rm y}^a(x_{\perp})\alpha_{\rm y}^a(x_{\perp})}\,,\nn
\bar W_\tau[\alpha]&=&\int D[\pi]\,\,{\rm e}^{-i\int d{\rm y}
\int d^2x_\perp\,\pi_{\rm y}^a(x_{\perp})\alpha_{\rm y}^a(x_{\perp})}\,
\tilde W_\tau[\pi]\,.\ee
By inserting this representation for $\bar W_\tau[\alpha]$
in eq.~(\ref{RGE-MFA}), and using
\be
{\delta \over {\delta \alpha_\tau^a(x_{\perp})}} 
\int d{\rm y}
\int d^2z_\perp\,\pi_{\rm y}^c(z_{\perp})\alpha_{\rm y}^c(z_{\perp})
\,=\,\pi_{\tau}^a(x_{\perp}),\ee
one obtains the following equation for $\tilde W_\tau[\pi]$ (compare
to eq.~(\ref{FPBMk})):
\be
{\del \tilde W_\tau[\pi] \over {\del \tau}}\,=\,-\,{1 \over 2}
\int_{x_\perp,y_\perp}\!\gamma_\tau(x_{\perp},y_\perp)\,
\pi_{\tau}^a(x_{\perp})\pi_{\tau}^a(y_{\perp})\tilde W_\tau[\pi]\,,
\ee
with the immediate solution 
(transverse coordinates are omitted, for simplicity):
\be
\tilde W_\tau[\pi]\,=\,{\rm e}^{-{1 \over 2}
\int_0^\tau d{\rm y} \gamma_{\rm y}\pi_{\rm y}^a \pi_{\rm y}^a}\,
\tilde W_0[\pi]\,.\ee
The argument $\pi_{\rm y}$ of the initial weight function $\tilde W_0[\pi]$
has support only at ${\rm y} < 0$.
After insertion in eq.~(\ref{barW-F}), this yields:
\be\label{barW}
\bar W_\tau[\alpha]=\!\int\! D[\pi]\,\,{\rm e}^{-i
\int_{-\infty}^\infty d{\rm y}\,
\pi_{\rm y}^a\alpha_{\rm y}^a}\,\,{\rm e}^{-{1 \over 2}
\int_0^\tau d{\rm y}\, \gamma_{\rm y}\pi_{\rm y}^a \pi_{\rm y}^a}
=\delta[\alpha^>]\,\bar {\cal W}_\tau[\alpha^<],\ee
with $\delta[\alpha^>]$ defined in eq.~(\ref{fdelta}) (this has been
generated by the functional 
integral over $ \pi_{\rm y}$ with ${\rm y}>\tau$) and
\be\label{barW<}
\bar {\cal W}_\tau[\alpha^<]\,=\,{\cal N}_\tau\,
{\rm exp}\!\left\{-\,{1 \over 2}
\int_0^\tau d{\rm y}
\int_{x_\perp,y_\perp}\!\frac{\alpha_{\rm y}^a(x_{\perp})
\alpha_{\rm y}^a(y_{\perp})}{\gamma_{\rm y}(x_{\perp},y_\perp)}
\right\}\,{\cal W}_{0}[\alpha]\,.\ee
In this equation, ${\cal W}_{0}[\alpha]$ is the original weight 
function at $\tau=0$, and is a functional of the field
$\alpha_{\rm y}$ with ${\rm y} \le 0$. 
(${\cal N}_\tau$ is an irrelevant  normalization factor.)

The solution (\ref{barW})--(\ref{barW<})
has the general structure anticipated in
eqs.~(\ref{longitW})--(\ref{calW}). If the initial conditions
are described by the MV model, 
or any other MFA, then ${\cal W}_{0}[\alpha]$ is a Gaussian too
(see, e.g., eq.~(\ref{MV-W})), and eq.~(\ref{barW<}) can be rewritten as:
\be\label{MFA-W}
\bar {\cal W}_\tau[\alpha^<]\,=\,{\cal N}_\tau\,
{\rm exp}\!\left\{-\,{1 \over 2}
\int_{-\infty}^\tau d{\rm y}
\int_{x_\perp,y_\perp}\!\frac{\alpha_{\rm y}^a(x_{\perp})
\alpha_{\rm y}^a(y_{\perp})}{\gamma_{\rm y}(x_{\perp},y_\perp)}
\right\}\,.\ee
For ${\rm y} \le 0$, the width $\gamma_{\rm y}$ is specified by the 
initial conditions, while at positive rapidities $0 <{\rm y} \le\tau$,
it is determined by the quantum evolution, as we shall see.

The fact that the weight function (\ref{MFA-W}) is a Gaussian
does not necessarily mean that the present approximation describes a 
system of independent colour sources (like the MV model). 
It just means that, in the MFA, all the correlations are encoded in 
the width of the Gaussian, or, equivalently, in
the 2-point function 
\be\label{alphaHM}
\langle \alpha_{\rm y}^{a}(x_\perp)\,
\alpha_{{\rm y}'}^{b}(y_\perp)\rangle_\tau&=&\delta^{ab}\delta({\rm y}
-{\rm y}')\theta(\tau-{\rm y})\,\gamma_{\rm y}(x_\perp,y_\perp).\ee
But this 2-point function contains also
information on the higher-point correlations,
although just in an averaged way, because
 it is determined
by the following, {\em non-linear}, self-consistency equation:
\be\label{gammaSC}
\gamma_\tau(x_{\perp},y_\perp)&=&
{1\over \pi}\!\int \!{d^2z_\perp\over (2\pi)^2}\,
{\cal K}(x_\perp,y_\perp,z_\perp)\\
&{}&\quad\times\,\Big(1+S_\tau(x_\perp,y_\perp)
-S_\tau(x_\perp,z_\perp)-S_\tau(z_\perp,y_\perp)\Big),
\nonumber\ee
which follows from eqs.~(\ref{gammatau}) and (\ref{eta}) together
with the fact that, for a Gaussian weight function\footnote{Note that,
as compared to eq.~(\ref{Stau}), $S_\tau$ is now written
in the adjoint representation.},
\be\label{StauADJ}
\langle (V^\dagger_x V_y)^{ab}\rangle_\tau\,=\,
\frac{\delta^{ab}}{N_c^2-1}\Big\langle {\rm Tr} \,
\big(V^\dagger(x_{\perp}) V(y_{\perp})\big)\Big\rangle_\tau\,\equiv\,
\delta^{ab} S_\tau(x_\perp,y_\perp),\ee
with $S_\tau$ a (non-linear) functional of $\gamma_{\rm y}$,
to be constructed shortly.

The correlation function (\ref{alphaHM})
is local in y : colour sources located at different 
space-time rapidities appear to be statistically independent.
This is, of course, just an artifact of the MFA. The
complete RGE generates correlations in rapidity, via the Wilson
lines in its coefficients. But
the only trace of these correlations in the MFA is the fact
that the self-consistency equation (\ref{gammaSC}) is non-local in y.

To perform the average in eq.~(\ref{StauADJ}), we first derive
an evolution equation for $S_\tau$, by using the corresponding equation
(\ref{RGE-MFA}) for $\bar W_\tau\,$:
\be\label{RG-Stau}
{\del \over {\del \tau}}S_\tau(x_\perp,y_\perp)\!&=&\!
\int D[\alpha]\,
{\del \bar W_\tau[\alpha] \over {\del \tau}}\,V^\dagger_x V_y\\
\!&=&\! \int D[\alpha]\,  \bar W_\tau[\alpha]
\int_{u_\perp,v_\perp} {1 \over 2}\,\gamma_\tau(u_{\perp},v_\perp)\,
{\delta^2 \over {\delta
\alpha_\tau^a(u_{\perp})\delta \alpha_\tau^a(v_{\perp})}}\,V^\dagger_x V_y, 
\nn
\!&=&\! -{g^2N_c \over 2}\Big[\gamma_\tau(x_{\perp},x_\perp)
+ \gamma_\tau(y_{\perp},y_\perp) - 2\gamma_\tau(x_{\perp},y_\perp)\Big]
S_\tau(x_\perp,y_\perp).\nonumber \ee
(The functional
derivatives of the Wilson lines have been evaluated as
\be\labe{diff21}
{\delta^2 \over {\delta
\alpha_\tau^a(u_{\perp})\delta \alpha_\tau^a(v_{\perp})} }
{\rm Tr} (V^\dagger_x V_y)=-g^2N_c\,
{\rm Tr} (V^\dagger_x V_y)(\delta_{xv}-\delta_{yv})
(\delta_{xu}-\delta_{yu}),\,\,\,\,\,\,\ee
where we have used eq.~(\ref{DIFFU}) and $T^aT^a=N_c$.)
Eq.~(\ref{RG-Stau}) can be trivially integrated. To simplify
the calculations, we assume homogeneity in the transverse plane
within the hadron disk of radius $R\,$; then
$\gamma_{\rm y}(x_\perp,y_\perp) =
\gamma_{\rm y}(x_\perp-y_\perp)$ and 
\be\label{Stau-MFA}
S_\tau(r_\perp)={\rm e}^{-g^2N_c\int_0^\tau d{\rm y}
[\gamma_{\rm y}(0_\perp)-\gamma_{\rm y}(r_\perp)]}S_0(r_\perp)
={\rm e}^{-g^2N_c[\xi_\tau(0_\perp)
-\xi_\tau(r_\perp)]},\ee
where $r_\perp= x_\perp-y_\perp$,
\be\label{xitau}
\xi_\tau(r_\perp) \equiv \xi_0(r_\perp)+\int_{0}^{\tau}d{\rm y}
\,\gamma_{\rm y}(r_\perp),\ee
and in writing the second equality in (\ref{Stau-MFA}) we have assumed
that the initial condition $S_0(r_\perp)$ can be written in the
form $S_0(r_\perp)={\rm e}^{-g^2N_c[\xi_0(0_\perp)
-\xi_0(r_\perp)]}$. This is indeed the case for the weight function 
in eq.~(\ref{MFA-W}) --- in particular, for
the MV model, cf. eq.~(\ref{SA}) ---, which yields :
\be \label{xi0}
\xi_0(r_\perp)\,=\,
\int_{-\infty}^0 d{\rm y}\,\gamma_{\rm y}(r_\perp).\ee

By combining eqs.~(\ref{gammaSC}), (\ref{Stau-MFA}) and (\ref{xitau}),
one can finally rewrite the self-consistency equation
as an evolution equation for $\xi_\tau(r_\perp)$ :
\be\label{RG-xi}
{\del \xi_\tau(x_\perp-y_\perp)\over {\del \tau}}&=&
{1\over \pi}\!\int \!{d^2z_\perp\over (2\pi)^2}\,
{\cal K}(x_\perp,y_\perp,z_\perp)\\
&{}&\quad\times\,\Big(1+S_\tau(x_\perp-y_\perp)
-S_\tau(x_\perp-z_\perp)-S_\tau(z_\perp-y_\perp)\Big),
\nonumber\ee
with  $S_\tau(r_\perp)={\rm e}^{-g^2N_c[\xi_\tau(0_\perp)
-\xi_\tau(r_\perp)]}\,$. As anticipated, this equation is highly non 
linear in $\xi_\tau$. It is furthermore non-local in the transverse 
coordinates, but local in the ``evolution time'' $\tau$.
(The original non-locality of eq. (\ref{gammaSC}) in y
has been now absorbed in the relation  (\ref{xitau}) between 
$ \xi_\tau$ and $\gamma_{\rm y}$.)

In the next sections, we shall develop further approximations, 
which rely on the kinematics and allow us to make 
progress with  eq.~(\ref{RG-xi}).

\subsection{Saturation scale and Kinematical Approximations}

Both the non-local and the non-linear structure
of the evolution equation (\ref{RG-xi}) 
depend crucially upon the behaviour of $S_\tau(r_\perp)$
with the transverse separation $r_\perp$. 
From its definition (\ref{StauADJ}), it is clear that 
$S_\tau(r_\perp)\to 1$ as $r_\perp\to 0$ for any $\tau$.
Moreover, since a large dipole is strongly absorbed
by a hadronic target, we expect that $S_\tau(r_\perp)\ll 1$ for
sufficiently large $r_\perp$, where what we mean by 
``sufficiently large'' will generally depend on $\tau$. For instance,
we have seen in Sect. \ref{MV-SAT}, within the MV model, that
$S_\tau(r_\perp)\ll 1$ for $r_\perp\gg 1/Q_s$, with 
$Q_s$ the {\it saturation scale} for gluons in the hadron wavefunction
(cf. eq.~(\ref{BDL})).
In that classical model, $Q_s$ was independent of energy, but in 
general we expect it to increase with $\tau$, because of the quantum evolution
(cf. the discussion in Sect. \ref{SECT-DIS} and  Sect. \ref{MFA-HIGH}
below).
At a formal level, this intimate connection between the strong 
absorbtion limit for a colour dipole and gluon saturation is
based on the fact that, in both problems, the non-linear effects are 
encoded in Wilson lines. 
So, let us introduce the correlation length $1/Q_s(\tau)$ of
$S_\tau(r_\perp)$ :
\be\label{NT0}
S_\tau(r_\perp)\,\approx\,
\left\{ \begin{array} {c@{\quad\rm for\quad}l}
 1, & r_{\perp} \ll 1/Q_s(\tau) \\
 0, & r_{\perp} \gg 1/Q_s(\tau)
\end{array}
\right.
\ee
which, as its notation suggests, will play also the role of the
saturation scale. This behaviour of $S_\tau(r_\perp)$, with an
unique separation scale between a short-range regime and
a long-range one, is confirmed by numerical studies
of the Kovchegov equation, which also show a rapid increase
of $Q_s$ with $\tau$  \cite{LT99,B00,LL,GB01}. 

Eq.~(\ref{NT0}), together with the expression (\ref{Stau-MFA}) for $S_\tau(r_\perp)$
in the MFA, imply the following condition:
\be\label{QS-MFA}
g^2N_c[\xi_\tau(0_\perp)
-\xi_\tau(r_\perp)]\,\sim\,1\quad{\rm for}\quad r_\perp\sim 1/Q_s(\tau),\ee
that we shall use later to obtain an estimate for $Q_s(\tau)$.

An external probe with transverse momentum $k_{\perp}$ will measure
correlations in the hadron over a typical transverse size $r_{\perp}
\sim 1/k_{\perp}$. Thus,  short distances $r_{\perp} \ll 1/Q_s(\tau)$
correspond to high momenta, $k_{\perp} \gg Q_s(\tau)$,
while large separations $r_{\perp} \gg 1/Q_s(\tau)$ correspond to
low momenta $k_{\perp} \ll Q_s(\tau)$.
In what follows, we shall not aim at a precise description 
of the physics around the
saturation scale, but rather focus on the two limiting regimes
---  high--$k_{\perp}$ and low--$k_{\perp}$ --- and perform appropriate
simplifications on the evolution equation (\ref{RG-xi}). 

\bigskip
{\it a) High--$k_{\perp}\,$. } 
It is convenient to rewrite eq.~(\ref{Stau-MFA}) 
in momentum space as:
\be\label{Stau-MOM}
S_\tau(r_\perp)={\rm exp}\left\{-g^2N_c
\int \!{d^2p_\perp\over (2\pi)^2}\,\xi_\tau(p_\perp)\Bigl[1-
{\rm e}^{ip_\perp\cdot r_\perp}\Bigr]\right\}.\ee
For $r_{\perp} \ll 1/Q_s(\tau)$, the integral over $p_\perp$
is dominated by momenta 
within the range $Q_s(\tau) \!\ll\! p_\perp \!\ll \! 1/r_{\perp}$. This holds
to leading {\em transverse}--log accuracy: In this range,
$\xi_\tau(p_\perp)\sim 1/p_\perp^4$ (up to logs), so the integral
over $p_\perp$ produces the large logarithm 
$\ln({1}/{r_{\perp}^2Q_s^2(\tau)})$. To the same logarithmic accuracy,
one can expand the exponential in (\ref{Stau-MOM}) 
in powers of $p_\perp\cdot r_\perp$, like in eq.~(\ref {xi1}),
and thus obtain:
\be\label{Stau-MOM1}
S_\tau(r_\perp)\,\simeq\, {\rm exp}\Bigg\{-\frac{g^2N_c}{4}\,
r_\perp^2 \int\limits^{1/r_\perp^2} {d^2p_\perp\over (2\pi)^2}\,\,
p_\perp^2\,\xi_\tau(p_\perp)\Bigg\}.\ee
When extrapolated to $r_{\perp} \sim 1/Q_s(\tau)$, this expression
gives us the following estimate for the correlation
length $1/Q_s(\tau)$ (cf. eq.~(\ref{QS-MFA})) :
\be\label{QS-XI}
Q_s^2(\tau)\,\simeq\,\frac{\alpha_s N_c}{4}\int\limits^{Q_s^2(\tau)} dp_\perp^2\,\,
p_\perp^2\,\xi_\tau(p_\perp).\ee
For $r_{\perp} \ll 1/Q_s(\tau)$, where eq.~(\ref{Stau-MOM1}) is strictly
valid, the exponential there can be expanded to lowest order:
\be\label{Stau-high}
S_\tau(r_\perp)\,\simeq\, 1\,-\,\frac{g^2N_c}{4}\,r_\perp^2 \left(
-\grad_\perp^2 \xi_\tau(r_\perp)\right)_{r_\perp=0}.\ee
(The ultraviolet cutoff $1/r_\perp$ is implicit in the
momentum representation of $\xi_\tau(0)$.)
By inserting this into
(\ref{RG-xi}), we obtain a {\it linear} evolution equation 
for $\xi_\tau(r_\perp)\,$:
\be\label{RG-xi-high}
{\del \xi_\tau(x_\perp-y_\perp)\over {\del \tau}}&=&\alpha_s N_c
\!\int \!{d^2z_\perp\over (2\pi)^2}\,
\frac{(x^i-z^i)(y^i-z^i)}{(x_\perp-z_\perp)^2(y_\perp-z_\perp)^2 }\\
&{}&\quad\times\,\Big((x_\perp-y_\perp)^2-
(x_\perp-z_\perp)^2-(z_\perp-y_\perp)^2\Big)
\grad_\perp^2 \xi_\tau(0)\,.
\nonumber\ee
Thus, the short-distance approximation is automatically a linear,
or weak-field, approximation. This is to be expected since, at high
$k_\perp$, the gluon density is low.

To perform the integral over $z_\perp$ in eq.~(\ref{RG-xi-high}),
it is useful to recall eq.~(\ref{id0}) and then notice that,
within the integrand of (\ref{RG-xi-high}), one can effectively replace:
\be
\frac{1}{(2\pi)^2}\,
\frac{(x^i-z^i)(y^i-z^i)}{(x_\perp-z_\perp)^2(y_\perp-z_\perp)^2 }\,
\longrightarrow\, \frac{1}{2}\grad^2_z\left(
\langle x_\perp|\,\frac{1}{-\grad^2_\perp}\,|z_\perp\rangle\,
\langle y_\perp|\,\frac{1}{-\grad^2_\perp}\,|z_\perp\rangle\right).
\nonumber\ee
(The additional terms in the r.h.s. are $\delta$-functions 
at $z_\perp=x_\perp$ or $z_\perp=y_\perp$, which vanish when multiplied
by the remaining factor in (\ref{RG-xi-high}).) By using this, together
with a couple
of integrations by parts w.r.t. $z_\perp$, and a Fourier transform 
to momentum space, we finally obtain the following
evolution equation:
\be\label{delmu}
{\partial \mu_\tau(k_\perp)\over \partial \tau}\,=\,\frac{\alpha_s N_c}
{\pi} \int\limits^{k_\perp^2}\frac{dp_\perp^2}{p_\perp^2}\,
\mu_\tau(p_\perp)\,,\ee 
for the quantity:
\be\label{mu-def}
\mu_\tau(k_\perp)\,\equiv\,k_\perp^4\,\xi_\tau(k_\perp),\ee
which, physically, is the 2-point function of the colour charge
density in the transverse plane $\rho^a(x_\perp)$ :
\be\label{rho-MFA}
\langle \rho^a(x_\perp)\rho^b(y_\perp)\rangle_\tau=
\delta^{ab}\mu_\tau(x_\perp-y_\perp),\qquad 
\rho^a(x_\perp) =\int d{\rm y}\,\rho^a_{\rm y}(x_\perp).\ee
The initial condition for eq.~(\ref{delmu}) can be taken from the MV
 model: $\mu_\tau(k_\perp)= \mu_A$ for $\tau= 0$,
cf.  eq.~(\ref{MV-corr}). This initial condition is independent
of $k_\perp$ and, together with eq.~(\ref{delmu}), it implies
that $\mu_\tau(k_\perp)$ remains a rather slowly varying function
of $k_\perp$ in this high momentum regime. This will be manifest
on the solutions to eq.~(\ref{delmu}) that we shall write
in the next subsection.

\bigskip
{\it b) Low--$k_\perp\,$.} For large distances
$r_{\perp} \gg 1/Q_s(\tau)$, $S_\tau(r_\perp)\ll 1$,
and the 2-point functions of the Wilson lines can be simply neglected
in the self-consistency equations (\ref{gammaSC}) or (\ref{RG-xi})
\cite{SAT,IIM}.
Eq.~(\ref{RG-xi}) then simplifies to (see also eq.~(\ref{id0}))
\be\label{RG-xi-low}
{\del \xi_\tau(x_\perp-y_\perp)\over {\del \tau}}&\approx &\frac{1}{\pi}
\int \!{d^2z_\perp}\,\partial^i_z
\langle x_\perp|\,\frac{1}{-\grad^2_\perp}\,|z_\perp\rangle\,
\partial^i_z
\langle y_\perp|\,\frac{1}{-\grad^2_\perp}\,|z_\perp\rangle\,\nn
&=& \frac{1}{\pi} \langle x_\perp|\,\frac{1}{-\grad^2_\perp}\,|y_\perp\rangle,
\ee
or in momentum space (cf. eq.~(\ref{xitau})):
\be\label{RG-gamma-low}
\gamma_\tau(k_\perp)\,\equiv\,
{{\del \xi_\tau(k_\perp)}\over {\del \tau}}\,=\,
\frac{1}{\pi}\,\frac{1}{k_\perp^2}\,.\ee
This is not an equation anylonger, but rather an explicit, and rather
simple, expression for the propagator $\gamma_\tau(k_\perp)$ of the
fields $\alpha\,$: this is just the 2-dimensional Coulomb propagator.

Remarkably, the QCD coupling constant $g$ has
dropped out from eqs.~(\ref{RG-xi-low}) and (\ref{RG-gamma-low}).
(This should be contrasted with the corresponding equation at high $k_\perp$,
eq.~(\ref{RG-xi-high}), whose r.h.s. is explicitly proportional
to $\alpha_s=g^2/4\pi$.) The same property holds then
for the corresponding mean-field
Hamiltonian (cf. eq.~(\ref{RGE-MFA})):
\be\label{HLM}
\bar H_{{\rm low}-k_\perp}&\approx& -\,{1\over 2\pi }
\int
{d^2k_\perp\over (2\pi)^2}\,\,\frac{1}{k_\perp^2}\,
{\delta^2 \over {\delta
\alpha_\tau^a(k_{\perp})\delta \alpha_\tau^a(-k_{\perp})}},\ee
which is quite remarkable since at low $k_\perp$
 we are effectively in a strong coupling regime
(in the sense that the COV-gauge fields are strong: $\alpha^a\sim 1/g\,$;
see Sect. \ref{MFA-LOW}). If $g$ 
nevertheless drops out in this limit, it is because of
the special way it enters the evolution Hamiltonian: via
the exponent of the Wilson lines. That is, the relevant degrees of
freedom in the non-linear regime are not the (strong) 
colour fields by themselves, 
but rather the Wilson lines built with these fields. The Wilson lines
are rapidly oscillating over distances $r_\perp\gg 1/Q_s(\tau)$ (since 
their exponent is of order one, and the typical scale for variations is
$1/Q_s(\tau)$), and thus average to zero (``random phase approximation'').

\bigskip
For what follows, it is useful to summarize the previous kinematical
approximations into the following, factorized, form for the weight function
(\ref{MFA-W}), which is most conveniently written as a weight function 
for\footnote{This is the colour charge density
in the COV-gauge, but we omit the tilde symbol on $\rho$, to simplify writing.}
$\rho_{\rm y}^a(k_\perp)=k_{\perp}^2\alpha_{\rm y}^a(k_\perp)$ :
\be\label{W-KIN}
\bar {\cal W}_\tau[\rho]
&\approx& {\cal W}_\tau^{\,{\rm high}}[\rho]\,
{\cal W}_\tau^{\,{\rm low}}[\rho],\\
\label{W-low}
{\cal W}_\tau^{\,{\rm low}}[\rho]&\equiv&{\cal N}_\tau\,
{\rm exp}\!\Bigg\{-\,{\pi \over 2}
\int\limits_{-\infty}^\tau d{\rm y}
\int\limits^{Q_s({\rm y})}{d^2k_\perp\over (2\pi)^2}\,
\frac{\rho_{\rm y}^a(k_{\perp})
\rho_{\rm y}^a(-k_{\perp})}{k_\perp^2}
\Bigg\},\\
\label{W-high}
{\cal W}_\tau^{\,{\rm high}}[\rho] &\equiv& {\cal N}_\tau\,
{\rm exp}\!\Bigg\{-\,{1 \over 2}
\int\limits_{-\infty}^\tau d{\rm y}
\int\limits_{Q_s({\rm y})}{d^2k_\perp\over (2\pi)^2}\,
\frac{\rho_{\rm y}^a(k_{\perp})
\rho_{\rm y}^a(-k_{\perp})}{\lambda_{\rm y}(k_\perp)}
\Bigg\}.\ee
In writing this equation, we have separated, for each rapidity y,
the low-momentum ($k_{\perp}<Q_s({\rm y})$) modes of $\rho$ from the
high-momentum ($k_{\perp}>Q_s({\rm y})$) ones, we have used the 
approximation (\ref{RG-gamma-low}) for the width of the Gaussian at low 
momenta, and we have written
$\lambda_{\rm y}(k_\perp)
\equiv \partial \mu_{\rm y}(k_\perp)/\partial {\rm y}$,
with  $\mu_\tau(k_\perp)$ determined by eq.~(\ref{delmu}), at high momenta.
Note that the modes with $k_\perp\sim Q_s({\rm y})$ are not correctly 
described by the present approximations, but we shall assume
that they give only small contributions to the quantities to
be computed below.

\subsection{High--$k_\perp\,$: Recovering the perturbative evolution}
\label{MFA-HIGH}

We now consider the implications of eqs.~(\ref{delmu}) and
(\ref{W-high}) for the physics
at high transverse momenta $k_\perp \gg Q_s(\tau)$. To this aim, we compute the
gluon density (\ref{TPS}) in this low density regime, where one can use
the linear approximation ${\cal F}^{+j}(\vec k)\simeq
(ik^j/ k_{\perp}^2){\rho(\vec k)}$. 
The calculation is similar to that already performed in 
eqs.~(\ref{aaimom})--(\ref{MV-LOW}). Specifically, by using (cf.
eq.~(\ref{W-high})) :
\be\label{rhoHM}
\langle \rho_{\rm y}^{a}(x_\perp)\,
\rho_{{\rm y}'}^{b}(y_\perp)\rangle_\tau=\delta^{ab}\delta({\rm y}
-{\rm y}')\lambda_{\rm y}(x_\perp-y_\perp),
\quad \lambda_{\rm y}(r_\perp)=\frac{\partial \mu_{\rm y}(r_\perp)}
{\partial {\rm y}},\qquad\, \ee
one eventually obtains:
\be\label{N0-LOW}
{\cal N}_\tau(k_\perp)&\simeq&
\frac{N_c^2-1}{4\pi^3}\,\frac{\mu_\tau(k_\perp)}{ k_{\perp}^2}\,,
\\\label{N-LOW}
\x G(\x,Q^2)&\simeq&\frac{N_c^2-1}{4\pi}\,\,R^2 \int\limits^{Q^2}_{Q_s^2(\tau)}
\frac{dk_\perp^2}{ k_{\perp}^2}\,\mu_\tau(k_\perp)\,.\ee
Note the lower limit $Q_s(\tau)$ in the integral
giving $\x G(\x,Q^2)$ : for $Q^2\gg Q_s^2(\tau)$, and to
leading {transverse}--log accuracy, it is sufficient to consider
the contribution of the high--$k_\perp$ modes of $\rho$ to the
gluon distribution. 
We shall check later that the corresponding contribution of the modes with
$k_\perp\ll Q_s(\tau)$ is infrared finite, although subleading as
compared to eq.~(\ref{N-LOW}) \cite{SAT,IIM}. 
This cures the infrared problem that we have faced 
in the classical calculation of Sects. \ref{MVmodel}--\ref{MV-SAT}.

Physically, $\mu_\tau(k_\perp)$ 
plays the same role as $\mu_A$ in the MV model:
It measures the density
of the colour sources in the transverse plane,
and, in the linear regime at high--$k_\perp$,
it is also  proportional to the unintegrated gluon distribution:
$\mu_\tau(Q^2)\propto
\partial \,\x G(\x,Q^2)/\partial \ln Q^2$.
But unlike $\mu_A$, which is constant for a given atomic number $A$,
$\mu_\tau(k_\perp)$ has non-trivial dependences
upon both $\tau$ and $k_\perp$, as determined by its quantum
evolution according to eq.~(\ref{delmu}). The dependence on
$\tau$ describes the increase in the density of the colour sources
via soft gluon radiation. The dependence on $k_\perp$
corresponds in coordinate space to correlations in the transverse 
plane, which occur via the exchange of quantum gluons
(see Fig. \ref{CHISIG-LIN}.a).

Eq.~(\ref{delmu}) can be recognized
as the standard, linear evolution equation in the
double-logarithmic approximation (DLA) \cite{DGLAP},
i.e., in the limit in which BFKL and DGLAP coincide with each other.
(In this limit, only the first, ``real'', term must be retained
in the r.h.s. of the BFKL equation (\ref{BFKL}); for $k_\perp
\gg p_\perp$, this term reduces indeed to that in eq.~(\ref{delmu}).)
The emergence of DLA is natural, given the approximations 
performed in deriving eq.~(\ref{delmu}): we have kept only 
terms of leading-log accuracy in both $\tau=\ln(1/\x)$ 
(in the construction of the 
effective theory), and $\ln(k_\perp^2/Q^2_s(\tau))$
(in the short-range expansion at high $k_\perp$).

\begin{figure} 
\begin{center} 
\includegraphics[width=0.9\textwidth]
{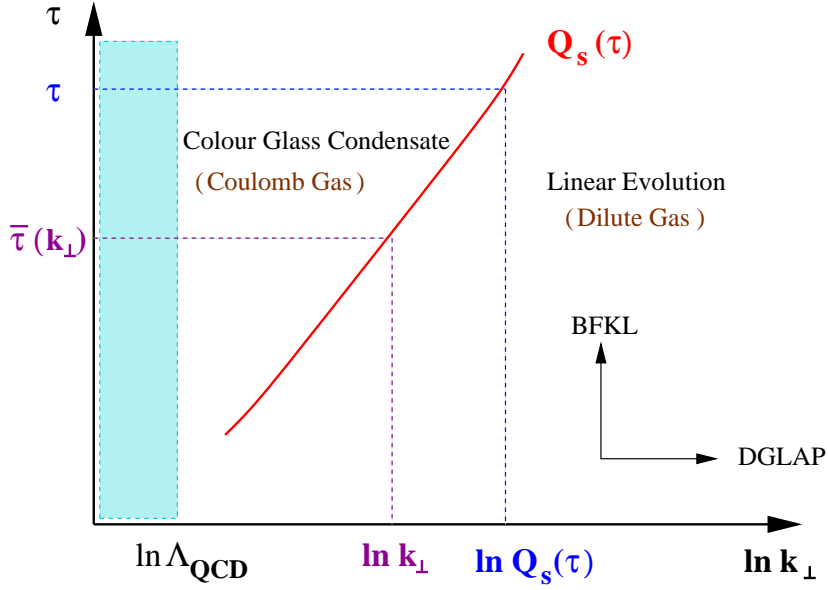} 
\caption{A ``phase-diagram'' of the various regions for evolution
in the $\tau-k_\perp$ plane.}
\label{phase-diagram} 
\end{center}
\end{figure}

Eqs.~(\ref{delmu}) and (\ref{N-LOW}) imply the 
 more standard form of the DLA equation \cite{DGLAP} :
\be\label{DL}
{\partial^2\over \partial \tau\,\partial\ln Q^2
}\,\x G(\x,Q^2)\,
=\,\frac{\alpha_s N_c}{\pi}\,\x G(\x,Q^2).\ee
At large $\tau$ and/or $Q^2$,
the solution to this equation increases like 
(with $\bar\alpha_s \equiv \alpha_s N_c/\pi$ and $Q_0^2$ some scale of reference)  \cite{DGLAP} 
\be\label{DLA}
\x G(\x,Q^2)\,\propto\,{\rm exp}\left\{2\sqrt{\bar\alpha_s\,
\tau\,\ln(Q^2/Q^2_0)}\right\},\ee
where we have assumed $\alpha_s$ to be independent of $Q^2$. If instead
one takes the running of the coupling into account,
by writing 
$\alpha_s(Q^2)=b_0 /\ln(Q^2/\Lambda^2_{QCD})$,
then the dependence of the solution upon $Q^2$ gets softer \cite{DGLAP} :
\be\label{DLA-RUN}
\x G(\x,Q^2)\,\propto\,{\rm exp}\left\{2\sqrt{b_0\,
\tau\ln\Big(\ln(Q^2/\Lambda^2_{QCD})\Big)}\right\}.\ee
In any case, eqs.~(\ref{DLA}) and (\ref{DLA-RUN}) show that, at
high transverse momenta $Q^2\gg Q_s^2(\tau)$, the gluon distribution 
$\x G(\x,Q^2)$ grows rapidly with $\tau$.
This is the standard picture
of parton evolution, which, if extrapolated to arbitrarily high energies,
would predict violations of the unitarity bound\footnote{Note that, 
although slower than for the BFKL solution (\ref{expgr}), the growth
with $\tau$ of the DLA solution (\ref{DLA}) or (\ref{DLA-RUN})
is still faster than that of any power of $\tau\sim \ln s$.}.
But from the previous analysis, we know that the approximations
leading to eq.~(\ref{DL}) will break down at sufficiently
large energies, where the non-linear
effects in the quantum evolution cannot be neglected anylonger.
Alternatively, for fixed rapidity $\tau$, the linear approximation
breaks down at low transverse momenta  $k_\perp\ll Q_s(\tau)$, with
$ Q_s(\tau)$ the saturation scale. An estimate for this scale has
been given in eq.~(\ref{QS-XI}), which, together with
eqs.~(\ref{mu-def}) and (\ref{N-LOW}), implies:
\be\label{Qsat}
Q^2_s(\tau)\,\simeq\,\frac{\alpha_s N_c}{4}\,
\int\limits^{Q^2_s(\tau)}\frac{dp_\perp^2}{p_\perp^2}\,\mu_\tau(p_\perp)\,
=\,\frac{\pi\alpha_s N_c}{N_c^2-1}\,\frac{1}{R^2}\,\x G(\x,Q^2_s(\tau)).\ee
By further combining this result
with eq.~(\ref{DLA}) or (\ref{DLA-RUN}), one can 
deduce the $\tau$-dependence of the saturation scale in the DLA. One 
thus obtains:
\be\label{Qstau}
Q^2_s(\tau)\,=\,Q^2_0\,{\rm e}^{\,4\bar\alpha_s \tau}\,,\qquad
({\rm fixed \,\,\,\,coupling}), \ee
and, respectively,
\be\label{Qstau-RUN}
Q^2_s(\tau)\,=\,\Lambda^2_{QCD}\, {\rm e}^{\,\sqrt{2b_0
\tau\ln\tau}}\,,\qquad
({\rm running\,\,\,\,coupling}). \ee
Eq.~(\ref{Qstau}) (or (\ref{Qstau-RUN})) defines a curve
in the $\tau-k_\perp$ plane, which divides this plane in two
(see Fig. ~\ref{phase-diagram}) : Points on
its right are effectively in the high momentum regime; they
correspond to a dilute gas of weakly correlated colour sources
whose density is rapidly increasing with $\tau$. Points on the
left of the saturation curve correspond to the low momentum regime,
to be discussed in the next subsection.

\subsection{Low--$k_\perp\,$: Coulomb gas and gluon saturation}
\label{MFA-LOW}

We finally turn to the most interesting physical regime, that of
the non-linear physics at small transverse momenta $k_\perp\ll Q_s(\tau)$
(with $k_\perp\gg \Lambda_{QCD}$, though), whose understanding was a 
main motivation for all the previous developments.

Within the effective theory, the low-momentum modes of the colour 
source are described by the weight function ${\cal W}_{\tau}^{\,{\rm low}}$,
eq.~(\ref{W-low}), which is equivalently rewritten as (cf. 
Fig. ~\ref{phase-diagram}):
\be\label{W-low1}
{\cal W}_{\tau}^{\,{\rm low}}
[\rho]\,=\,{\cal N}_\tau\,
{\rm exp}\!\Biggl\{-\,{\pi \over 2}
\int\limits^{Q_s^2(\tau)}{d^2k_\perp\over (2\pi)^2}\!
\int\limits_{\bar\tau(k_\perp)}^\tau \! d{\rm y}\,
\frac{\rho_{\rm y}^a(k_{\perp})
\rho_{\rm y}^a(-k_{\perp})}{k_\perp^2}
\Bigg\},\,\,\,\,\,\ee
with $\bar\tau(k_\perp)=$ the rapidity at which the
saturation momentum is equal to $k_\perp\,$:
\be\label{barQs}
Q_s^2(\bar\tau(k_\perp))\,=\,k_\perp^2\,.\ee
There are several noteworthy features about eq.~(\ref{W-low1}) :

i) This describes a {\it Coulomb gas}, i.e., a system of colour charges
interacting via long-range Coulomb forces. The colour source
$\rho_{\rm y}^a(x_{\perp})$ at $x_{\perp}$ feels the Coulomb field 
$\alpha_{\rm y}^a(x_{\perp})$ created at $x_{\perp}$ by all the other sources:
\be
\int_{k_{\perp}}\!\!\frac{\rho_{\rm y}^a(k_{\perp})
\rho_{\rm y}^a(-k_{\perp})}{k_\perp^2}=\!
\int_{x_{\perp},y_{\perp}}\! \!\!\rho_{\rm y}^a(x_{\perp}) 
\langle x_\perp|\,\frac{1}{-\grad^2_\perp}\,|y_\perp\rangle
\rho_{\rm y}^a(y_{\perp})=\!\int_{x_{\perp}}\!\!
\rho_{\rm y}^a(x_{\perp})\alpha_{\rm y}^a(x_{\perp}).\nonumber\ee
The fact that the charge-charge correlator appears to vanish when 
$k_{\perp}\to 0$ is in agreement with gauge symmetry: The 
colour source $\rho_{\rm y}^a(k_{\perp})$ at low $k_{\perp}$
is an {\it induced} source, whose global strength must vanish:
\be
\langle {\cal Q}^2\rangle \equiv
\int_{x_{\perp},y_{\perp}}\! \!\int_{{\rm y},{\rm y'}}
\langle \rho_{\rm y}^a(x_{\perp}) \rho_{\rm y'}^a(y_{\perp})\rangle 
\propto \langle \rho_{\rm y}^a(k_{\perp})
\rho_{\rm y}^a(-k_{\perp})\rangle\Big |_{k_{\perp}=0}=0.\ee
ii) The colour charge correlations are {\it local in rapidity} : the 
Coulomb forces couple only sources located 
in the same layer of y (or $x^-$).  At low--$k_\perp$, this property
is not just an artifact of the MFA, but rather has a deep physical
meaning: In the quantum evolution, the colour sources at different 
rapidities get correlated with each other because of the presence of Wilson
lines in the evolution Hamiltonian (\ref{H}). But these correlations
are washed out on a large scale $r_\perp \gg 1/ Q_s(\tau)$, on
which the Wilson lines average to zero. In particular, this 
explains why the width $\propto k_\perp^2$ of the
Gaussian (\ref{W-low1}) is independent of the initial conditions
at ${\tau}\simeq 0$. (By contrast, at high momenta, the width 
$\lambda_{\rm y}(k_\perp)= 
\partial \mu_{\rm y}(k_\perp)/\partial {\rm y}$ in eq.~(\ref{W-high})
is sensitive to the initial conditions, since
determined by solving eq.~(\ref{delmu}).)

iii) According to eq.~(\ref{W-low1}), the low-momentum modes
of $\rho$ are {\it uniformly} distributed in rapidity, within the
interval $\bar\tau(k_\perp) < {\rm y} <\tau$. It follows that
the integrated quantity:
\be\label{mu-sat}
\mu_\tau(k_\perp)=
\int\limits_{\bar\tau(k_\perp)}^\tau \! d{\rm y \,} \,\frac{k_\perp^2}{\pi}\,
= \Big(\tau-\bar\tau(k_\perp)\Big)\frac{k_\perp^2}{\pi}\,,\ee
which measures the density of sources (with given $k_\perp$)
in the transverse plane,
grows only {\it linearly} with $\tau$, that is, logarithmically with
the energy. This is to be contrasted with the strong, quasi-exponential,
increase of $\mu_\tau(k_\perp)$
in the high-momentum regime (cf. eqs.~(\ref{DLA}) and (\ref{DLA-RUN})).
We conclude that, at low momenta $k_\perp\ll Q_s(\tau)$, the colour
sources {\it saturate}, because of the strong non-linear effects
in the quantum evolution. 

iv) The saturated sources form the outermost layers of the hadron 
in the longitudinal direction: for given $k_\perp$, they
are located at $x^-\ge x^-_0{\rm e}^{\bar\tau(k_\perp)}$. In particular,
\be\label{tbt}
\tau\,-\,\bar\tau(k_\perp)\,\simeq\,{1\over 4\bar\alpha_s}\,
\ln{Q_s^2(\tau)\over k_\perp^2}\,,\ee
is the longitudinal extent of the saturated part of the hadron,
in units of rapidity (for modes with tranverse momentum $k_\perp$). 
In writing (\ref{tbt}), we have used the DLA
estimate (\ref{Qstau}) for the $\tau$-dependence of the saturation 
scale.

v) Note the factor $1/\alpha_s$ in the r.h.s. of (\ref{tbt}); 
this implies that,  at saturation, 
the {\it integrated} charge density $\rho^a(x_\perp)$ 
has typically large amplitudes: $\bar\rho \sim \sqrt{\langle \rho\rho\rangle}
\sim 1/g$. The same is therefore true for the COV-gauge 
field $\alpha^a(x_\perp)$ : $\bar\alpha \sim 1/g$.

\bigskip
Since the colour sources at low--$k_\perp$ are
saturated, there should be no surprise that the gluons  emitted
by these sources are saturated as well, and this independently of 
their mutual
interactions (i.e., of the non-linear effects in the classical 
Yang-Mills equations).
Indeed, a quasi-Abelian calculation of the gluon distribution,
based on the linearized solution ${\cal F}^{+j}(k)\approx
(ik^j/k_\perp^2)\rho$, yields the following gluon density (cf.
eqs.~(\ref{N0-LOW}) and eq.~(\ref{mu-sat})) :
\be\label{SAT}
{\cal N}_\tau(k_\perp)\,\simeq\,
{N_c^2-1\over 4 \pi^4 c}\,
\Bigl( \tau-\bar\tau(k_\perp)\Bigr)\,\simeq\,
{N^2_c-1\over 16 \pi^4 c}\,{1\over \bar\alpha_s}\,
\ln{Q_s^2(\tau)\over k_\perp^2}\,,\ee
which already exhibits saturation ! In fact, as argued 
in Refs. \cite{SAT}, 
the only effect of the non-linearities in the classical Yang-Mills
equations in this low--$k_\perp$ regime is to modify the overall
normalization of the linear-order result. In anticipation of this,
we have inserted in eq.~(\ref{SAT}) a corrective factor $c$, which
cannot be accurately determined in the present approximations
(since sensitive to the physics around $Q_s$), but should 
be smaller than one (although not much smaller).

\begin{figure} 
\begin{center} 
\includegraphics[width=0.9\textwidth]
{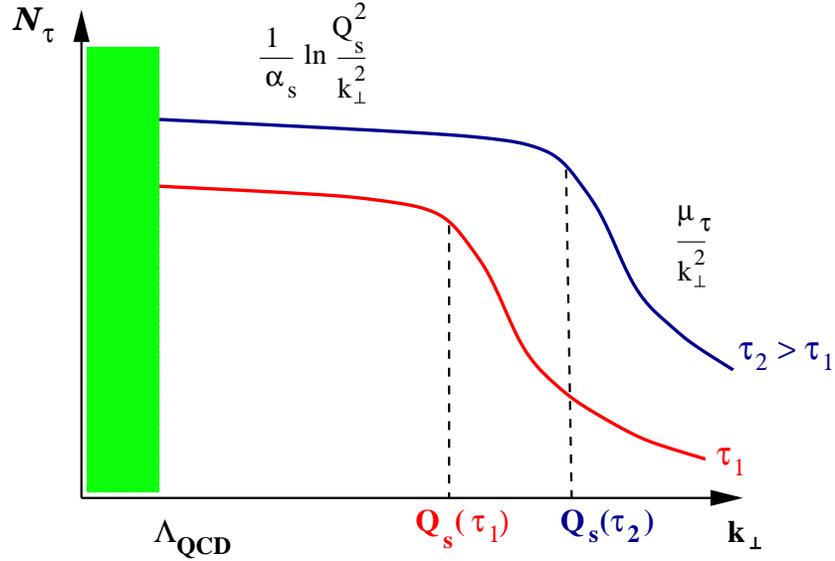} 
\caption{The gluon phase-density ${\cal N}_\tau(k_\perp)$
in the effective theory plotted as a function of $k_\perp$ for
two values of $\tau=\ln(1/\x)$.}
\label{SAT-RGE} 
\end{center}
\end{figure}

Note the striking similarity between eq.~(\ref{SAT}) and the
corresponding prediction (\ref{Low-N}) of the classical MV model.
Despite of the differences in the physical mechanism leading to
saturation --- non-linear quantum evolution for eq.~(\ref{SAT}),
as opposed to non-linear classical dynamics for
eq.~(\ref{Low-N}) ---, the final results look very much the same.
So, the earlier discussion of eq.~(\ref{Low-N}) can be immediately
adapted to eq.~(\ref{SAT}), after replacing $A\to s$ :
Eq.~(\ref{SAT}) shows {\it marginal saturation} (in the sense
of a logarithmic increase only) with both $s$ and $1/k_\perp^2$,
with a typical amplitude of order $1/\alpha_s$. This is illustrated
in Fig. \ref{SAT-RGE}, which should be compared to Fig. 
\ref{SATURATION-MV}. (The high--$k_\perp$ behaviour in
Fig. \ref{SAT-RGE} is taken from eq.~(\ref{N0-LOW}).)

\bigskip
Aside from saturation, eq.~(\ref{SAT}) has also other 
important consequences, which
all reflect the proportionality to the rapidity 
window\footnote{These properties are therefore generic:
They hold for any quantity which receives his dominant contributions
from the saturated gluons.}
$\tau-\bar\tau(k_\perp)$, eq.~(\ref{tbt}):

{\it a) Scaling.} The gluon density at saturation
depends upon the energy $s$
and the transverse momentum $k_\perp$ only via the scaling variable
\be
{\cal T} \equiv {Q_s^2(\tau)/k_\perp^2}\,.\ee
A similar scaling is observed in the solutions to the
Kovchegov equation \cite{LT99,Lub01,GB01}.
As mentioned in the Introduction, such a
scaling has been actually observed in DIS at HERA \cite{gb}.

{\it b) Universality}. Eq.~(\ref{SAT})
is only weakly sensitive --- via its logarithmic dependence
upon the saturation scale ---
 to the initial conditions for quantum evolution, 
and therefore to the specific properties of the hadron under 
consideration (e.g., its size and atomic number).
Thus, eq.~(\ref{SAT}) not only provides arguments in the favour
of hadron universality at high energy, 
but also predicts what should be the pattern of its violation.

The gluon distribution $\x G(\x,Q^2)$ at $Q^2\ll Q_s^2(\tau)$
is immediately obtained by integration in eq.~(\ref{SAT}):
\be\label{GDSAT}
\x G(\x,Q^2)
&\simeq&{N^2_c-1\over 16 \pi^3 c}\,{1\over \bar\alpha_s}
\,\,\pi R^2 \int\limits^{Q^2}_{0}
{dk_\perp^2}\,\ln{Q_s^2(\tau)\over k_\perp^2}\nn
&=&{N^2_c-1\over 16 \pi^2 c}\,{1\over \bar\alpha_s}\,
R^2 Q^2\left[\ln(Q_s^2(\tau)/Q^2)+1\right].\ee
Note that, since $\mu_\tau(k_\perp)\sim k_\perp^2$ in the
saturation regime (cf. eq.~(\ref{mu-sat})), the above integral 
is almost insensitive to the soft modes $k_\perp\simle \Lambda_{QCD}$.
This has allowed us to extend the integration
down to $k_\perp =0$ without loss of accuracy.
As anticipated, the phenomenon of saturation reduces the sensitivity
of physical quantities to the infrared gauge fields,
thus making the weak coupling expansion reliable.
(In Ref. \cite{GB01} a similar conclusion is drawn on the basis
of Kovchegov equation.)
If extrapolated up to $Q\sim Q_s$,  eq.~(\ref{GDSAT}) yields
\be\label{GD-Qs}
\x G(\x,Q_s^2(\tau))\,\simeq\,
{N^2_c-1\over 16 \pi^2 c}\,{1\over \bar\alpha_s}\,
R^2 Q_s^2(\tau),\ee
in rough agreement with the corresponding extrapolation
from the high momentum regime, eq.~(\ref{Qsat}). 
Eq.~(\ref{GD-Qs}) gives also the contribution of the saturated
modes to the gluon distribution at momenta $Q> Q_s(\tau)$.
But for very high momenta, $Q\gg Q_s(\tau)$, the dominant contribution
comes form the hard modes ($Q_s\ll k_\perp\simle Q$), and is given
by eq.~(\ref{N-LOW}). 

As a final application, let us compute the 2-point function
$S_\tau(r_\perp)$ of the Wilson lines for large distances
$r_{\perp} \gg 1/Q_s(\tau)$. This is interesting for at least two
reasons: It shows how the unitarity limit is reached for the scattering
of a large colour dipole off the hadron, and it allows us to check a
posteriori the consistency of the ``random phase approximation'' that
we have used at low $k_\perp$.

To this aim, we rewrite  eq.~(\ref{Stau-MOM}) as
\be\label{Stau-low}
S_\tau(r_\perp)\simeq {\rm exp}\Bigg\{-\frac{g^2N_c}{\pi}
\int\limits_{-\infty}^\tau d{\rm y}
\int\limits^{Q_s({\rm y})}
 {d^2p_\perp\over (2\pi)^2}\,\frac{1}{p_\perp^2}\,\Bigl[1-
{\rm e}^{ip_\perp\cdot r_\perp}\Bigr]\Bigg\},\ee
where we have anticipated that the main contribution comes
from the saturated modes, for which $\gamma_\tau=1/(\pi p_\perp^2)$,
cf. eq.~(\ref{RG-gamma-low}). The integral over $ p_\perp$ is now
infrared finite (as opposed to the MV model: compare to
eq.~(\ref{xi1})), and to leading log accuracy can be evaluated as:
 \be \int\limits^{Q_s({\rm y})} 
{d^2p_\perp\over (2\pi)^2}\,\frac{1}{p_\perp^2}\,\Bigl[1-
{\rm e}^{ip_\perp\cdot r_\perp}\Bigr]\,\simeq\,
\theta({\rm y} -\bar\tau(r_\perp))\,
{1\over 4\pi}\ln\Bigl(Q_s^2({\rm y})r_\perp^2\Bigr).\ee
The result can be understood as follows: as long as 
$1/r_\perp\gg Q_s({\rm y})$, or ${\rm y}<\bar\tau(r_\perp)$,
${\rm e}^{ip_\perp\cdot r_\perp}\approx 1$, and the integral
vanishes. Bur for ${\rm y}>\bar\tau(r_\perp)$, or
$1/r_\perp\ll Q_s({\rm y})$, the integrals corresponding to
the two terms in the brackets are cut off at different
 ultraviolet scales: $Q_s({\rm y})$  for the
first term, and $1/r_\perp$ for the second one.
Their difference gives the log in the r.h.s.
By also using $\ln(Q_s^2({\rm y})r_\perp^2)=4\bar\alpha_s({\rm y}
-\bar\tau(r_\perp))$, cf. eq.~(\ref{tbt}), and performing
the integral over y, we finally deduce:
\be\label{VVLM}
S_\tau(r_\perp)\simeq \,{\rm exp}\Bigl\{
-2\bar\alpha_s^2(\tau-\bar\tau(r_\perp))^2\Bigr\}\,=\,
{\rm exp}\bigg\{-{1\over 8}\Bigl[\ln
(Q_s^2(\tau)r_\perp^2)\Bigr]^2\bigg\},\ee
which coincides with the result obtained from the
Kovchegov equation \cite{LT99,AMCARGESE}. Eq.~(\ref{VVLM})
shows that the correlator of the Wilson lines is rapidly
decreasing when $Q_s^2(\tau)r_\perp^2\gg 1$, so that the
RPA is indeed justified, at least as a mean field approximation.

More details and further applications of the mean field approximation
will be presented in Ref. \cite{IIM}, where the results obtained in
this way will be also compared to the corresponding predictions of
the Kovchegov equation. It would be also interesting (especially in
view of applications to phenomenology) to take into account
the transverse inhomogeneity of the hadron (i.e., the dependence
upon the impact parameter in the transverse plane). This can be
done already in the framework of the MFA, but, more generally,
it would be important to understand the limitations of the latter, 
and to be able to solve the complete RGE. This might be done,
for instance, via numerical simulations on a lattice.

\section{Acknowledgments}

We would like to thank our colleagues Alejandro Ayala-Mercado,
Jean-Paul Blaizot, Elena Ferreiro, Kazunori Itakura, Dima Kharzeev,
Yuri Kovchegov, Alex Kovner, Jamal Jalilian-Marian, Genya Levin, Al Mueller,
Robi Peschanski,
Raju Venugopalan and Heribert Weigert with whom many of the ideas presented
in these lectures were developed, or discussed.  
We particularly thank Kazunori Itakura and Michaela Oswald for
a careful and critical reading of the manuscript.

This manuscript has been authorized under Contract No. DE-AC02-98H10886 with
the U. S. Department of Energy.

\end{document}